\documentclass{article}
\usepackage{authblk}
\usepackage[title]{appendix}
\usepackage{amsmath,amsfonts}

\usepackage{tcolorbox}
\usepackage{lmodern}
\usepackage{subfigure}
\usepackage{caption}
\usepackage{cite}
\usepackage{xparse}
\usepackage{graphicx}
\usepackage{float} 
\usepackage{subfigure}
\newcommand{\bunit}{\mathbf{b}_0}
\newcommand{\hdelta}{\hat{\delta}}
\NewDocumentCommand{\gpgc}{o}{
    \IfNoValueTF{#1}
    {\left\langle\phi\right\rangle}
    {\left\langle\phi_{#1}\right\rangle}
}
\newcommand{\avperpgc}{\left\langle\delta\mathbf{A}_{\perp}\cdot\mathbf{v}_\perp\right\rangle}
\newcommand{\savg}[1]{\left\langle\left\{S_1,#1\right\}\right\rangle}

\newcommand{\dbeq}[1]{\delta\mathbf{B}_{eq,#1}}
\newcommand{\curlBu}[1]{\left(\nabla\times\mathbf{B}_0\right)^{#1}}
\newcommand{\curlBd}[1]{\left(\nabla\times\mathbf{B}_0\right)_{#1}}
\newcommand{\phibparaangle}[1][]{\left(\langle\phi\rangle_{#1}+\frac{\mu}{Z_s}\langle\langle\delta B_\parallel\rangle\rangle\right)}
\newcommand{\phibparaanglept}{\left(\langle\delta\phi\rangle+\frac{\mu}{Z_s}\langle\langle\delta B_\parallel\rangle\rangle\right)}

\newcommand{\phibpara}[1]{\left(\phi_{#1}+\frac{\mu}{Z_s}\delta B_\parallel\right)}
\newcommand{\phibparaele}[1]{\left(\delta\phi_{#1}-\frac{\mu}{e}\delta B_\parallel\right)}

\usepackage{geometry}
\usepackage{esint}
\author[1]{Xishuo Wei\thanks{xishuow@uci.edu}}
\author[2]{Pengfei Liu}
\author[3]{Gyungjin Choi}
\author[4] {Guillaume Brochard}
\author[2] {Jian Bao}
\author[5] {\\Javier H. Nicolau}
\author[6]{Yuehao Ma}
\author[7]{Haotian Chen}
\author[1]{Handi Huang}
\author[7]{Shuying Sun}
\author[1]{\\Yangyang Yu}
\author[1]{Ethan Green}
\author[1]{Fernando Eizaguirre}
\author[ ]{Zhihong Lin\thanks{zhihongl@uci.edu} $^1$}

\affil[1] {Department of Physics and Astronomy, UC Irvine, CA 92617, USA}
\affil[2] {Institute of Physics, Chinese Academy of Sciences, Beijing 100190, China}
\affil[3] {Department of Nuclear and Quantum Engineering, KAIST, Daejeon 34141, \authorcr Korea, Republic Of}
\affil[4] {Aix-Marseille Université, CNRS PIIM, UMR 7345, Marseille, France}
\affil[5] {San Diego Supercomputer Center, University of California, San Diego, \authorcr CA 92093, USA}
\affil[6] {School of Nuclear Sciences and Technology, University of Science and Technology \authorcr of China, Hefei 230026, China}
\affil[7] {Fusion Simulation Center, Peking University, Beijing 100871, China}
\title{Formulation and verification of multiscale gyrokinetic simulation of kinetic-MHD processes in toroidal plasmas}
\makeatletter
\usepackage{titling}
\usepackage{xcolor}
\usepackage{hyperref}
\hypersetup{colorlinks=true,linkcolor=red,citecolor=blue,urlcolor=blue}
\makeatother

\begin{document}
\bibliographystyle{unsrt}
\maketitle
\begin{abstract}

A comprehensive gyrokinetic simulation model has been implemented in the global toroidal gyrokinetic code (GTC) and verified for studying low-frequency waves and turbulence in magnetic fusion plasmas by treating all kinetic-MHD processes on an equal footing. A theoretical framework has been formulated to unify various methods for efficiently solving the electron drift kinetic equation in multiscale simulations by separating electron responses into analytic and non-analytic parts based on the smallness parameter of electron-to-ion mass ratio. The model can be reduced to the ideal MHD model with both the linear dispersion relation and the nonlinear ponderomotive force in theory and simulation. The model is used for the verification and validation of simulating internal kink modes in the DIII-D tokamak with accurate calculations of equilibrium parallel current and compressible magnetic perturbation. A large simulation database has been generated to train a surrogate model to predict the kink instability. Statistical analysis shows that the radial location of safety factor q=1 flux-surface, the pressure gradient, the minimum q value, and the plasma beta inside the q=1 surface are the most important parameters for predicting the kink instability.
\end{abstract}

\section{Introduction}

Gyrokinetic particle simulation model\cite{Lee1983,WWLee1987} was first developed as a comprehensive and efficient tool for realistic simulations of low-frequency (below cyclotron frequency) microscopic driftwave turbulence\cite{Tang_1978,Horton1999} in magnetized plasmas. With the development of low-noise perturbative ($\delta f$) simulation method\cite{Dimits1993,Parker1993} and the introduction of massively parallel computing to fusion simulation\cite{Lin_1998}, gyrokinetic simulations have significantly advanced the fundamental understanding of microturbulence that leads to the paradigm of the turbulence self-regulation by zonal flows\cite{Lin_1998,Diamond_2005}. Validated for quantitative prediction of turbulent transport in fusion plasmas\cite{Garbet_2010}, gyrokinetic simulation is an indispensable tool for studying turbulent transport in magnetic fusion plasmas using many particle codes including GTC\cite{Xiao2015}, GEM\cite{CHEN2007GEM}, XGC\cite{Ku2018XGC}, Orb5\cite{JOLLIET2007ORB5}, Euterpe\cite{KLEIBER2024EUTERPE}, and NLT\cite{SJWang2024NLT}, continuum codes including GYRO\cite{CANDY2003GYRO},{CGYRO\cite{CANDY2016CGYRO}}, GENE\cite{Jenko2000GENE}, GT5D\cite{IDOMURA2008GT5D}, GKV\cite{Watanabe_2006}, GKW\cite{PEETERS2009GKW}, and GKNET\cite{OBREJAN2017GKNET},{GS2\cite{KOTSCHENREUTHER1995GS2}, stella\cite{BARNES2019STELLA}}, and semi-Lagrangian code GYSELA\cite{GRANDGIRARD2006GYSELA}.

In addition to turbulent transport that determines thermal plasma confinement, the confinement of energetic particles (EPs)\cite{Salewski_2025} is also a critical physics issue in burning plasmas. These EPs need to be well-confined to heat the thermal plasmas to sustain ignition and avoid damage to the reactor wall. However, they can excite low-frequency meso-scale Alfv\'en eigenmodes (AE) and macroscopic MHD instabilities that induce the loss of EPs\cite{ChenZona2016RMP_AE}. Cross-scale interactions among microturbulence, AE, and MHD can be mediated by self-generated zonal flows \cite{Bass2010PoP, Liu2022PRL, Brochard2024PRL} and phase-space zonal structures\cite{Zonca_2015}, EPs' scatterings by microturbulence\cite{ZhangWL2008PRL}, and mode couplings\cite{Chen_2023}. These cross-scale interactions ultimately determine the overall performance of the fusion reactors\cite{Na2025Nature} and require multiscale simulation models that treat all kinetic-MHD processes on an equal footing based on the nonlinear gyrokinetic formulation\cite{BrizardHahm_2007}. Such multiscale, multiphysics gyrokinetic simulations are computational grand challenges because of the immense ranges of spatial and temporal scales due to the small electron-to-ion mass ratio. Thanks to the interdisciplinary collaborations, multiscale gyrokinetic simulation has now been developed in the global GTC code by incorporating physical effects typically neglected in the simulations of microturbulence but which are important for the simulations of AEs and MHD instabilities including electron drift kinetic equation with equilibrium currents\cite{Lin_Chen_2001,Holod_2009,Deng_2012, Bao_2017} and compressible magnetic perturbations \cite{Dong_2017}, by implementing efficient numerical methods such as global field-aligned mesh\cite{Lin2002PRL, Ethier_2005} in realistic toroidal geometry \cite{Xiao2015}, and by efficiently utilizing exascale computers\cite{ZhangWL2019}. 

These multiscale gyrokinetic simulations have been rigorously verified and validated using GTC\cite{WangZX2013PRL, Taimourzadeh_2019, Brochard_2022, Liu2022PRL, Brochard2024PRL} {and other gyrokinetic codes\cite{EUTERPE_2025_GK_IN_STELLARATOR, MISHCHENKO2019, Startsev_GTS,GENE_TEM_NTM}}, which provide a powerful tool for the prediction and optimization of the burning plasma experiments such as ITER\cite{LinIAEA2023}. Recently, GTC has been used to simulate the MHD modes, including internal kink\cite{Brochard_2022} driven by the equilibrium current and fishbone \cite{Brochard2024PRL} excited by the EPs in the DIII-D tokamak, and compared with the hybrid-MHD codes GAM-solver\cite{BaoGAMSolver}, M3D-C1\cite{JARDIN2004133}, NOVA\cite{GorelenkovNOVA}, XTOR-K\cite{LUTJENSXTOR}. Simulations from all codes agree well in the long wavelength limit and agree reasonably with experimental measurements. Building on these verification and validation, GTC has been extensively utilized to study instability, turbulence, and transport using realistic geometry and parameters with real electron mass in magnetic fusion experiments including DIII-D\cite{Liu_2024}, NSTX-U\cite{Singh_2025}, ITER\cite{Brochard2025_NF}, JET\cite{Fil2021}, EAST\cite{Ma_2023}, KSTAR\cite{Wei2025_NF}, HL-2A \cite{Li2024}, MAST\cite{Wong2024}, ADITYA-U \cite{Singh_2024_ADITYA}, and ST40 \cite{Huang2025} tokamaks, W7-X\cite{Nicolau_2025} and LHD\cite{WangHY2020,Singh_2022} stellarators, and C2 field-reversed configuration\cite{WangWH2024}.

In this paper, we describe the comprehensive physics model implemented in the GTC for multiscale simulations of low-frequency kinetic-MHD processes in magnetized plasmas in a general 3D geometry. We present a single framework with various methods for solving the electron drift kinetic equation (DKE) developed in \cite{Chen_2001,Lin_2007,Holod_2009,Bao_2017}. The electron fluid continuity equation, parallel Ampere's law, and the parallel electron momentum equation are used to solve electron density, electron parallel flow velocity, and the parallel vector potential. The electron kinetic effect only appears in the pressure tensor in the momentum equation. Different definitions of the analytic part of electron responses can lead to the \textit{hybrid} scheme\cite{Holod_2009} and the \textit{conservative} scheme\cite{Bao_2017}. In this work, we also found that the accurate calculation of the equilibrium parallel current $J_{\parallel0}$ and $\delta B_\parallel$ is required for the correct simulation results for kink modes. The GTC electromagnetic model with the parallel current was first considered in the Ref \cite{Deng_2012}, but now we find that the radial component of the parallel current in Boozer coordinates is critical for the current-driven MHD instability. We have developed a method to accurately calculate the parallel current in the axisymmetric equilibrium. Furthermore, the compressible magnetic perturbation $\delta B_\parallel$ has been thought to be unimportant in the low-$\beta$ regime. But in the GTC simulations of kinetic ballooning mode(KBM)\cite{Dong_2017}, low-frequency Alfv\'enic modes\cite{Choi_2021}, and the internal kink and fishbone modes\cite{Brochard_2022, Brochard2024PRL}, the compressible magnetic perturbation $\delta B_\parallel$ has important effects on the linear instability and nonlinear dynamics. In the kink simulation, $\delta B_\parallel$ has a significant effect on the kink growth rate and a magnitude comparable to the shear magnetic perturbation $\delta B_\perp$. Notably, the importance of $\delta B_\parallel$ in the gyrokinetic simulations has been recently pointed out in other works as well. \cite{Kennedy_2024, Sheffield_2025,Li_2025_dbpara} Utilizing these physics capabilities, we have simulated more than 5000 DIII-D experiments with kink instability to build a database. The data analysis shows that the q=1 surface location, minimum q value, pressure gradient at q=1 surface, and the thermal energy stored inside the q=1 surface have strong correlations to the instability. We have used this database to train a surrogate model to predict the linear kink drive using experimental parameters, the results of which are presented in a separate paper\cite{Dong_2021}. The successful applications of the comprehensive model on the kink mode and the large amount of kink mode simulations demonstrate the capability of this gyrokinetic model for kinetic-MHD simulations.

This paper is organized as follows. In Section 2, the model equations are given, in which the gyrokinetic ion model, the drift kinetic electron models, and the gyrokinetic Maxwell equations are presented. In the long wavelength limit, this model reduces to the ideal MHD model with both the linear ideal MHD dispersion relation (the vorticity equation) and the nonlinear ponderomotive force. In Section 3, we describe the simulation results from the kink mode benchmark and the results from over 5000 simulations of kink instability. The conclusion is given in Section 4. Some detailed formulations and numerical implementations are given in Appendices. The expressions of the equation of particle motion and field equations in Boozer coordinates are given in Appendix \ref{sec:eqs_in_boozer}. The method to construct Boozer coordinates is given in Appendix \ref{sec:XMAP}. A complete form of the perpendicular Laplacian operator is given in Appendix \ref{sec:LaplacionOp}. In Appendix \ref{sec:GKenergy}, we show the gyrokinetic energy conservation and the way to calculate the transfer between kinetic and field energy. The electron parallel momentum equation is used in our model, and the detailed expression of each term is listed in Appendix \ref{sec:EMomt}. The model to simulate neoclassical tearing mode (NTM) is shown in Appendix \ref{sec:NTMmodel}. Finally, the numerical methods to calculate the equilibrium parallel current, which is essential to drive the internal kink mode, are shown in Appendix \ref{sec:Jpara}.

\section{Simulation Model}
The formulation of GTC simulation model is introduced in this section. We start from the gyrokinetic Vlasov equation, followed by the perturbative $\delta f$ simulation method to evolve the perturbed distribution function in Subsection \ref{sec:GK_Vlasov}. In Subsection \ref{sec:GK_Maxwell}, the gyrokinetic form of Maxwell equations is introduced, which connects the distribution function in guiding center space and the electromagnetic fields in particle space. The Section \ref{sec:numerical_implementations} presents a unified formulation to efficiently solve the electron drift kinetic equation, which separates electron responses into an analytic part described by the continuity equation and a non-analytic part described by the drift kinetic equation.
This electron model incorporates the equilibrium parallel current, reduces electron particle noise, and avoids the so-called ``cancellation problem" in electromagnetic simulation. In Section \ref{sec:MHD_eqns}, it is demonstrated that this set of gyrokinetic equations in the limit of long wavelength and small electron-to-ion mass ratio can be reduced to the ideal MHD model, including both linear dispersion relation and nonlinear ponderomotive force. {The formulations in this section have been derived independently with somewhat simplified forms in several earlier GTC papers. This section serves as a review that presents a complete model with comprehensive formulas, unified notations, and detailed implementations. There are also three novel features compared to previous papers: 1) The two electron models are unified in a single fluid-kinetic framework, with the different definition of a term $\delta\phi_{ind}$ (See \ref{sec:conservative_scheme} and \ref{sec:hybrid_scheme}); 2) The reduction from gyrokinetic models to MHD dispersion relation in the presence of $\delta B_\parallel$ has been shown(See \ref{sec:MHD_eqns} and Appendix \ref{sec:MHD_DR}); and 3) The accurate implementation of $J_{\parallel0}$ in Boozer coordinates, especially the $m=1$ component caused by $\hdelta$ term, is found to be very important for the current driven kink mode. (See Appendix \ref{sec:Jpara} and the results in Section 3)}
\subsection{Gyrokinetic simulation model}\label{sec:GK_eqns}
\subsubsection{Gyrokinetic Vlasov equation} \label{sec:GK_Vlasov}

The gyrokinetic Vlasov equation \cite{Lee1983,BrizardHahm_2007} is used to describe the evolution of distribution
function for each species in five-dimensional gyrocenter phase space, $f = f\left(\mathbf{R},v_{\parallel},\mu,t\right)$
\begin{equation}
Lf\equiv\left(\frac{\partial}{\partial t}+\dot{\mathbf{R}}\cdot\nabla+\dot{v}_{\parallel}\frac{\partial}{\partial v_{\parallel}} - \mathcal{C}\right)f\left(\mathbf{R},\mu,v_{\parallel}\right)=0,
\end{equation}
where $\mathbf{R}$ is the gyrocenter position, $v_{\parallel}$ the gyrocenter parallel velocity along field line, $\mu$ the magnetic moment. $\mathcal{C}$ stands for the collision operator. Note that the operator $\nabla$ means $\partial/\partial\mathbf{R}$ while keeping $\mu$ and $v_\parallel$ constant. The time evolution of $\mathbf{R}$ and $v_{\parallel}$ are given
by
\begin{equation}
\begin{aligned}
\dot{\mathbf{R}}=v_{\parallel}\frac{\mathbf{B}^{*}}{B_{\parallel}^{*}}+\mathbf{v}_{E}+\mathbf{v}_{g}+\mathbf{v}_{b\parallel}\label{eq:Rdot}
\end{aligned}
\end{equation}
\begin{equation}
\begin{aligned}
\dot{v}_{\parallel} =&-\frac{1}{m_{s}}\frac{\mathbf{B}^{*}}{B_{\parallel}^{*}}\cdot\left(\mu\nabla B_{0}+Z\nabla\left\langle \phi\right\rangle -Z\nabla\left\langle \delta \mathbf{A}_{\perp}\cdot\mathbf{v}_{\perp}\right\rangle \right),\\
 & -\frac{Z}{m}\frac{\partial\left\langle \delta A_{\parallel}\right\rangle }{\partial t}.\label{eq:vparadot}
\end{aligned}
\end{equation}
Here, we have utilized the parallel-symplectic representation of the
modern gyrokinetic formula\cite{Brizard_1992,BrizardHahm_2007}, where $Z$ and $m$ are the particle charge and mass, respectively. $B_{0}$ is the amplitude
of equilibrium magnetic field, $\mathbf{B}^{*}=\mathbf{B}_{0}^{*}+\delta\mathbf{B}_{\perp}$, ${\mathbf{B}_0^*}=\mathbf{B}_{0}+\frac{mv_{\parallel}}{Z}\nabla\times\mathbf{b}_{0}$, $\mathbf{b}_0=\mathbf{B}_0/B_0$ is the unit vector along the equilibrium magnetic field line, and
$B_{\parallel}^{*}=\mathbf{B}_{0}^{*}\cdot\mathbf{b}_{0}$. $\phi$
is the electrostatic potential, while $\delta\mathbf{A}_{\perp}$ and $\delta A_\parallel$
stand for the perpendicular and parallel components of the vector
potential, and $\delta B_{\parallel}=\mathbf{b}_{0}\cdot\left(\nabla\times\delta\mathbf{A}_{\perp}\right)$, $\delta\mathbf{B}_{\perp}=\nabla\times\left(\delta A_{\parallel}\mathbf{b}_0\right) + \left(\nabla\times\delta \mathbf{A}_\perp\right)_\perp$. 
The $\mathbf{E}\times\mathbf{B}$ velocity, the grad-B drift velocity
and the drift velocity from $\delta B_\parallel$ are
\begin{align}
\begin{split}
    \mathbf{v}_{E}&=\frac{\mathbf{b}_{0}\times\nabla\left\langle \phi\right\rangle }{B_{\parallel}^{*}}\\
    \mathbf{v}_{g}&=\frac{\mu}{ZB_{\parallel}^{*}}\mathbf{b}_{0}\times\nabla B_{0}\\
    v_{\mathbf{b}\parallel}&=-\frac{\mathbf{b}_{0}\times\nabla\left\langle \delta\mathbf{A}_{\perp}\cdot\mathbf{v}_{\perp}\right\rangle }{B_{\parallel}^{*}}.
\end{split}
\end{align}
Note that the curvature drift velocity $\mathbf{v}_c$ appears in the first term of the right-hand side in Eq \eqref{eq:Rdot}. In the above expressions, the operator $\left\langle \cdots\right\rangle $
denotes gyro-averaging, $\left\langle a\right\rangle \left(\mathbf{R}\right)=\frac{1}{2\pi}\oint d\zeta\int d\mathbf{x}a\left(\mathbf{x}\right)\delta\left(\mathbf{x}-\mathbf{R}-\boldsymbol{\rho}\right)$,
with $\zeta$ being the phase angle and $\boldsymbol{\rho}=m\bunit\times\mathbf{v}/(ZB_0)$ being the gyro radius. Note that $\phi$ consists of the perturbed electrostatic potential $\delta \phi$ and the time-static equilibrium potential $\phi_{eq}$. But  $\delta A_\parallel$ and $\delta \mathbf{A}_\perp$ only stand for the time varying part of the vector potential. All equilibrium fields, including the RMP(resonant magnetic perturbation) field\cite{Evans_2005_RMP}, are contained in $\mathbf{B}_0$.

The $\delta f$ scheme\cite{Parker1993} is used for GTC simulation, in which we split
the distribution function to an equilibrium part and a perturbed part,
$f=f_{0}+\delta f$. The equilibrium part $f_{0}$ is the
solution of the unperturbed Vlasov equation,
\begin{equation}
L_{0}f_{0}\equiv\frac{\partial f_{0}}{\partial t}+\left(v_{\parallel}\frac{\mathbf{B}_{0}^{*}}{B_{\parallel}^{*}}+\mathbf{v}_{E,eq}+\mathbf{v}_{g}\right)\cdot\nabla f_{0}-\frac{\mu}{m}\frac{\mathbf{B}_{0}^{*}}{B_{\parallel}^{*}}\cdot\nabla B_{0}\frac{\partial f_{0}}{\partial v_{\parallel}}-\frac{\mathbf{B}^*_0}{B_\parallel^*}\cdot\nabla\left\langle\phi_{eq}\right\rangle\frac{Z}{m}\frac{\partial f_{0}}{\partial v_\parallel}-\mathcal{C}f_{0}=0.\label{eq:eqVlasov}
\end{equation}
 And the governing equation of $\delta f$ is given by
\begin{equation}
\begin{aligned}
    L\delta f  = &-\delta L f_{0} =  -(L-L_0)f_{0}\\
    =&-\left(v_{\parallel}\frac{\delta\mathbf{B}_{\perp}}{B_{\parallel}^{*}}+\delta\mathbf{v}_{E}+\mathbf{v}_{b\parallel}\right)\cdot\nabla f_{0} \\
    & +\frac{1}{m}\left[\frac{\mu\delta\mathbf{B}_{\perp}\cdot\nabla B_{0}}{B_{\parallel}^{*}}+\frac{\mathbf{B}^{*}_0}{B_{\parallel}^{*}}\cdot\left(Z\nabla\left\langle\delta\phi\right\rangle -Z\nabla\left\langle \delta\mathbf{A}_{\perp}\cdot\mathbf{v}_{\perp}\right\rangle \right)+\frac{\delta\mathbf{B}_\perp}{B_{\parallel}^{*}}\cdot\left(Z\nabla\left\langle\phi\right\rangle -Z\nabla\left\langle \delta\mathbf{A}_{\perp}\cdot\mathbf{v}_{\perp}\right\rangle \right)\right.\\
    &\left.+Z\frac{\partial\left\langle \delta A_{\parallel}\right\rangle }{\partial t}\right]\frac{\partial}{\partial v_{\parallel}}f_{0}.\label{eq:Ldfs}
\end{aligned}
\end{equation}
$L$ involves time derivatives of the phase space coordinates $(\dot{\mathbf{R}}, \dot{v_\parallel})$ given in Eqs \eqref{eq:Rdot} and \eqref{eq:vparadot} solved from the Euler-Lagrange equation in gyrocenter space\cite{BrizardHahm_2007}. Instead of using the vector form equations, Eqs \eqref{eq:Rdot} and \eqref{eq:vparadot}, the implementation in GTC uses $(\dot{\psi},\dot{\theta},\dot{\zeta},\dot{\rho_\parallel})$ to update the particle coordinate in gyrocenter phasespace. Here $(\psi,\theta,\zeta)$ is the Boozer coordinates, and $\rho_\parallel=mv_\parallel/(ZB_0)$. The scalar form equations can be obtained directly from the chain rule.
\begin{equation}
\begin{split}
    \dot{\alpha} &= \dot{\mathbf{R}}\cdot\nabla\alpha, \quad \alpha=(\psi,\theta,\zeta),\\
    \dot{\rho_\parallel} &= \frac{m}{Z B_0} \dot{v}_\parallel - \frac{mv_\parallel}{ZB_0}\frac{\nabla B_0}{B_0}\cdot\dot{\mathbf{R}}.\label{eq:scalar_eq_motion}
\end{split}
\end{equation}
The reason to use Boozer coordinates is originally to follow the Hamiltonian equations of motion in Boozer coordinates derived by White and Chance\cite{White_1984}. But we can prove that after a transformation, the scalar form equations shown in Eq \eqref{eq:scalar_eq_motion} are identical to the Hamiltonian equations of motion. The detailed implementation can be seen in Appendix \ref{sec:eqs_in_boozer}, where the relation between Eq \eqref{eq:scalar_eq_motion} and the Hamiltonian equations of motion\cite{White_1984} in the Boozer coordinate system is also explained. Using Boozer coordinates has other conveniences, for example, the concept and method for optimizing stellarator geometry relies on the Boozer coordinates\cite{Helander2014_3Dgeometry}.

Define the particle weight as $w=\delta f/f$, then
the evolution of $w$ is given by 
\begin{equation}
    \frac{dw}{dt}=-(1-w)\frac{1}{f_{0}}(L-L_0)f_{0}.
\end{equation}
The detailed formula of the $w_s$ equation can be found in Appendix \ref{sec:eqs_in_boozer}. This standard formulation of the gyrokinetic simulation model can be applied to all species in the GTC. For the drift-Alfv\'enic turbulence with $k_\perp\rho_e\ll1$, the electron equations can be simplified to the drift kinetic equation (DKE) by neglecting gyro-averaging.

In the presence of magnetic islands, the form of $f_{0s}$ is difficult to write out. So we further split equilibrium magnetic field to $\mathbf{B}_{0}+\mathbf{B}_{IS}$, where $\mathbf{B}_{IS}$ is the magnetic field contribution from the islands, split the equilibrium distribution function to $f_{0}+f_{IS}$, and split the original $L_0$ operator to $L_{0}+L_{IS}$. The new $L_0$ operator, magnetic field, and distribution function in the previous equations should be replaced accordingly. $L_{0}$ and $L_{IS}$ are given by
\begin{equation}
\begin{split}
    L_{0}&\equiv\frac{\partial}{\partial t}+\left(v_{\parallel}\frac{\mathbf{B}_{0}^{*}}{B_{\parallel}^{*}}+\mathbf{v}_{E,eq}+\mathbf{v}_{g}\right)\cdot\nabla -\frac{\mu}{m}\frac{\mathbf{B}_{0}^{*}}{B_{\parallel}^{*}}\cdot\nabla B_{0}\frac{\partial}{\partial v_{\parallel}}-\frac{\mathbf{B}^*_{0}}{B_{\parallel}^*}\cdot\nabla\left\langle\phi_{eq}\right\rangle\frac{Z}{m}\frac{\partial}{\partial v_\parallel}-\mathcal{C},\\
    L_{IS} &= \left(\frac{B_{\parallel}^*}{(B_0+B_{IS})_\parallel^*} - 1\right) L_{0,eq} + v_\parallel\frac{\mathbf{B}_{IS}}{(B_0+B_{IS})_\parallel^*}\cdot\nabla + \frac{\mu}{m}\frac{\mathbf{B}_{IS}}{(B_0+B_{IS})_\parallel^*}\cdot\frac{\partial}{\partial v_\parallel} - \frac{\mathbf{B}_{IS}}{(B_0+B_{IS})_\parallel^*}\cdot\langle\phi_{eq}\rangle\frac{Z}{m}\frac{\partial}{\partial v_\parallel},\label{deltaLisland}
\end{split}
\end{equation}
where $(\mathbf{B}_0+\mathbf{B}_{IS})^* = (\mathbf{B}_0+\mathbf{B}_{IS}) + \frac{m}{Z}v_\parallel\nabla\times [(\mathbf{B}_0+\mathbf{B}_{IS})/|(\mathbf{B}_0+\mathbf{B}_{IS})|]$, $(B_0+B_{IS})_\parallel^* = (\mathbf{B}_0+\mathbf{B}_{IS})^*\cdot[(\mathbf{B}_0+\mathbf{B}_{IS})/|(\mathbf{B}_0+\mathbf{B}_{IS})|]$. In Eq \eqref{deltaLisland}, the term $\left(\frac{B_{\parallel}^*}{(B_0+B_{IS})_\parallel^*} - 1\right)(\partial_t - \mathcal{C})$ is assumed to 0, and $(\mathbf{B}_0+\mathbf{B}_{IS})^*$ is approximated to $\mathbf{B}_{0}^* + \mathbf{B}_{IS}$ in the second equation.

Similar to the original $\delta f$ method, by defining $f_{0}$ which satisfying $L_{0} f_{0} = 0$, we can obtain the equilibrium distribution function by solving
\begin{equation}
\begin{split}
    L_{0} f_0 = 0 \\
    (L_{0} + L_{IS}) f_{IS} &= -L_{IS} f_{0}. \label{eq:island_df}
\end{split}
\end{equation}
The argument of these approximations and the electrostatic turbulence simulation with magnetic island can be seen in \cite{Fang2019a,Wei2025_NF}.

From Eq \eqref{eq:island_df} and Eq \eqref{eq:Ldfs}, we can also alternatively solve
\begin{equation}
\begin{split}
    L_{0} f_{0} &= 0,\\
    (L_{0} + L_{IS} + \delta L) (\delta f + f_{IS}) &= -(\delta L + L_{IS}) f_{0}\label{eq:alt_islanddf}.
\end{split}
\end{equation}
 Using the Fokker-Planck collision operator for $\mathcal{C}$, we can find that the solution of $L_{0} f_{0} = 0$ is the neoclassical distribution function $f_{0} = f_{nc}$. The numerical way to find the neoclassical solution can be found in \cite{Lin1995_NC}. In the collisionless limit, the equilibrium distribution $f_{0}$ should be a function of constants of motion. However, when simulating the turbulent dominant cases, we will usually use the local shifted Maxwellian $f_M$ to approximate the real equilibrium distribution function when calculating the perturbed distribution function, and ignore the difference between $(\delta L+L_{IS})f_{M}$ and $(\delta L + L_{IS}) f_{nc}$, i.e.,
\begin{equation*}
    f_{0}\approx f_{M}=\frac{n_{0}}{(2\pi T/m)^{3/2}} \exp \left(-\frac{m(v_\parallel-u_{\parallel0})^2}{2T}-\frac{\mu B_0}{T}\right).
\end{equation*}
$u_{\parallel0}$ is the equilibrium parallel flow velocity, $n_{0}=n_{eq0}*\exp\left(-q\phi_{eq}/T\right)$ is the equilibrium density, and $n_{eq0}$ is the equilibrium density when $\phi_{eq}=0$. {Note that The RHS of the second equation of Eq\eqref{eq:alt_islanddf} is actually $-L_0f_0-(\delta L+L_{IS})f_0$, the approximation of replacing $f_{nc}$ with $f_M$ is in the second term, while the first term is using the exact $f_0$(or $f_{nc}$) since $L_0f_0$ is explicitly set to 0. So, this approximation is at a higher order since $f_{nc}/f_M-1\sim L_{orbit}/L_p$. Where $L_{orbit}$ is the drift orbit width, and $L_p$ is the pressure scale length. The approximation is valid with the parameters we are interested in.}

In practice, the analytical form of $f_{IS}+f_0$ is unknown, and the evolution of $f_{IS,i}$, $f_{IS,e}$, and $\delta f$ all have different time scales. To study the turbulence in the presence of islands, we will use the second equation in Eq \eqref{eq:island_df} to get a steady-state $f_{IS}$ for ions, and {set the position, energy and $\mu$ of electrons to be identical to ions, $(\mathbf{R,\mu,\mathcal{E}})_{m,e}=(\mathbf{R},\mu,\mathcal{E})_{m,i}$, for $m=1,...,M$, where $M$ is the total numerical marker number for ions and electrons. The energy and $\mu$ of the electrons are scaled from those of the ions according to the temperature ratio}. Therefore, the quasi-nutrality is satisfied, and $f_{IS,i}$ and $f_{IS,e}$ do not evolve significantly due to the large electrostatic field. Then we can turn to solve the second equation of Eq \eqref{eq:alt_islanddf} to get the evolution of $\delta f$ and the associated turbulence behaviors. This procedure also guarantees the quasi-neutrality during the turbulence simulation. 

For energetic particles, we need to consider the particle source term in the unperturbed Vlasov equation, and the steady-state slowing-down distribution can be chosen to model $f_{0EP}$\cite{YChen_2020,Gaffey_1976,Brochard2025_NF},
\begin{equation}
    f_{0EP}\approx f_{SD}=c\frac{n_{0EP}H(v_0-v)}{v^3+v_c^3}\exp\left[-\left(\frac{\Lambda-\Lambda_0}{\Delta\Lambda}\right)^2\right].\label{eq:sd_dist}
\end{equation}
Here $v_c$ is the critical velocity, $v_0$ is the birth velocity, $H$ is the Heaviside step function, and $c$ is the normalization factor. $\Lambda=\mu B_a/E$ is the pitch angle, $E=\mu B_0+mv_\parallel^2/2$ is the kinetic energy, $B_a$ is the on-axis magnetic field strength. $\Lambda_0$ is the peak of the pitch angle, and $\Delta\Lambda$ is the width of the pitch angle distribution.

\subsubsection{Maxwell equations}\label{sec:GK_Maxwell}
The particle distribution function is solved in gyrocenter space. To solve the Maxwell equations in real space, we need to transform the distribution function for each species $f_s$ from gyrocenter space to particle space. The transformation up to the first order is given by \cite{BrizardHahm_2007, Brizard_1992}
\begin{equation}
\begin{split}
    \delta F_s (\mathbf{x},v_\parallel, \mu, t) = &\int{d\mathbf{R}}\left[\delta f_s (\mathbf{R},v_\parallel, \mu, t)\right.\\
    &\left.+\left(\mathbf{G}_1^R\cdot \frac{\partial}{\partial\mathbf{R}} + G_1^{v_\parallel} \frac{\partial}{\partial v_\parallel} + G_1^\mu\frac{\partial}{\partial\mu}\right) f_s\left(\mathbf{R},v_\parallel, \mu, t\right)\right] \delta (\mathbf{x} - \mathbf{R} - \boldsymbol{\rho}), \label{eq:deltaf_compact}
\end{split}
\end{equation}
where $\delta F_s (\mathbf{x},v_\parallel, \mu, t)$ is the perturbed distribution function in particle space, the subscript `s' stands for a certain particle species, $\delta f_s$ and $f_s$ are the perturbed and total distribution function in gyrocenter space of species `s'. $G_1^\alpha$ is the first-order generating vector along the $\alpha$-direction in the Lie transformation method, where $\alpha$ can be $\mathbf{R}, {v_\parallel}$, and $\mu$. The generating vectors are the components of the pull-back operator that transforms the distribution function from gyrocenter space to guiding center space, and the operator $\int d\mathbf{R}\cdot \delta(\mathbf{x}-\mathbf{R}-\boldsymbol{\rho})$ further transforms the distribution to particle space. The detailed expression of the generating vectors can be found in \cite{BrizardHahm_2007}. $\mathbf{x}$ is the particle position in real space. With $\delta F_s$, we can obtain the particle density, flow, pressure, etc.
The velocity moments of the particle distribution function consist of the gyro-averaged guiding center part and the polarization part,
\begin{equation}
\begin{split}
    \delta\mathcal{M}_s\left(\mathbf{x}\right)=  \delta \bar{\mathcal{M}}_s\left(\mathbf{x}\right) + \delta\mathcal{M}_{pol,s}\left(\mathbf{x}\right)
\end{split}
\end{equation}
where
\begin{equation}
\begin{split}
    \delta\bar{\mathcal{M}}_s &= \int d\mathbf{v} \int d\mathbf{R} \mathcal{V} \delta f_s(\mathbf{R},v_\parallel,\mu,t)\delta(\mathbf{x}-\mathbf{R}-\boldsymbol{\rho}),\\
    \delta \mathcal{M}_{pol,s} &= \int d \mathbf{v} \int d\mathbf{R}\mathcal{V}\left(\mathbf{G}_1^\mathbf{R}\cdot\frac{\partial}{\partial\mathbf{R}} + G_1^{v_\parallel}\frac{\partial}{\partial v_\parallel} + G_1^\mu \frac{\partial}{\partial\mu} \right) f_s(\mathbf{R},v_\parallel,\mu,t)\delta(\mathbf{x}-\mathbf{R}-\boldsymbol{\rho}).
\end{split}
\end{equation}
Where $\mathcal{V}$ is a function of $\mathbf{v}$. When $\mathcal{V}=1$, $\delta\mathcal{M}$ is the perturbed density $\delta n_s$. When $\mathcal{V}=\mathbf{v}_\perp$, the gyrocenter perpendicular flow $\delta \bar{\mathcal{M}}_s = 0$, and the perpendicular flow in particle space $n_0\delta\mathbf{u}_{\perp,s}$ is given from the polarization term $\delta \mathcal{M}_{pol,s}$. And when $\mathcal{V}=v_\parallel$, the parallel flow $n_0\delta u_{\parallel,s}$ in particle space is approximately given by the gyrocenter term $\delta \bar{\mathcal{M}}_s$, while the polarization term $\delta \mathcal{M}_{pol,s}$ only gives high order correction. The detailed expressions of generating vectors $G_1^\alpha$ can be found in \cite{BrizardHahm_2007}.

The field equations in GTC include the parallel Ampere's law, perpendicular Ampere's law, and the gyrokinetic Poisson equation. The parallel Ampere's law reads
\begin{equation}
    \nabla_\perp^2\delta A_\parallel = -\mu_0\sum_s Z_s n_{0s} \delta u_{\parallel,s},\label{eq:Ampere}
\end{equation}
where $\mu_0$ is the vacuum permeability.
The coupled equations of electrostatic potential $\phi$ and $\delta B_\parallel$ are given by the gyrokinetic Poisson equation (quasi-neutrality condition) and the perpendicular Ampere's law,
\begin{equation}
\begin{split}
    0 = & \sum_s Z_s \delta n_{pol,s} + \sum_s Z_s \delta \bar{n}_s,\\
    \nabla \delta B_\parallel \times \bunit =& \mu_0 \sum_s Z_s\delta \mathbf{u}_{\perp pol, s}.
\end{split}
\end{equation}

In the fusion-related parameter region, the most important mechanism to cause large transport is often the ion-scale turbulence or meso-scale and macro-scale coherent modes. These modes satisfy the condition $1/\left(k_\perp L_p\right) \ll 1$, and $k_\perp \rho_e \ll 1$. The electron equations reduce to the drift kinetic equation without gyro-averaging, and thus the polarization and magnetization terms reduce to simpler forms. The two equations for $\delta\phi$ and $\delta B_\parallel$ are given by \cite{Dong_2017},
\begin{equation}
\begin{split}
    \sum_{s\neq e}\frac{Z_{s}^{2}n_{s}}{T_{s}}\left(\delta \phi-\delta \tilde{\phi}_{s}\right)-\frac{1}{B_{0}}\left(\sum_{s\neq e}Z_{s}n_{s0}\left\{ \delta B_{\parallel}\right\} _{s}-en_{0}\left\{ \delta B_{\parallel}\right\} _{e}\right)=\sum_{s\neq e}Z_{s}\delta\bar{n}_{s}-e\delta n_{e},\label{eq:Poisson_orig}
\end{split}
\end{equation}
\begin{equation}
\begin{split}
    & \frac{1}{\mu_{0}}\delta B_{\parallel}B_{0}+2\pi\Omega_{e}^{2}\int d\mu dv_{\parallel}\left[B_{0}\left\langle \int_{0}^{\rho_{e}}\delta f_{e}rdr\right\rangle \right.\\
    & \left.+\frac{f_{0e}}{\rho_{e}^{2}}\left\langle \int_{0}^{\rho_{e}}\left\langle \int_{0}^{\rho_{e}}\delta B_{\parallel}r'dr'\right\rangle rdr\right\rangle \right]\\
   = & -\sum_{s\neq e}2\pi\Omega_{s}^{2}\int d\mu dv_{\parallel}\left[B_{0}\left\langle \int_{0}^{\rho_{s}}\left(\delta f_{s}+\frac{Z_{s}\left\langle\delta\phi\right\rangle -Z_{s}\delta \phi}{T_{s}}f_{0s}\right)rdr\right\rangle \right.\\
    & \left.+\frac{f_{0s}}{\rho_{s}^{2}}\left\langle \int_{0}^{\rho_{s}}\left\langle \int_{0}^{\rho_{s}}\delta B_{\parallel}r'dr'\right\rangle rdr\right\rangle \right].\label{eq:Bpara_orig}
\end{split}
\end{equation}
Where $\delta\tilde{\phi}_s$ is the double gyro-averaged potential\cite{WWLee1987,Lin_1995},
\begin{equation}
\delta\tilde{\phi}_s\left(\mathbf{x}\right)=\int d\mathbf{v}\int d\mathbf{R}\left\langle\delta\phi\right\rangle \left(\mathbf{R}\right)f_{0s}\left(\mathbf{R}\right)\delta\left(\mathbf{x}-\mathbf{R}-\boldsymbol{\rho}\right).
\end{equation}
$\left\{ \delta B_{\parallel}\right\} _{s}$ denotes the double-gyro-averaged
$\delta B_{\parallel}$,
\begin{align}
\begin{split}
    \left\{ \delta B_{\parallel}\right\} _{s}\left(\mathbf{x}\right) =& \frac{m\Omega_{s}^{2}}{2\pi n_{0s}T_{s}}\int d\mathbf{v}\int d\mathbf{R}\int_{0}^{\rho}r'dr'\int_{0}^{2\pi}d\zeta'\int d\mathbf{x'}\delta B_{\parallel}\left(\mathbf{x}'\right)\\
 & \times\delta\left(\mathbf{x}'-\mathbf{R}-\mathbf{r'}\right)f_{0s}\delta\left(\mathbf{x}-\mathbf{R}-\boldsymbol{\rho}\right).
\end{split}
\end{align}
The term $\langle\delta\mathbf{A}_\perp\cdot\mathbf{v}\rangle$ in Eq \eqref{eq:Ldfs} stands for the ``perturbed mirror potential" felt by the particle gyrocenter, and can be calculated from $\delta B_\parallel$\cite{Dong_2017, Porazik_2011},
\begin{equation*}
\left\langle \delta\mathbf{A}_{\perp}\cdot\mathbf{v}_{\perp}\right\rangle =-\frac{\mu}{Z_{s}}\left\langle \left\langle \delta B_{\parallel}\right\rangle \right\rangle =-\frac{\mu}{Z_{s}}\frac{1}{\pi}\int_{0}^{1}\xi d\xi\int_{0}^{2\pi}\delta B_{\parallel}\left(\mathbf{R}+\xi\boldsymbol{\rho}\right)d\zeta,
\end{equation*}
where the integral is performed on the plane surface enclosed by the gyro-orbit around the gyrocenter $\mathbf{R}$. $\zeta$ is the gyro phase angle, and $\mathbf{R} + \xi\boldsymbol{\rho}$ indicates a position between the gyrocenter and the gyro-orbit on the surface. When $\xi=0$, $\delta B_\parallel$ takes the value on the gyrocenter position, and when $\xi=1$, $\delta B_\parallel$ takes the value at the point on the gyro orbit with gyrophase $\zeta$. The numerical implementation to calculate these integrals can be found in \cite{Porazik_2011}.

The two equations can be decoupled in the low-$\beta$ limit\cite{Dong_2017}, and the gyrokinetic Poisson equation becomes
\begin{equation}
\begin{split}
    \sum_{s\neq e}\frac{Z_{s}^{2}n_{s}}{T_{s}}\left(\delta \phi-\delta \tilde{\phi}_{s}\right)=\sum_{s\neq e}Z_{s}\delta\bar{n}_{s}-e\delta n_{e}.\label{eq:Poisson_reduced}
\end{split}
\end{equation}
The $\delta B_\parallel$ equation can be further simplified via an expansion in terms of $k_\perp\rho_s$,
\begin{equation}
\begin{split}
    \frac{\delta B_\parallel}{B_0}
    =&-\frac{\beta_e}{2+\beta_e+2\sum_{s\neq e}\beta_s}\left[\frac{3}{2}\nabla_\perp^2\delta\phi\sum_{s\neq e}\frac{\beta_s}{\beta_e}\frac{Z_s}{T_s}\rho_s^2+\frac{5}{4}\nabla_\perp^4\delta\phi\sum_{s\neq e}\frac{\beta_s}{\beta_e}\frac{Z_s}{T_s}\rho_s^4\right.\\
    &\left.+\frac{1}{P_{\perp0e}}\left(\delta P_{\perp e}+\sum_{s\neq e}\delta \tilde{P}_{\perp s}\right)\right],\label{eq:Bpara_reduced}
\end{split}
\end{equation}
where $\beta_s = (2\mu_0n_{0s}T_s)/B_0^2$, $P_{\perp0e}=n_{0e}T_e$, $\rho_s = \sqrt{m_sT_s}/(Z_s B_0)$, $\delta\tilde{P}_{\perp s}=2\pi\Omega_i^2\int d\mu dv_\parallel B_0 \langle\int_0^{\rho_s}\delta f_s rdr\rangle$. The electron polarization density has been neglected since $k_\perp\rho_e\ll1$ for the parameters of interest.

In GTC, the Vlasov equation and Maxwell equations are solved in Boozer coordinates. A toolkit to construct Boozer coordinates from EFIT GEQDSK data file is developed as explained in Appendix \ref{sec:XMAP}. In the Poisson equation and Ampere's law, the perpendicular Laplacian operator is used. The numerical implementation of solving this type of equation is introduced in \cite{Xiao2015}. Here we give a more complete form of the perpendicular Laplacian operator in Boozer coordinates in Appendix \ref{sec:LaplacionOp}. The gyrokinetic energy conservation and energy transfer between kinetic and field energy are explained in Appendix \ref{sec:GKenergy}, following the derivations in \cite{BrizardHahm_2007, Brizard_2017}.

\subsection{Simulation models for solving electron DKE}\label{sec:numerical_implementations}
Solving the electron DKE accurately together with gyrokinetic ions in multiscale electromagnetic simulations is numerically challenging due to the small electron-to-ion mass ratio. One effect of the small electron mass is the short collisionless skin depth, which shields out the parallel electric field $E_\parallel$ and leads to the ideal Alfv\'enic state with $E_\parallel=0$ in the long wavelength limit and uniform plasmas. In the standard gyrokinetic formulation, $E_\parallel$ is calculated from both electrostatic potential $\delta \phi$ using Poisson equation and parallel vector potential $\delta A_\parallel$ using Ampere's law, which cancel out with each other in the ideal Alfv\'enic state. Small errors in calculating $\delta \phi$ and $\delta A_\parallel$ due to inconsistency between electron density and flow perturbations\cite{Bao_2017} can lead to a large error in $E_\parallel$, leading to the ``cancellation problem" exacerbated by the choice of using canonical momentum as an independent variable to avoid taking explicit time derivative of $\delta A_\parallel$\cite{CHEN2007GEM}. To circumvent this difficulty, most {gyrokinetic continuum simulations of drift-Alfv\'enic turbulence numerically resolve the cancellation problem arising from $p_\parallel$ formulation by using the exact same discretization for the perturbed electron current and the skin current.\cite{CANDY2003GYRO, GENE_CANCELLATION}}

Another effect of the small electron mass is that its thermal velocity is much larger than the electron parallel flow velocity that carries the equilibrium parallel current, making it difficult to incorporate the equilibrium current in the electron distribution function. Consequently, nearly all gyrokinetic codes in the toroidal geometry neglect the equilibrium current and therefore can not simulate current-drive instabilities such as kink and resistive tearing modes.

The kinetic manifestation of the small electron mass is that electron response to drift-Alfv\'enic turbulence is mostly adiabatic, which motivates the development of the `split-weight' scheme\cite{Manuilskiy2000PoPSplitWeight} that solves only the non-adiabatic responses,therebyreducing electron particle noise by using an analytic solution for the adiabatic response. However, this scheme does not address the issue of the equilibrium current.  

These two problems are overcome using the new formulation in GTC \cite{Lin_Chen_2001, Lin_2007, Holod_2009, Deng_2012, Bao_2017} where the $E_\parallel$ is directly calculated from the electron parallel force balance to avoid the ``cancellation problem". The $E_\parallel$ is subsequently used for calculating the $\delta A_\parallel$. The electron flow perturbation is calculated from Ampere's law and used in the continuity equation to calculate the electron density perturbation, ensuring consistency between density and flow perturbations. The density perturbation is then used to define an analytic part of the electron fluid response, and the non-analytic part of the perturbed distribution function is dynamically solved by the DKE. The fluid response already contains the equilibrium current and non-resonant part of the perturbed parallel current, incorporating kinetic shear Alfv\'en wave, ion acoustic wave, and driftwave, even without the need to solve the kinetic response of the non-analytic part of the perturbed distribution function. Two versions of this electron formulation have been implemented in GTC: In the fluid–kinetic hybrid electron model \cite{Lin_Chen_2001, Holod_2009, Deng_2012}, the fluid response is defined as electron adiabatic response, and the kinetic response can be calculated using an expansion of the DKE by removing the collisionless tearing mode{(The expansion is based on small $\omega/k_\parallel v_\parallel$, and only finite $k_\parallel$ is kept in the linear response)}\cite{Lin_Chen_2001}. Meanwhile, in the conservative scheme \cite{Bao_2017}, the fluid response is defined for the total density perturbation and the kinetic response is calculated from the exact DKE that preserves the collisionless tearing mode. The two methods are identical in the simulations that only keep the fluid response, including the kink and resistive tearing modes. This new GTC formulation enables multiscale gyrokinetic simulations for cross-scale coupling between microturbulence, AE, and MHD modes\cite{Liu2022PRL, MaYuehao2025}.

The idea \cite{Lin_Chen_2001, Holod_2009, Deng_2012} of calculating $E_\parallel$ directly from electron fluid response and using it to calculate the fluid (MHD) part of the $\delta A_\parallel$ has inspired the development of the mixed-variable algorithm\cite{Mishchenko2004MixtureVariable}. However, the electron fluid response in this algorithm only contains the ideal shear Alfv\'en wave (i.e., {when $E_\parallel=0$}), which is efficient for simulating macroscopic MHD modes but still needs to address the cancellation problem when simulating drift-Alfv\'enic turbulence\cite{Mishchenko2004MixtureVariable}. {A similar approach, "re-splitting method", is developed and used in GEM.\cite{GEM_resplitting}}

In this subsection, we present a unified GTC formulation solving the electron DKE starting from the fluid response, followed by the kinetic response using both the fluid–kinetic hybrid electron model and the conservative scheme, and finally reduced to the electrostatic limit. Finally, a model to simulate neoclassical tearing mode (NTM) is also implemented in GTC and presented in Appendix \ref{sec:NTMmodel}, where the pressure perturbation $\delta P$ is calculated from a diffusion equation with given diffusivity.

\subsubsection{Electron fluid equation}
We start from the electron continuity equation\cite{Dong_2017} by integrating nonlinear the DKE,
\begin{align}
\begin{split}
	&\frac{\partial\delta n_{e}}{\partial t}+\mathbf{B}_{0}\cdot\nabla\left(\frac{n_{0e}\delta u_{\parallel e}}{B_{0}}\right)+B_{0}\delta\mathbf{v}_{E}\cdot\nabla\left(\frac{n_{0e}}{B_{0}}\right) -n_{0}\left(\delta\mathbf{v}_{*}+\delta\mathbf{v}_{E}\right)\cdot\frac{\nabla B_{0}}{B_{0}}+\delta\mathbf{B}_{\perp}\cdot\nabla\left(\frac{n_{0e}u_{\parallel0e}}{B_{0}}\right)\\
	&-\frac{\nabla\times\mathbf{B_{0}}}{eB_{0}^{2}}\cdot\left(\nabla\delta P_{\parallel e}+\frac{\left(\delta P_{\perp e}-\delta P_{\parallel e}\right)\nabla B_{0}}{B_{0}}-n_{0e}e\nabla\delta\phi\right) +\nabla\cdot\left(\frac{\delta P_{\parallel e}\mathbf{b}_{0}\nabla\times\mathbf{b}_{0}\cdot\mathbf{b}_{0}}{eB_{0}}\right)\\
	&+\delta\mathbf{B}_{\perp}\cdot\nabla\left(\frac{n_{0e}\delta u_{\parallel e}}{B_{0}}\right)+B_{0}\mathbf{v}_{E}\cdot\nabla\left(\frac{\delta n_{e}}{B_{0}}\right)+\frac{\delta n_{e}}{B_{0}^{2}}\mathbf{b}_{0}\times\nabla B_{0}\cdot\nabla\phi+\frac{\delta n_{e}}{B_{0}^{2}}\nabla\times B_{0}\cdot\nabla\phi\\
	&-\frac{\mathbf{b}_{0}\times\nabla\delta B_{\parallel}}{e}\cdot\nabla\left(\frac{\delta P_{\perp e}+P_{\perp0e}}{B_{0}^{2}}\right)-\frac{\nabla\times\mathbf{b}_{0}\cdot\nabla\delta B_{\parallel}}{eB_{0}^{2}}\left(\delta P_{\perp e}+P_{\perp0e}\right)=0,\label{eq:Electron_Continuity}
\end{split}
\end{align}
with $\delta\mathbf{v}_{*} = \mathbf{b}_{0}\times\nabla\left(\delta P_{\parallel e}+\delta P_{\perp e}\right)/\left(n_{0e}m_{0e}\Omega_{e}\right)$, and $n_{0e}u_{\parallel0e}=-\nabla\times\mathbf{B}_0/(e\mu_0) + \sum_{s\neq e}Z_s n_{0s} u_{\parallel0s}/e$ denotes the electron equilibrium parallel flow. It should be pointed out that the continuity equation directly integrated from the electron DKE is in gyrocenter space, and $\delta n_e$ is the gyrocenter density instead of the electron particle density in the MHD equations. The difference between these two densities, i.e., the polarization density, includes two parts. The first part is caused by electrostatic potential and has the order of $k_\perp^2\rho_e^2$, which is usually negligible. While the second part is approximately $n_{0e}\delta B_\parallel/B_0$, and cannot be neglected. The electron parallel flow velocity $\delta u_{\parallel e}$ used in the continuity equation is solved from Ampere's law, Eq \eqref{eq:Ampere}.
\begin{equation}
    e n_{0e}\delta u_{\parallel e} = \frac{1}{\mu_0}\nabla_\perp^2\delta A_\parallel + \sum_{s\neq e}Z_s n_{0s} \delta u_{\parallel s},\label{eq:uepara}
\end{equation}
The $\delta \phi$ solved from Poisson equation, Eq \eqref{eq:Poisson_orig}, and the $\delta B_\parallel$ solved from Eq \eqref{eq:Bpara_orig} are also used in the continuity equation. The ion density and parallel flow velocity in these equations are solved from the ion Vlasov equation. $\delta A_\parallel$ is solved from the electron parallel momentum equation by using Eq \eqref{eq:uepara}
\begin{equation}
    en_{0e}\frac{\partial}{\partial t}\left[\left(-\frac{c^2}{\omega_{pe}^2}\nabla_\perp^2 + \frac{n_{0e} + \delta n_e}{n_{0e}}\right)\delta A_\parallel - \sum_{s\neq e}\frac{m_eZ_s}{e^2n_{0e}}n_{0s}\delta u_{\parallel s}\right] = \nabla\cdot\delta\mathbb{P} + \delta\Xi. \label{eq:Apara_dynamic}
\end{equation}
Where $\delta\mathbb{P}$ is the electron pressure tensor, and $-\delta\Xi + e(n_0+\delta n_e)\partial_t\delta A_\parallel$ is the parallel electromagnetic force acting on the electron fluid element, which includes the magnetic mirror force, the force from $\delta E_\parallel$, and the nonlinear ponderomotive force $\delta\mathbf{B}_\perp\cdot\nabla\delta\phi$ due to deviation from the Alfv\'enic state. The ponderomotive force is responsible for the generation of convective cells and zonal currents (i.e., the `dynamo' effect). The detailed expression of these two terms can be found in Appendix \ref{sec:EMomt}. Note that if we have the expression for pressure terms (including the pressure terms in $\delta \mathbb{P}$ and $\delta \Xi$), the system of Eqs \eqref{eq:Poisson_orig}, \eqref{eq:Bpara_orig}, \eqref{eq:Electron_Continuity}, \eqref{eq:uepara}, \eqref{eq:Apara_dynamic} is closed. For example, if the isothermal condition is used for electron $\delta P_{\perp e} = \delta P_{\parallel e} = \delta n_e T_{e}$, the electron kinetic effects are neglected from the system. Next, we show how the fluid and kinetic electron responses are included in the GTC simulation model.
\subsubsection{Electron drift kinetic equation}
We separate the electron distribution function into an analytic part and a non-analytic part, $\delta f_e = \delta f_e^a + \delta h_e$. Correspondingly, we separate the $\delta A_\parallel$ into an analytic part $\delta A_\parallel^a$ and a non-analytic part $\delta A_\parallel^{na}$, while $\delta A_\parallel^a$ is the leading order portion of $\delta A_\parallel$ with $k_\parallel\neq 0$. In addition, any perturbed field can be separated into a flux surface-averaged zonal part and the remaining non-zonal part, $\delta U(\psi,\theta,\zeta) = \delta U_{00}(\psi) + \delta U_{nz}(\psi,\theta,\zeta)$.  $\delta A_\parallel^a$ can be linked to a \textit{inductive} potential,
\begin{equation}
    \frac{\partial}{\partial t}\delta A_\parallel^a = \bunit\cdot\nabla\delta\phi_{ind}.
\end{equation}
And the remaining $\delta A_\parallel^{na}$ can be calculated using momentum equation,  Eq \eqref{eq:Apara_dynamic}
\begin{equation}
\begin{split}
    en_{0e}\frac{\partial}{\partial t}\left[\left(-\frac{c^2}{\omega_{pe}^2}\nabla_\perp^2+\frac{n_{0e}+\delta n_e}{n_{0e}}\right)\delta A_\parallel^{na} - \sum_{s\neq e} \frac{m_e Z_s}{e^2 n_{0e}} n_{0s}\delta u_{\parallel s} \right] =& \nabla\cdot\mathbb{P}^{na} + \delta\Xi^{na} \\
    &+ \frac{en_{0e}c^2}{\omega_{pe}^2}\nabla_\perp^2\left(\bunit\cdot\nabla\delta \phi_{ind}\right), \label{eq:Apara_na}
\end{split}
\end{equation}
where the expressions of $\delta\mathbb{P}^{na}$ and $\delta\Xi^{na}$ can be found in Appendix \ref{sec:EMomt}. In practice, the Laplacian terms and ion parallel flow terms in the left-hand side can be neglected due to small electron mass, and Eq \eqref{eq:Apara_na} becomes the electron parallel force balance equation.

We can now use the $\delta A_\parallel^a$ to define $\delta f_e^a$ from the leading order terms of the electron drift-kinetic equation
\begin{equation}
\begin{split}
    \frac{v_\parallel\mathbf{B_0}\cdot\nabla\delta f_e^{a}}{B_0} \equiv & -v_\parallel\frac{\delta\mathbf{B}_\perp^{a}}{B_0}\cdot\nabla f_{0e} - \frac{\mu v_\parallel}{B_0 T_e}\delta \mathbf{B}_\perp^{a}\cdot\nabla B_0 f_{0e}\\ &+ \frac{ev_\parallel}{T_e}\bunit\cdot\nabla\phi_{eff} f_{0e} - \frac{\mu v_\parallel}{T_e}\bunit\cdot\nabla\delta B_\parallel f_{0e} + e \frac{\delta\mathbf{B}_\perp^{a}}{B_0}\cdot\nabla\phi_{eq}\frac{v_\parallel}{T_e}f_{0e},\label{eq:0thVlasov}
\end{split}
\end{equation}
where we have assumed Maxwellian for the electron equilibrium distribution function, and the difference between $B_0$ and $B_\parallel^*$ has been dropped for the electron since $m_ev_\parallel\nabla\times\bunit/(eB_0)\ll1$. $\delta \mathbf{B}^a_\perp\equiv \bunit\times\nabla\delta A_\parallel^a$, $\phi_{eff}\equiv\delta\phi_{ind}+\delta\phi_{nz}$. The name $\phi_{eff}$ comes from the relation $E_\parallel=-\bunit\cdot\nabla\phi_{eff} - \partial_t\delta A_\parallel^{na}$, so $\phi_{eff}$ acts as an \textit{effective} analytic potential.
The solution of Eq \eqref{eq:0thVlasov} is
\begin{equation}
\begin{split}
    \delta f^{a}_e = \frac{e\phi_{eff}}{T_e} f_{0e} - \frac{\mu}{T_e}\delta B_{\parallel,nz} f_{0e} + \frac{\partial f_{0e}}{\partial\psi_0} \delta\psi^{a} + \frac{\partial f_{0e}}{\partial\alpha_0}\delta\alpha^{a} - \frac{e}{T_e}\frac{\partial\phi_{eq}}{\partial\psi_0} f_{0e}\delta\psi^{a},\label{eq:adiae_solution}
\end{split}
\end{equation}
where $\delta \psi^a$ and $\delta \alpha^{a}$ are defined through $\mathbf{B}_0\cdot\nabla\delta\psi ^a = -\delta\mathbf{B}_\perp^a\cdot\nabla\psi_0$, $\mathbf{B}_0\cdot\nabla\alpha^a=-\delta\mathbf{B}_\perp^a\cdot\nabla\alpha_0$, and $\psi_0$ and $\alpha_0$ comes from the Clebsch representation $\mathbf{B}_0=\nabla\psi_0\times\nabla\alpha_0$. $\delta\psi^a$ and $\delta\alpha^a$ can be solved from $\partial_t\delta\psi^a=-\partial_{\alpha_0}\delta\phi_{ind}$, $\partial_t\delta\alpha^a = \partial_{\psi_0}\delta\phi_{ind}$ once $\delta\phi_{ind}$ is determined. 

By integrating the above equation in velocity space, we can obtain $\delta\phi_{ind}$,

\begin{equation}
\begin{split}
    \frac{e\delta\phi_{ind}}{T_e} = \frac{\delta n_{e}^a}{n_{0e}} - \frac{e\delta\phi_{nz}}{T_e}+\frac{\delta B_{\parallel,nz}}{B_0} - \frac{\partial\ln n_{0e}}{\partial\psi_0}\delta\psi^{a} - \frac{\partial \ln n_{0e}}{\partial \alpha_0}\delta\alpha^{a} + \frac{e}{T_e}\frac{\partial\phi_{eq}}{\partial\psi_0}\delta\psi^{a}.\label{eq:phi_ind_general}
\end{split}
\end{equation}
Where $\delta n_e^a = \int d\mathbf{v}\delta f_e^a$ is the analytic electron density. In the case where electron kinetic effects are not important, $\delta h_e=0$ can be assumed, and $\delta n_e^a$ can be directly approximated by $\delta n_{e,nz}$. The pressure terms can also be calculated from $\delta f_e$, $\delta P_{\perp e} = \delta P_{\perp e}^{a} = \int d\mathbf{v}\mu B_0 \delta f_e$, $\delta P_{\parallel e} = \delta P_{\parallel e}^a = \int d\mathbf{v} mv_\parallel^2 \delta f_e^a$. Therefore, the system is closed, and no kinetic electron effect is included in the system.

To incorporate the kinetic response, $\delta h_e$ needs to be evaluated. The governing equation of $\delta h_e$ can be obtained by subtracting $\delta f_e^a$ from the original DKE equation,
\begin{equation*}
L\delta h_e = -\delta L f_{0e} - L \delta f_e^a.
\end{equation*}
Defining electron particle weight $w_e=\delta h_e/f_e$, we can get the equation for $w_e$, \begin{equation}
\begin{split}
    \frac{dw_e}{dt}=&L\frac{\delta h_e}{f_e} = -\left(1-\frac{\delta f_e^{a}}{f_{0e}}-w_e\right) \left(\frac{1+\delta f_{e}^{a}/f_{0e}}{f_{0e}}\delta Lf_{0e}+L\frac{\delta f_{e}^{a}}{f_{0e}}\right)\\
    =& -\left(1-\frac{\delta f_e^{a}}{f_{0e}}-w_e\right)\frac{1+\delta f_e^{a}/f_{0e}}{f_{0e}}\times\left\{\left(v_\parallel\frac{\delta \mathbf{B}_\perp}{B_\parallel^*} + \delta\mathbf{v}_E + \mathbf{v}_{b\parallel}\right)\cdot\nabla|_\mu f_{0e}\right.\\
    &\left.-\frac{1}{m_{e}}\left[\frac{\mu\delta\mathbf{B}_{\perp}\cdot\nabla\left(B_{0}-\frac{e}{\mu}\phi_{eq}\right)}{B_{\parallel}^{*}}-e\frac{\mathbf{B}^{*}}{B_{\parallel}^{*}}\cdot\nabla\left(\delta\phi-\frac{\mu}{e}\delta B_\parallel\right)-e\frac{\partial\delta A_{\parallel}}{\partial t}\right]\frac{\partial}{\partial v_{\parallel}}f_{0e}\right\}\\
    & -\left(1-\frac{\delta f_e^{a}}{f_{0e}}-w_e\right)\left(\frac{\partial}{\partial t}\frac{\delta f_{e}^{a}}{f_{0e}} + \dot{\mathbf{R}}\cdot \nabla|_\mu \frac{\delta f_{e}^{a}}{f_{0e}}+\dot{v}_{\parallel}\frac{\partial}{\partial v_\parallel}\frac{\delta f_e^{a}}{f_{0e}}\right).\label{eq:dwe_equation}
\end{split}
\end{equation}
After solving $\delta h_e$, the non-analytic part of the pressure terms are calculated following $\delta P_{\perp,e}^{na} = \int d\mathbf{v} \mu B_0\delta h_e$, and $\delta P_{\parallel, e}^{na} = \int d \mathbf{v} mv_\parallel^2 \delta h_e$. The integration can be done numerically in velocity space.

Note that to include the kinetic response, one needs to calculate $\delta n_e^a$ in Eq \eqref{eq:phi_ind_general}, $\delta n_e^a = \delta n_e - \int d\mathbf{v} \delta h_e$. So the equations of $\delta h_e$ and $\delta \phi_{ind}$ are coupled. In addition, $\delta\phi_{ind}$ is not formally defined. By properly choosing the definition of $\delta \phi_{ind}$ or $\delta A_\parallel^a$, we can solve the kinetic response. In the next two sections, we introduce two schemes that are implemented in GTC, which correspond to two ways to define $\delta \phi_{ind}$ or $\delta A_\parallel^a$.

\subsubsection{Conservative scheme}\label{sec:conservative_scheme}
The conservative scheme was developed in \cite{Bao_2017}, where the exact DKE is solved to preserve the full electron dynamics, including the collisionless tearing mode. {Note that the superscript `a' in the conservative scheme means \textit{analytic}}. In the conservative scheme, $\delta\phi_{ind}$ is chosen such that the analytic density is \textit{exactly} the non-zonal density,
\begin{equation}
    \int\delta f_{e}^{a}d\mathbf{v}=\delta n_{e,nz}\equiv\delta n_e^a.
\end{equation}
From Eq \eqref{eq:phi_ind_general}, we obtain the expression of $\delta \phi_{ind}$,
\begin{equation}
\begin{split}
    \frac{e\delta\phi_{ind}}{T_e} = \frac{\delta n_{e,nz}}{n_{0e}} - \frac{e\delta\phi_{nz}}{T_e}+\frac{\delta B_{\parallel,nz}}{B_0} - \frac{\partial\ln n_{0e}}{\partial\psi_0}\delta\psi^{a} - \frac{\partial \ln n_{0e}}{\partial \alpha_0}\delta\alpha^{a} + \frac{e}{T_e}\frac{\partial\phi_{eq}}{\partial\psi_0}\delta\psi^{a}.\label{eq:phi_ind_cons}
\end{split}
\end{equation}

We can solve the time derivative of $\delta f_e^a$ in Eq \eqref{eq:dwe_equation} by using
\begin{equation}
\begin{split}
    \frac{\partial \delta f_e^a}{f_{0e}\partial t} = &\frac{1}{n_{0e}}\left(\frac{\partial \delta n_e}{\partial t} - \frac{\partial \delta n_{e,00}}{\partial t}\right) + \frac{1}{B_0}\left(1-\frac{\mu B_0}{T_e}\right)\frac{\partial \delta B_{\parallel,nz}}{\partial t}+\left[\frac{1}{f_{0e}}\frac{\partial f_{0e}}{\partial T_e}\frac{\partial T_e}{\partial \psi_0} +\right.\\
    &\left.\frac{1}{f_{0e}}\frac{\partial f_{0e}}{\partial u_{\parallel0e}}\frac{\partial u_{\parallel0e}}{\partial \psi_0}\right]\frac{\partial \delta\psi^a}{\partial t} + \left[\frac{1}{f_{0e}}\frac{\partial f_{0e}}{\partial T_e}\frac{\partial T_e}{\partial\alpha_0}+\frac{1}{f_{0e}}\frac{\partial f_{0e}}{\partial u_{\parallel0e}}\frac{\partial u_{\parallel0e}}{\partial\alpha_0}\right]\frac{\partial \delta\alpha^a}{\partial t}. \label{eq:dfad_dt_cons}
\end{split}
\end{equation}
The terms in the first bracket can be evaluated from the continuity equation, $\partial_t\delta\alpha^a$ and $\partial_t\delta\psi^a$ can be replaced by $\partial_{\psi_0} \delta\phi_{ind}$ and $-\partial_{\alpha_0}\delta\phi_{ind}$. The $\partial_t\delta n_{e,00}$ term can be easily evaluated after calculating $\partial _t\delta n_e$ when solving the continuity equation. The detailed implementation to solve $w_e$ can be seen in Appendix \ref{sec:eqs_in_boozer}. Note that the $\delta\mathbf{B}_\perp$ in Eqs \eqref{eq:Electron_Continuity}, \eqref{eq:Apara_na}, \eqref{eq:dwe_equation} is the total magnetic perturbation.

The non-zonal density in Eq \eqref{eq:phi_ind_cons} includes the contribution from wave-particle resonance, so the superscript `a' in the conservative model actually represents an \textit{analytic} contribution instead of the \textit{adiabatic} one.

\subsubsection{fluid–kinetic
hybrid electron model }\label{sec:hybrid_scheme}
The fluid-kinetic hybrid scheme has been developed\cite{Lin_Chen_2001, Holod_2009, Deng_2012, Dong_2017} by expanding the electron DKE using the small electron-to-ion mass ratio to reduce the particle noise and to overcome the electron Courant condition. {Note that in the hybrid scheme, all superscript `a' means \textit{adiabatic}, which will be explained later}.Here we separate $\delta A_\parallel$ according to the $k_\parallel$ component, $\delta A_\parallel = \delta A_\parallel^{a} + \delta A_\parallel^{na}$, where the $\delta A_\parallel^{a}$ are defined through Eq \eqref{eq:Apara_dynamic} by only keeping the linear terms with $k_\parallel\neq 0$ while taking the limit $m_e=0$, and the $\delta A_\parallel^{na}$ includes all remaining nonlinear terms, terms with $k_\parallel=0$, and the kinetic terms caused by finite $m_e$.
In turn, $\delta \phi_{ind}$ is defined though $\bunit\cdot\nabla\phi_{ind}=\partial_t \delta A_\parallel^{a}$.

Since we have defined $\delta A_\parallel^a$, there is no freedom to adjust the definition of $\delta\phi_{ind}$, and we only have the relation Eq \eqref{eq:phi_ind_general}. 
We can notice that, unlike the conservative scheme, here we have $\delta f_e = \delta f_e^a + \delta h_e$, and $\delta n_e^a=\int \delta f_e^a$ only contains the adiabatic part, and has no contribution from wave-particle resonance. So the superscript `$a$' means \textit{adiabatic} in the hybrid scheme. All perturbed quantities in Eq \eqref{eq:phi_ind_general} have no $k_\parallel=0$ components. Because the equations of $\delta h_e$ and $\delta \phi_{ind}$ are coupled, the equations must be solved order by order based on a smallness parameter of $\sqrt{m_e/(m_i\beta_e)}$\cite{Lin_Chen_2001}. We expand $\delta \phi^{(0)}_{ind}$ to $\delta\phi_{ind}^{(0)}+\delta\phi_{ind}^{(1)}+\cdots$. For $k$-th order, a corresponding adiabatic response $\delta f_{e}^{a,{(k)}}$ can be defined to satisfy Eq \eqref{eq:adiae_solution}.

In the 0-th order, $\delta h_e^{(0)}$ is assumed to be 0, we can solve the 0-th order $\delta\phi_{ind}^{(0)}$,
\begin{equation*}
\begin{split}
    \frac{e\delta\phi_{ind}^{(0)}}{T_e} = \frac{\delta n_{e,nz}}{n_{0e}} - \frac{e\delta\phi_{nz}}{T_e}+\frac{\delta B_{\parallel,nz}}{B_0} - \frac{\partial\ln n_{0e}}{\partial\psi_0}\delta\psi^{a} - \frac{\partial \ln n_{0e}}{\partial \alpha_0}\delta\alpha^{a} + \frac{e}{T_e}\frac{\partial\phi_{eq}}{\partial\psi_0}\delta\psi^{a}.
\end{split}
\end{equation*}
Then $\delta h_e^{(0)}$ and $\delta\phi_{ind}^{(0)}$ are used in the governing equations for $\delta \psi^a$ and $\delta\alpha^a$ and the pressure terms in continuity equation. The adiabatic parallel potential is solved from $\partial _t\delta A_\parallel^a = \bunit\cdot\nabla\delta\phi_{ind}^{(0)}$.  Note that even in the 0-th order, $\delta A_\parallel^{na}$ is finite due to nonlinear effects. The important non-resonant parallel flow led by the ponderomotive force is included, $\partial_t\delta A_\parallel^{na}\sim \{\delta\mathbf{B}_\perp\cdot\nabla[\delta\phi_{ind}-T_e\delta B_\parallel/(eB_0)]/B_0\}_{k_\parallel=0}$. The exact $\delta A_\parallel^{na}$ should be solved from Eq \eqref{eq:Apara_na}. In the hybrid scheme, the linear part of $\delta A_\parallel$ with $k_\parallel=0$ is intentionally dropped, thus removing the linear parallel acceleration from $E_\parallel$ with $k_\parallel=0$ component. Therefore, the hybrid scheme does not include the tearing parity, and the linear tearing mode is excluded from the system. While the passive tearing component in $\delta A_\parallel$ can be observed due to nonlinear interaction \cite{Dong_2019}. At the 0-th order, the definition of $\delta \phi_{ind}$ in the conservative scheme is the same as $\delta \phi_{ind}^{(0)}$ in the hybrid scheme, and the two schemes are identical.

In \cite{Lin_Chen_2001}, the linear dispersion relation is discussed, and the nonlinear $\delta A_\parallel^{na}$ does not appear in the linear model. In \cite{Holod_2009}, the zonal part of Ampere's law is calculated, but the nonlinear non-resonance parallel flow associated with $\delta\mathbf{B}_\perp\cdot\nabla(\delta\phi-\frac{T_e}{e}\delta B_\parallel/B_0)/B_0$ term in $\delta A_\parallel^{na}$ is attributed to $\delta h_e$. As a result, the nonlinear non-resonance parallel flow can only be retained after calculating $\delta h_e$, leading to the inconsistency between the fluid electron model in GTC and the common MHD theory. In \cite{Dong_2019}, several important nonlinear terms are retained in $\delta A_\parallel^{na}$. {The complete $\delta A_\parallel ^{na}$ presented in this paper (solved from Eq \eqref{eq:Apara_na}) has incorporated higher order nonlinear terms compared to previous works.}

Next, we show how to include the higher-order kinetic corrections $\delta h_e$. The higher-order $\delta\phi_{ind}$ and $\delta f_e^{a}$ are needed before being inserted into any other equations. $\delta\phi_{ind}^{(0)}$ and the corresponding $\delta f_{e}^{a,(0)}$ are used in the electron weight equation Eq \eqref{eq:dwe_equation} to calculate the first order electron kinetic response $\delta h_e^{(1)}$. In Eq \eqref{eq:dwe_equation}, the nonlinear ponderomotive force term and the non-adiabatic $\partial_tA_\parallel^{na}$ term are removed together, since the parallel acceleration from ponderomotive force are mostly balanced with the nonlinear $\delta A_\parallel^{na}$, $[\delta\mathbf{B}_\perp/B_0\cdot\nabla(\delta\phi_{ind}-\frac{\mu}{e}\delta B_\parallel)]_{k_\parallel=0}+\partial_t \delta A_\parallel^{na}\approx 0$.

Then $\delta h_e^{(1)}$ is used in Eq \eqref{eq:phi_ind_general} to calculate $\delta\phi_{ind}^{(1)}$, and accordingly $\delta f_e^{a,(1)}$ can be obtained. This procedure can be repeated until the desired accuracy is obtained. A tricky part is the $\partial_t \delta f_{e}^{a} / f_{0e}$ term in Eq \eqref{eq:dwe_equation}. Due to the different definition of $\delta\phi_{ind}$ than that in the conservative scheme, the time derivative of $\delta f_{e}^a$ in the hybrid scheme can be written as
\begin{equation}
\begin{split}
    \frac{\partial \delta f_e^a}{f_{0e}\partial t} = &\frac{1}{n_{0e}}\frac{\partial \delta n_{e,nz}^a}{\partial t} + \frac{1}{B_0}\left(1-\frac{\mu B_0}{T_e}\right)\frac{\partial \delta B_{\parallel,nz}}{\partial t}+\left[\frac{1}{f_{0e}}\frac{\partial f_{0e}}{\partial T_e}\frac{\partial T_e}{\partial \psi_0} +\right.\\
    &\left.\frac{1}{f_{0e}}\frac{\partial f_{0e}}{\partial u_{\parallel0e}}\frac{\partial u_{\parallel0e}}{\partial \psi_0}\right]\frac{\partial \delta\psi^a}{\partial t} + \left[\frac{1}{f_{0e}}\frac{\partial f_{0e}}{\partial T_e}\frac{\partial T_e}{\partial\alpha_0}+\frac{1}{f_{0e}}\frac{\partial f_{0e}}{\partial u_{\parallel0e}}\frac{\partial u_{\parallel0e}}{\partial\alpha_0}\right]\frac{\partial \delta\alpha^a}{\partial t}. \label{eq:dfad_dt_hybr}
\end{split}
\end{equation}
To simplify the numerical operation, we choose to use the final value of $\partial_t n_{e,nz}^a$, $\partial_{\alpha_0}\delta\phi_{ind}$, $\partial_{\psi_0}\delta\phi_{ind}$ instead of using the $k$-th order values. To calculate $\partial_t\delta n_{e,nz}^a$ and $\delta B_{\parallel,nz}$, we need to store the $\delta n_{e,nz}^{a}$ and $\delta B_{\parallel,nz}$ values for current step and previous step, and directly take the time derivative. However, this will cause a time step mismatch between $dh_e/dt$ and $\partial_t\delta f_e^a$. The simulation time step must be small enough to overcome the numerical error due to this operation. {The time step size can usually be set through a convergence study.} After the iterations, the complete $\delta \phi_{ind}$ and $\delta h_e$ can be used for the next step equations.

\subsubsection{Reduction to electrostatic simulation model}
The fluid–kinetic hybrid electron model for electromagnetic scenario is not valid when $\beta_e < {m_e}/{m_i}$, where the shear-Alfv\'{e}n wave phase velocity is faster than the electron thermal velocity. In this low-$\beta_e$ regime, the electrostatic perturbations dominate the turbulence dynamics, so the hybrid model can be reduced to the electrostatic model. This scheme was developed in \cite{Lin_2007}. Similar to the split-weight algorithm, we also separate the electron response into adiabatic and non-adiabatic responses. However, using the iterative hybrid algorithm, we can overcome the electron Courant condition. In the electrostatic simulation, the analytic part of the electron response is defined as the adiabatic response:
\begin{equation}
    \delta f_e^{a} \equiv \frac{e\delta\phi_{nz}}{T_e} f_{0e}. \label{eq:fe_ad_es}
\end{equation}
The governing equation of the electron weight equation becomes
\begin{equation}
\begin{split}
    \frac{dw_e}{dt}
    =& -\left(1-\frac{e\delta\phi_{nz}}{T_e}-\delta h_e\right)\frac{1+e\delta\phi_{nz}/T_e}{f_{0e}}\times\left(\delta\mathbf{v}_E \cdot\nabla|_\mu f_{0e}\right.\\
    &\left.+\frac{e}{m_{e}}\frac{\mathbf{B}^{*}}{B_{\parallel}^{*}}\cdot\nabla\delta\phi\frac{\partial}{\partial v_{\parallel}}f_{0e}\right)\\
    & -\left(1-\frac{e\delta\phi_{nz}}{T_e}-\delta h_e\right)\left(\frac{\partial}{\partial t}\frac{e\delta\phi_{nz}}{T_e} + \dot{\mathbf{R}}\cdot \nabla|_\mu \frac{e\delta\phi_{nz}}{T_e}\right).\label{eq:dw_es}
\end{split}
\end{equation}
In the leading order, we assume $\delta h_e^{(0)} = 0$, so $\delta n_{e,nz}^{(0)} = n_{0e} e\delta\phi_{nz}^{(0)}/T_e$, and the Poisson equation can be solved. Note that this leading order assumption is only for the non-zonal component of electrons, so only the non-zonal potential is solved.
\begin{equation}
    \sum_{s\neq e}\frac{Z_sn_{0s}}{T_s}\left(\delta\phi^{(0)}_{nz} - \delta\tilde{\phi}_s^{(0)}\right) + \frac{e^2\delta\phi^{(0)}_{nz}}{T_e}n_{0e} = \sum_{s\neq e} Z_s \delta \bar{n}_{nz,s}. \label{eq:Poisson_es_ad}
\end{equation}
Then this $\phi^{(0)}_{nz}$ is substituted in Eqs \eqref{eq:fe_ad_es} and \eqref{eq:dw_es} to solve $\delta f_e^{a,(1)}$ and $\delta h_e^{(1)}$. The non-adiabatic density can be added to Eq \eqref{eq:Poisson_es_ad} to solve the next order potential
\begin{equation}
    \sum_{s\neq e}\frac{Z_sn_{0s}}{T_s}\left(\delta\phi^{(1)}_{nz} - \delta\tilde{\phi}_s^{(1)}\right) + \frac{e^2\delta\phi^{(1)}_{nz}}{T_e}n_{0e} = \sum_{s\neq e} Z_s \delta \bar{n}_{nz,s} - e\left(\int d\mathbf{v}\delta h_e^{(1)}\right)_{nz}. \label{eq:Poisson_es_ki}
\end{equation}
This iteration process can be repeated until the desired accuracy is reached. Then the zonal component potential can be solved using the flux-averaged Poisson equation.
\begin{equation}
\sum_{s\neq e} \frac{Z_s n_{0s}}{T_e} \left(\delta \phi_{00} - \delta \tilde{\phi}_{00,s}\right) = \sum_{s\neq e}Z_s\delta \bar{n}_{00,s} - e \left(\int d\mathbf{v}\delta h_e\right)_{00}
\end{equation}

Besides the iterative hybrid algorithm, the direct drift-kinetic electron equation solver is implemented in GTC. The electron equation will be the same as the ion equations, except that all gyro-averaging is ignored. This model has been used for the simulations with static magnetic islands\cite{Fang2019a}. Another different regime is the short wavelength limit $k_\perp\rho_i \gg 1$, where ions can be regarded as adiabatic, $\delta n_i=-n_{0i}Z_i\phi/T_e$, and only the electron equations are solved using the complete gyrokinetic equation. This model has been used for the simulations of the electron temperature gradient(ETG) mode\cite{Lin_2007}.

\subsection{Reduction to fluid models}\label{sec:MHD_eqns}
{The idea of separating the spatial-temporal scales of gyromotion to obtain the MHD equations can trace back to the early work of Chew \textit{et al}\cite{CGL_1956}, followed by the work of Frieman \textit{et al}\cite{Frieman_1966} and Kulsrud \cite{Kulsrud_1980}. Lee\cite{Lee_Qin_2003,Lee_2016} has also derived the MHD equations based on the gyrokinetic equations.} To demonstrate that the GTC gyrokinetic model contains the MHD physics and to delineate various fluid and kinetic physics in the gyrokinetic simulation, we reduce the GTC formulation to the two-fluid model first and then to the single-fluid MHD model in the limit of long wavelength and when the thermal ion kinetic effect can be neglected. To simplify the fluid model in GTC, we take a limit of low ion temperature, where the ion continuity equation and ion parallel momentum equations become:
\begin{align}
\begin{split}
    & \frac{\partial\delta n_{i}}{\partial t}+\nabla\cdot\left[\left(n_{0i}+\delta n_i\right)\mathbf{v}_{E}+n_{0i}\delta u_{\parallel i}\frac{\mathbf{B_0}+\delta\mathbf{B_\perp}}{B_0}\right]
    =0,\\
    & n_{0i}\frac{\partial \delta u_{\parallel i}}{\partial t} + \nabla\cdot\left(n_{0i}\delta u_{\parallel i}\mathbf{v}_E \right) + \frac{Z_i}{m_i}n_{0i}{\mathbf{b}_0}\cdot\nabla\phi_{eff}=0.
 \label{eq:Ion_Continuity}
\end{split}
\end{align}
where we have transformed to the ion frame such that $u_{\parallel0i}=0$,$u_{\parallel0e}=-J_{\parallel0}/(en_{0e})$. $\partial _t\delta u_{\parallel i} \ll \partial_t\delta u_{\parallel e}$ is assumed due to the large mass ratio. So we can merely use Eq \eqref{eq:Ion_Continuity} to replace the ion Vlasov equation to calculate $\delta n_i$ and $\delta u_{\parallel i}$. Note that the parallel acceleration from the electric field in \eqref{eq:Ion_Continuity} is essential to the ion acoustic wave. The {closed two-fluid model} is then formed by 
Eqs \eqref{eq:Ion_Continuity}, \eqref{eq:Ampere}, \eqref{eq:Poisson_orig}, \eqref{eq:Bpara_orig}, \eqref{eq:0thVlasov}, \eqref{eq:phi_ind_general} or \eqref{eq:phi_ind_cons}, \eqref{eq:Electron_Continuity}, \eqref{eq:delpperp_orig}, \eqref{eq:delppara_orig},  \eqref{eq:Ampere}, and \eqref{eq:Apara_na}, assuming $\delta h_e=0$. Note that here we need to assign all the pressure to the electron pressure to keep the equilibrium force balance. The ion pressure gradient terms are neglected in the ion model to avoid the need for an ion equation of state and to derive the single-fluid ideal MHD equation. Note that GTC does not intend to implement a comprehensive two-fluid model. A more comprehensive ion fluid equation can be found in \cite{Deng_2012}.

The two-fluid simulation model can be further simplified to a single-fluid ideal MHD model by assuming that $E_{\parallel}$ and accordingly $\phi_{eff}$ can be neglected. By defining $\delta n=\delta n_{e}-\sum_{s\neq e}Z_{s}\delta n_{s}/e,$ we get the continuity equation for the gyrocenter charge density,
\begin{align}
\begin{split}
 & \frac{\partial\delta n}{\partial t}+\mathbf{B}_{0}\cdot\nabla\left(\frac{n_{0}\delta u_{\parallel}}{B_{0}}\right)-n_{0}\delta \mathbf{v}_{*}\cdot\frac{\nabla B_{0}}{B_{0}}+\delta\mathbf{B}_{\perp}\cdot\nabla\left(\frac{n_{0}u_{\parallel0}}{B_{0}}\right)\\
 & -\frac{\nabla\times\mathbf{B_{0}}}{eB_{0}^{2}}\cdot\left(\nabla\delta P_{\parallel}+\frac{\left(\delta P_{\perp}-\delta P_{\parallel}\right)\nabla B_{0}}{B_{0}}\right)+\nabla\cdot\left(\frac{\delta P_{\parallel}\mathbf{b}_{0}\nabla\times\mathbf{b}_{0}\cdot\mathbf{b}_{0}}{eB_{0}}\right) + \delta \mathbf{B}_\perp\cdot\nabla\left(\frac{n_0 \delta u_\parallel}{B_0}\right)\\
 & -\frac{\mathbf{b}_{0}\times\nabla\delta B_{\parallel}}{e}\cdot\nabla\left(\frac{\delta P_\perp + P_{\perp0}}{B_{0}^{2}}\right)-\frac{\nabla\times\mathbf{b}_{0}\cdot\nabla\delta B_{\parallel}}{eB_{0}^{2}}\left(\delta P_\perp + P_{\perp0}\right)=0,\label{eq:MHD_continuity}
\end{split}
\end{align}
where $\delta P=\delta P_{e},P_{0}=P_{0e}$, $n_{0}=n_{0e}$, and
$e\delta u_{\parallel}=e\delta u_{\parallel e}-\sum_{s\neq e} n_{0s}Z_s\delta u_{\parallel s}/{n_{0e}},eu_{\parallel0}=eu_{\parallel0e}-\sum_{s\neq e}n_{0s}Z_{s}u_{\parallel0s}/n_{0e}$. $\delta u_{\parallel}$ is still solved from parallel Ampere's law,
\begin{equation}
    \delta u_{\parallel}=\frac{1}{\mu_{0}en_{0e}}\nabla_{\perp}^{2}\delta A_{\parallel}.\label{eq:Ampere_MHD}
\end{equation}
$\phi_{ind}=-\phi$ is used when calculating $\delta A_\parallel^{a}$ and $\delta A_\parallel^{na}$, assuming $\delta E_\parallel = 0$ in the ideal MHD limit.

The quasi-neutrality condition effectively reduces to
\begin{equation}
    \frac{c^{2}}{v_{A}^{2}}\nabla_{\perp}^{2}\phi=\frac{e\delta n}{\epsilon_{0}}\label{eq:poisson_MHD}
\end{equation}
in the low ion temperature limit, where $c$ is the speed of light, $v_{A}$ the Alfv\'{e}n velocity, and $\epsilon_{0}$ the dielectric constant of vacuum.
In the ideal MHD limit, Eq \eqref{eq:Bpara_reduced} is given passively by the perpendicular force balance.
\begin{equation}
\frac{\delta B_{\parallel}}{B_{0}}=-\frac{\beta_e}{2}\frac{\delta P_\perp}{P_{\perp 0}}=-\frac{\beta_{e}}{2}\frac{\partial P_{\perp0}}{\partial\psi_{0}}\frac{\delta\psi}{P_{\perp0}}\label{eq:bpara_MHD}
\end{equation}

The {single fluid simulation model} is composed by Eqs \eqref{eq:MHD_continuity}, \eqref{eq:Ampere_MHD}, \eqref{eq:poisson_MHD}, \eqref{eq:bpara_MHD}, \eqref{eq:0thVlasov}, \eqref{eq:phi_ind_general} or \eqref{eq:phi_ind_cons}, \eqref{eq:delpperp_orig}, \eqref{eq:delppara_orig},  and \eqref{eq:Apara_na}, assuming $\delta h_e=0$. We should note that $\phi_{eff}$ in these equations should be explicitly set to 0, and the perturbed electron pressure in Eqs \eqref{eq:delpperp_orig} and \eqref{eq:delppara_orig} stands for the total perturbed pressure.

We can recover the linear ideal MHD dispersion relation from this single-fluid simulation model. Combining the quasi-neutrality condition in long wavelength limit, Eq \eqref{eq:poisson_MHD}, the Ampere's law, Eq \eqref{eq:Ampere_MHD}, the single-fluid continuity equation, Eq \eqref{eq:MHD_continuity}, the $B_\parallel$ equation, Eq \eqref{eq:bpara_MHD}, and by further assuming $k_\parallel\ll k_\perp$, $k_\perp L_B \gg 1$,the commonly used ideal MHD dispersion relation\cite{Fu_2006, Bao_2021} can be found,
\begin{equation*}
    \begin{aligned}
        0=&\frac{\omega^{2}}{v_{A}^{2}}\nabla_{\perp}^{2}\delta\phi+i\mathbf{B}_{0}\cdot\mathbf{\nabla}\left(\frac{\nabla_{\perp}^{2}\left(k_{\parallel}\phi\right)}{B_{0}}\right)+i\mathbf{b}_{0}\times\nabla\left(k_{\parallel}\phi\right)\cdot\nabla\left(\frac{\mu_0J_{\parallel0}}{B_{0}}\right)\\
        &-i\omega\mu_{0}\frac{2\mathbf{b}_{0}\times\boldsymbol{\kappa}}{B_{0}}\cdot\nabla\delta P.
    \end{aligned}
\end{equation*}
A brief derivation is presented in Appendix \ref{sec:MHD_DR}. Note that the kinetic effect from the ion diamagnetic frequency $\omega^{*}_{Pi}$ does not appear in the dispersion relation, since we have dropped the ion pressure terms in the ion fluid equation.

In the ideal MHD simulation, we have shown that the incorporation of $\delta B_\parallel$ component in the continuity equation is critical to the kink instability calculation. This important effect of $\delta B_\parallel$ is also found for other instabilities.\cite{Berk_1977, Tang_1980, Dong_2017}. Another important parameter for kink modes is the equilibrium parallel current. In particular, we show that the poloidal variation of $J_{\parallel0}$, which is normally ignored in other gyrokinetic simulations, needs to be calculated accurately. In Appendix \ref{sec:Jpara} we present the numerical method used in GTC for $J_{\parallel0}$ calculation.

The fluid simulation model is verified by the simulation of RSAE. For this verification, we reduce the thermal ion temperature to $T_i=T_e/10000\approx0$. Both the thermal ions and electrons can be well described by the fluid model. We carried out three simulations, the first of which used the gyrokinetic thermal ion model and fluid electron model, the second one used the fluid thermal ion model and fluid electron model, and the third one used the single fluid model for thermal ions and electrons. The energetic particles are described by the gyrokinetic model in all three simulations. Table \ref{fluid_verification} shows that the three simulations agree well on the real frequency and growth rate. We have also carried out simulations on other modes, like the internal kink mode and NTM, to verify the implementation of the fluid model.

\begin{table}[h!]
\centering
\caption{Verification of two fluid model (TF) and single fluid (SF) model using gyrokinetic ion model with $T_i\approx0$}\label{fluid_verification}
\begin{tabular}{|c|c|c|}
\hline
\textbf{Simulation model} & \textbf{Real Frequency (kHz)} & \textbf{Growth Rate ($\times10^3/$s)} \\ \hline
GK-ion & 43.90 & 29.49 \\ \hline
Two Fluid & 44.00 & 29.58 \\ \hline
Single Fluid & 44.03 & 29.82 \\ \hline
\end{tabular}
\end{table}

\section{Internal kink mode simulation}

\begin{figure}[H]
    \includegraphics[scale=0.5]{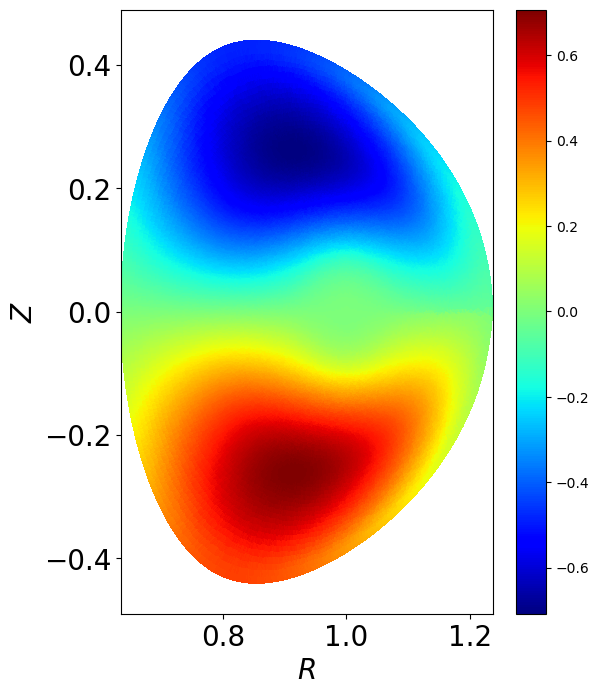}
    \centering
    \caption{$\hat{\delta}$ current. Normalized by $1/R_0$, where $R_0$ is the major radius.}
    \label{delta_function}
\end{figure}

{The verifications of the simulation models described in Section 2 have been performed for many physics previously, e.g., in \cite{REWOLDT2007,Taimourzadeh_2019}. While the successful verification and validation of the current-driven MHD modes have been carried out for the first time recently\cite{Brochard_2022}.} The DIII-D experimental shot \#141216 is selected, which exhibits the characteristics of kink instability. Using the equilibrium constructed from the experiments, several codes including GTC, GAM-solver, M3D-C1, NOVA-K, XTOR-K have been benchmarked on the kink instability with and without kinetic effects. Then, GTC is used to conduct more than 5000 simulations in which the equilibria are constructed from DIII-D shots. The simulation data are used to build a database and train a surrogate model based on deep learning methods \cite{Dong_2021}. In this section, we show the important physical parameters for simulating kink modes and the physical insights on the excitation of kink modes we have learned from the simulations.

Since kink instability is driven primarily by parallel current and pressure gradient, it is necessary to evaluate $J_{\parallel0}$ accurately. In the Boozer coordinate system, the equilibrium magnetic field and the equilibrium parallel current can be expressed as 
\begin{equation}
\begin{aligned}
    B_{0} &= \hat{\delta}\nabla\psi + I\nabla\theta + g\nabla\zeta\\
    J_{\parallel0} &= \frac{1}{\mu_0}\frac{1}{\mathcal{J} B_0}\left[(I'-\partial_\theta\hat{\delta})g-g'I\right],\label{eq:J_from_delta}
\end{aligned}
\end{equation}
where $\mathcal{J}=(gq+I)/B_0^2$ is the Jacobian of Boozer coordinates, $\hat{\delta}$ represents the non-orthogonality of the basis vectors of Boozer coordinates,
\begin{equation*}
    \hat{\delta} = -\frac{I\nabla\psi\cdot\nabla\theta+g\nabla\psi\cdot\nabla\zeta}{\nabla\psi\cdot\nabla\psi}.
\end{equation*}

\begin{figure}[H]
\subfigure[$m=0$ component]{
\begin{minipage}[t]{0.5\linewidth}
    \centering
    \includegraphics[scale=0.5]{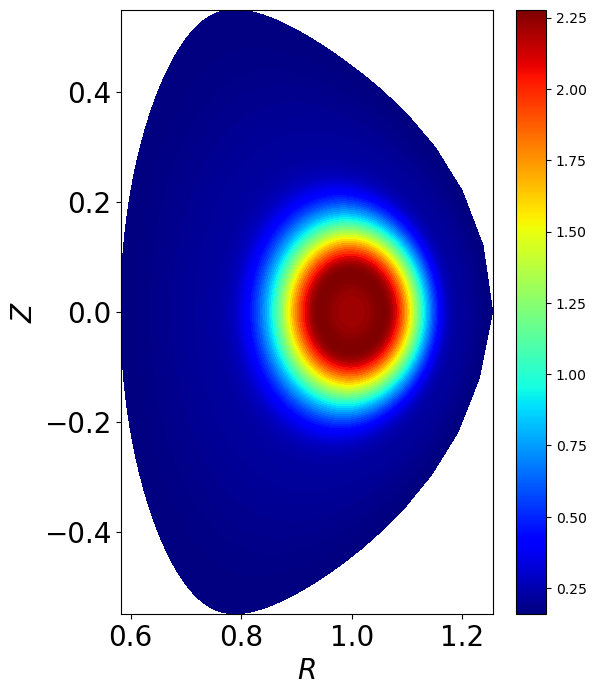}
\end{minipage}
}
\subfigure[$m\neq0$ component]{
\begin{minipage}[t]{0.5\linewidth}
    \centering
    \includegraphics[scale=0.5]{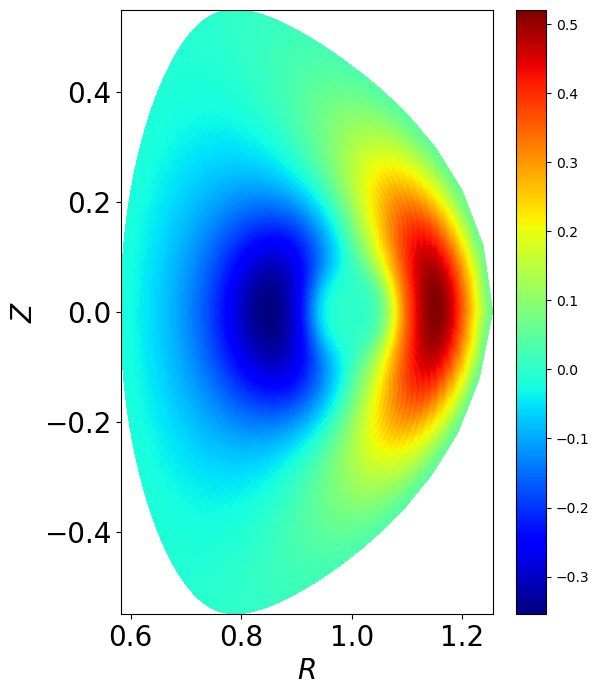}
\end{minipage}
}
\caption{Parallel current term $\mu_0 J_{\parallel0}/B_0$. Normalized by $1/R_0$.}
\label{J_para}
\end{figure}

\begin{figure}[H]
    \includegraphics[scale=0.5]{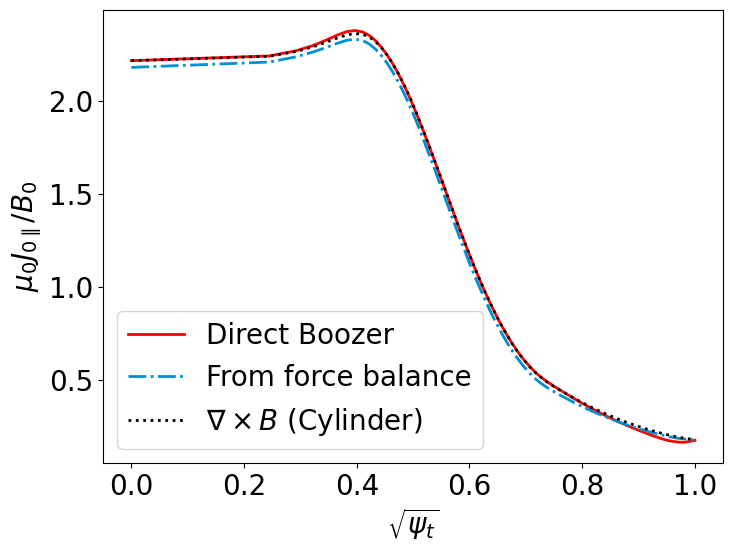}
    \centering
    \caption{Parallel current term from different methods implemented for Boozer coordinate (See Appendix \ref{sec:Jpara}). The reference value shown is calculated from $\nabla\times \mathbf{B}_0$ in cylindrical coordinates. The poloidal angle $\theta=0$. The value is normalized by $1/R_0$. }
    \label{J_calculation}
\end{figure}

The $\hat{\delta}$-current in the benchmark case is shown in Figure \ref{delta_function} for DIII-D \#141216. It can be seen that $\hat{\delta}$ is small near the magnetic axis, and has a large m=1 component when $\varepsilon$ is large. The current components from $g'$ and $I'$ do not vary along the flux surface, since $g$, $I$, and $q$ are functions of $\psi$ only. In previous simulations using Boozer coordinates, $\hat{\delta}$ is often neglected for simplicity. However, our simulation benchmark for DIII-D equilibrium shows that the neglect of $\hat{\delta}$ will lead the growth rate of kink instabilities to increase from $4.2\times 10^{4} \text{s}^{-1}$ to $22\times10^{4} \text{s}^{-1}$.\cite{Brochard_2022} The result shows the current contribution from $\hat{\delta}$ must be considered. Figure \ref{J_para} shows the equilibrium parallel current on the poloidal cross-section. The poloidally varying part in Figure \ref{J_para}b comes from $\partial_\theta \hat{\delta}$. The poloidally varying parallel current can be in the same order as or even larger than the m=0 component, especially when $\varepsilon$ is large. In Figure \ref{J_calculation}, we show that given the benchmark equilibrium, the parallel current term calculated from different numerical methods matches very well. Meanwhile, the kink instability is extremely sensitive to the parallel current. We have implemented three different numerical methods to calculate $J_{\parallel0}$ in Boozer coordinates. The first one is directly using $\bunit\cdot\nabla\times\mathbf{B}_0/\mu_0$ in the cylindrical coordinates to get $J_{\parallel0}\left(R,Z\right)$, and use the relation $R(\psi,\theta)$ and $Z(\psi,\theta)$ to map $J_{\parallel0}$ to Boozer coordinates. The second method is to use Eq \eqref{eq:J_from_delta}, which requires calculating the higher-order terms $\hdelta$ and $(\phi-\zeta)$ accurately. The third method makes use of the force balance in the axisymmetric geometry to express $J_{\parallel0}$ using $p(\psi)$ and $g(\psi)$, which requires the least computations and gives high accuracy easily. The equilibrium currents from these three methods are shown in Figure \ref{J_calculation}, and are labeled as '$\nabla\times \mathbf{B}$(Cylinder)', 'Direct Boozer', and 'From force balance', respectively. The second and third methods are described in Appendix \ref{sec:Jpara}. The simulations show that even using the 3 current profiles, which are very close to each other, would cause about a 20\% difference in the linear growth rate of the kink mode.

\begin{figure}[H]
    \subfigure[Relative amplitude of $\delta B_\parallel$]{
    \begin{minipage}[t]{0.5\linewidth}
        \centering
        \includegraphics[scale=0.25]{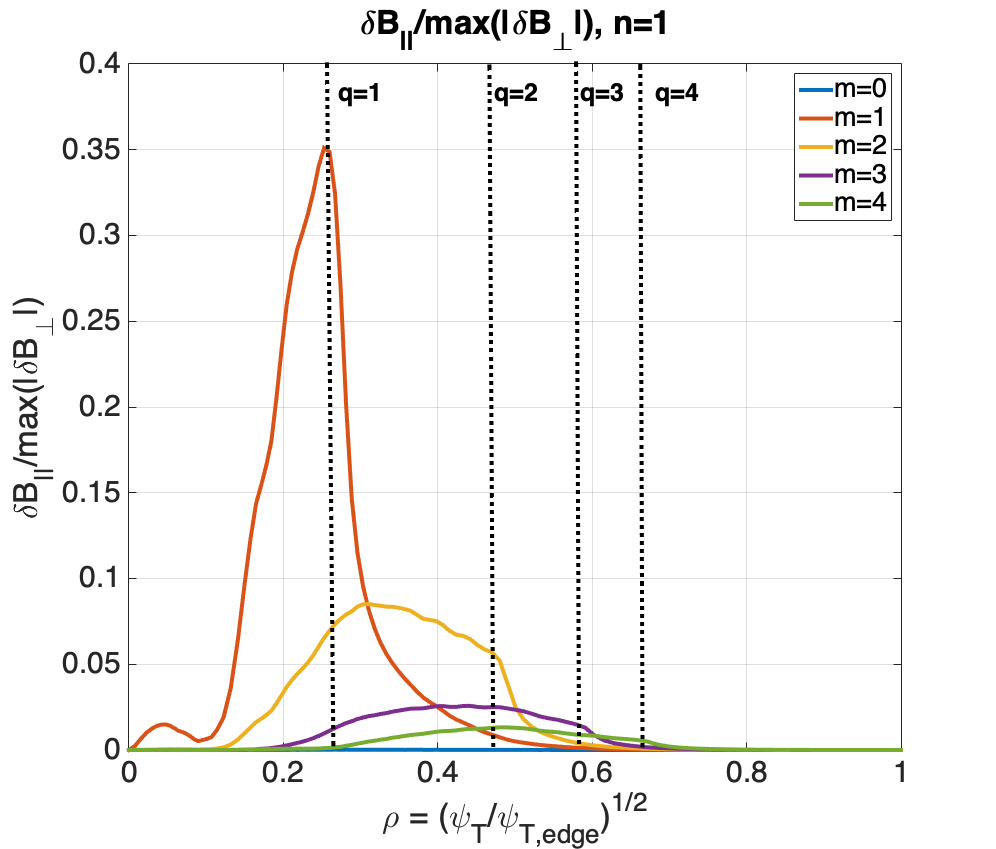}
    \end{minipage}
    }
    \subfigure[Relative amplitude of $\delta B_\perp$]{
    \begin{minipage}[t]{0.5\linewidth}
        \centering
        \includegraphics[scale=0.25]{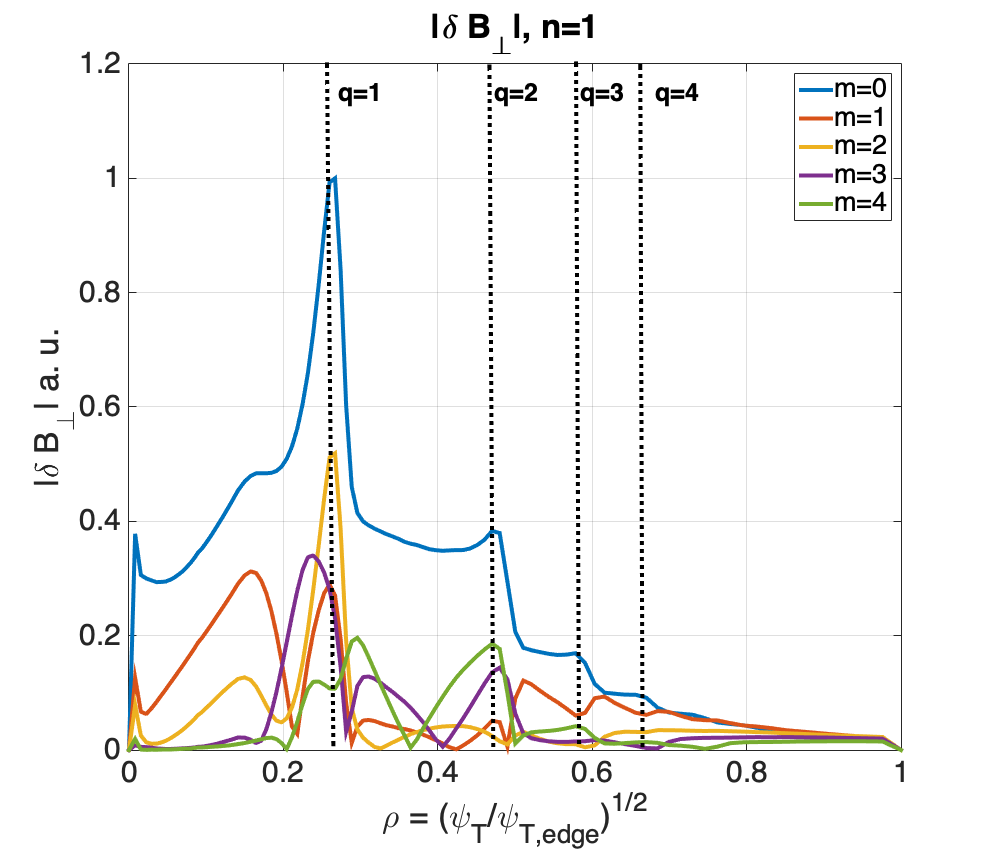}
    \end{minipage}
    }
    \caption{Poloidal harmonics of $\delta B_\parallel$ and $\delta B_\perp$ for $n=1$ kink mode}
    \label{dbperp}
    \end{figure}

In Section \ref{sec:MHD_eqns}, we have discussed the physical effect of compressional magnetic perturbation $\delta B_\parallel$ in the ideal MHD dispersion relation. In the simulation, significant effects from $\delta B_\parallel$ are also observed. $\delta B_\parallel$ appears in the equations of motion, the expressions of adiabatic pressures, and explicitly in the continuity equation. The simulations show that removing explicit $\delta B_\parallel$ terms in the continuity equation changes the growth rate from $4.2\times 10^{4} \text{s}^{-1}$ to 0 (full stabilization). {While the $\delta B_\parallel$ effect in the equation of motion and the expression of adiabatic pressure is negligible}. On the other hand, the ratio of $\delta B_\parallel$ to $\delta B_\perp$ is significant, as shown in Figure \ref{dbperp}. For the kink eigenmode structure, $\delta B_\parallel/\delta B_\perp$ can be as large as 0.35, which also indicates that $\delta B_\parallel$ can not be simply neglected.

\begin{figure}[H]
    \subfigure{
    \begin{minipage}[t]{0.5\linewidth}
        \centering
        \includegraphics[scale=0.3]{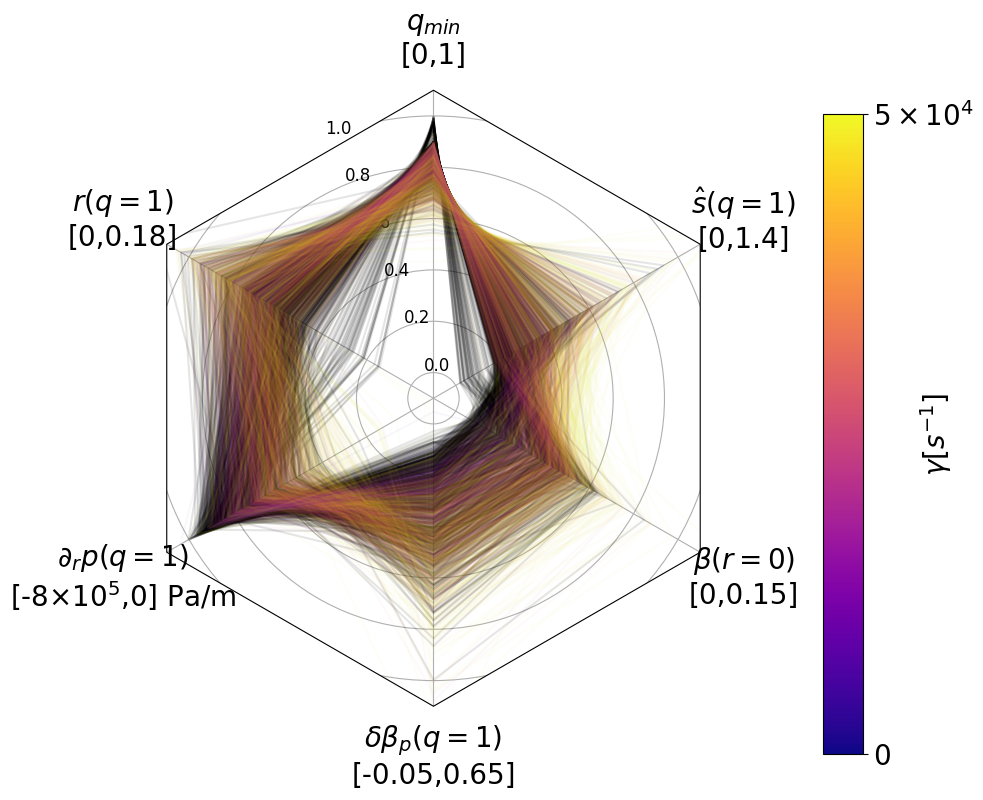}
    \end{minipage}
    }
    \subfigure{
    \begin{minipage}[t]{0.5\linewidth}
        \centering
        \includegraphics[scale=0.3]{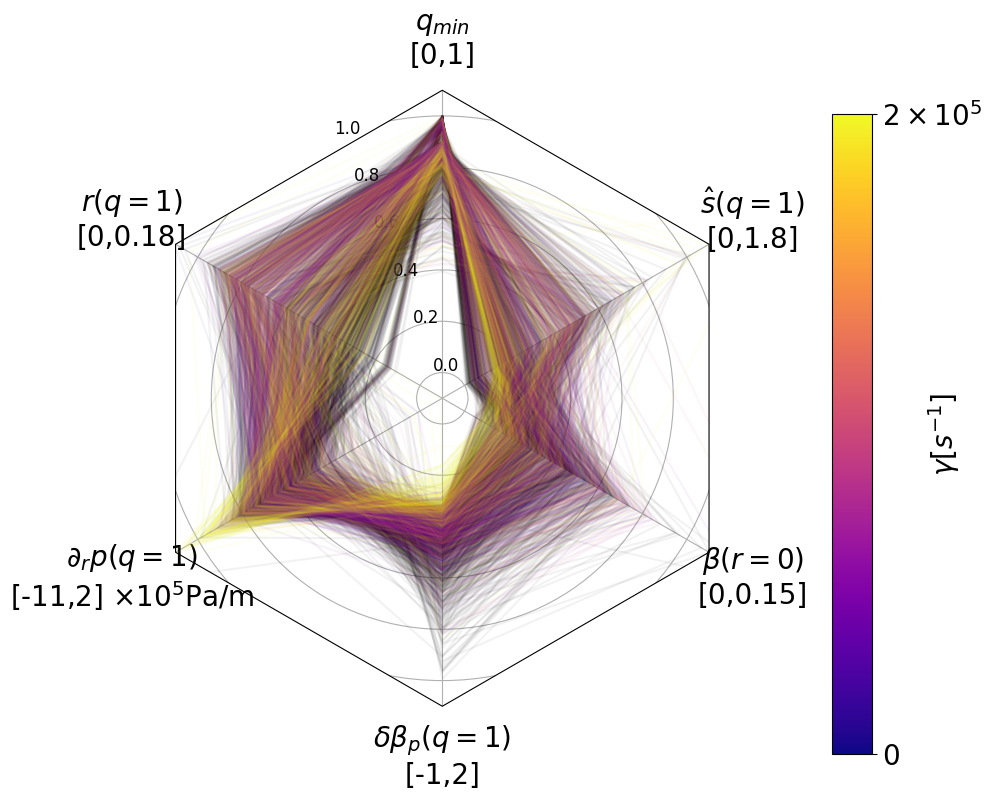}
    \end{minipage}
    }
    \caption{Spider plots of kink mode instability from GTC simulations with (a) EFIT01 equilibria and (b) EFIT02 equilibria.}
    \label{spider_plot}
\end{figure}

In order to build a surrogate GTC simulation model for future real-time plasma control systems, we used GTC to simulate 5758 equilibria selected from DIII-D experiments, generating a database. {The simulated time slices are selected randomly from shot \#139520 to shot \#180844. From the measurements of Mironov coils and the safety factor profile from EFIT, the n=1 signal is present and $q_{min}<1$ at the time slice of interest. In the simulation, about half of the cases show kink instability.} GTC has the capability to carry out a large number of MHD single fluid simulations very efficiently. These simulations have been performed in 12 GTC runs, each of which simulates 500 experiments in 30 minutes using 2000 nodes of the Summit supercomputer(which has about 4700 nodes in total). {The efficient single fluid model and the optimization for GPUs and I/O in GTC make the code suitable for producing a large database from ideal MHD simulations. Meanwhile, the simulation model can be easily extended to include the kinetic effects, and this database serves as a foundation for a future database of kinetic simulations of MHD modes.} {All the simulations are using 100 radial grids and 24 parallel grids, while the poloidal grid number depends on the specific flux surface shape (typically around 200). The time step size is 0.01$R_0/C_s$, where $C_s$ is the ion acoustic wave speed. The convergence study is done on the benchmark case, which has similar physical parameters to the whole dataset. The single fluid simulation model is used, and there are no numerical particles used in the simulations. More detailed data analysis can be seen in \cite{Dong_2021}. }Thanks to the database, we are able to investigate the kink stability dependence on different physical parameters for realistic experimental equilibria. In Figure \ref{spider_plot}, we show the spider plots of kink instability. The line colors correspond to the simulated linear growth rates of these cases. All the cases show strong $n=1$ signals on experiments, but it's difficult to know if the measured signals are kink modes, since the poloidal mode structure is unknown. In the simulation, 1972 of those cases are unstable and show typical kink mode structures. The other cases could be tearing modes or saw-teeth signals. Six important physical parameters are shown in the spider plots, including the minimum value of q-profile($q_{min}$), the minor radius of $q=1$ surface normalized to $R_0$ ($r(q=1)$), pressure gradient at $q=1$ surface ($\partial_r p$), the magnetic shear at $q=1$ surface ($\hat{s}(q=1)$), the $\beta$ value on magnetic axis($\beta_0$), and $\delta\beta_p$ at $q=1$ surface, which is defined by $\delta\beta_p(q=1)=-2R_0^2\int_0^{r_1}\partial_rpr^2dr/(B_0^2r_1^4)$, where $r_1$ represents $r(q=1)$. $\delta\beta_p$ can be regarded as the thermal energy stored inside the q=1 surface. Figure \ref{spider_plot}a and Figure \ref{spider_plot}b correspond to different sets of equilibrium, constructed by EFIT01 and EFIT02, respectively. The $q$ profile of EFIT02 data is generally more accurate since the motional Stark effect(MSE) diagnostics information is considered. The Pearson correlation coefficients between these parameters and the linear growth rate are calculated to show the importance of these parameters. It turns out that for the EFIT01 data, the stabilities (`stable' or `unstable') are most sensitive to $q_{min}$ values($corr=-0.40$), while the linear growth rates are most sensitive to $\partial_r p$ at $q=1$ surface ($corr=-0.30$). For the EFIT02 data, the stabilities are most sensitive to $r(q=1)$ ($corr=0.28$), while the growth rates are most sensitive to $\delta \beta_p$ ($corr=-0.37$). These dependencies are qualitatively consistent with the ideal MHD theory. The differences between the two datasets may be attributed to the inaccurate reconstruction of the EFIT01 data and the fact that the EFIT01 data only show normal shear. In contrast, the EFIT02 data include more reversed-shear cases. The low correlation coefficients suggest that prediction based on several physical parameters can be inaccurate, and a more complicated machine learning model\cite{Dong_2021} is needed.

\section{Conclusion}
In this paper, we formulate the comprehensive nonlinear electromagnetic gyrokinetic model in GTC. We showed that the hybrid scheme and the conservative can be unified, with a different definition of $\delta\phi_{ind}$.  In the long wavelength limit, the gyrokinetic equations reduce to the two-fluid model. And if $E_\parallel$ is neglected, the model further reduces to the single-fluid MHD model, from which the ideal MHD dispersion relation is recovered. The implementation of this simulation model in the general 3D geometry is presented, including the numerical methods to calculate the parallel equilibrium current and the method to construct Boozer coordinates. In the benchmark between GTC and other kinetic-MHD codes, it is found that incorporating accurate perturbed magnetic compression and equilibrium parallel current is important to recover the linear and nonlinear properties of kink modes. A large number of kink cases chosen from DIII-D experiments have been successfully simulated to build a database for machine learning. The q=1 surface location, $q_{min}$, pressure gradient at q=1 surface, and $\delta\beta_p(q=1)$ are found to be related to the linear instability of kink modes.

\section* {Data Avaibility}
The data that support the findings of this study are available from the corresponding author upon reasonable request.

\section* {Acknowledgements}
This work is supported by the US Department of
Energy (DOE) SciDAC project ISEP and INCITE project, and used resources of the
Oak Ridge Leadership Computing Facility at Oak Ridge National Laboratory (DOE Contract No. DE-AC05-00OR22725)
and the National Energy Research Scientific Computing Center (DOE Contract No. DE-AC02-05CH11231). The work reported in this paper was completed before July 1, 2025. This report was prepared as
an account of work sponsored by an agency of the United
States Government. Neither the United States Government nor
any agency thereof, nor any of their employees, makes any
warranty, express or implied, or assumes any legal liability
or responsibility for the accuracy, completeness, or usefulness of any information, apparatus, product, or process disclosed, or represents that its use would not infringe privately
owned rights. Reference herein to any specific commercial
product, process, or service by trade name, trademark, manufacturer, or otherwise does not necessarily constitute or imply
its endorsement, recommendation, or favoring by the United States Government or any agency thereof. The views and opinions of authors expressed herein do not necessarily state or
reflect those of the United States Government or any agency
thereof.

\begin{appendices}
\numberwithin{equation}{section}
\renewcommand{\theequation}{\thesection\arabic{equation}}

\section{Numerical implementations in Boozer coordinate system}\label{sec:eqs_in_boozer}

The covariant and contravariant forms of the magnetic field in the Boozer coordinate system are given by
\begin{equation}
\begin{aligned}
\mathbf{B}_{0}=&q\left(\psi\right) \nabla\psi\times\nabla\theta-\nabla\psi\times\nabla\zeta\\
=&\hat{\delta}\left(\psi,\theta,\zeta\right)\nabla\psi+I\left(\psi\right)\nabla\theta+g\left(\psi\right)\nabla\zeta,\label{eq:B_Boozer}
\end{aligned}
\end{equation}
which are frequently used in deriving the following equations. In Eq \eqref{eq:B_Boozer}, $q$ is the safety factor, $I$ the toroidal current, $g$ the poloidal current. In this section, $\psi,\theta,\zeta$ stand for the flux coordinate in the equilibrium B field. $\hat{\delta} $ current comes from the non-orthogonality of the coordinate system. In GTC we use $\dot{\psi}$, $\dot{\theta}$, $\dot{\zeta}$ and $\dot{\rho}_{\parallel}=d/dt(mv_{\parallel}/\left(ZB_{0}\right))$
to update the gyrocenter location in phase space,
\begin{align}
\begin{split}
    \dot{\zeta} =&\rho_{\parallel}B_{0}^{2}\frac{Z}{m}\frac{q}{D}+\frac{\rho_{\parallel}B_{0}^{2}}{D}\frac{Z}{m}\left(\rho_{\parallel}+\lambda\right)\left(I'-\partial_{\theta}\hat{\delta}\right)+\rho_{\parallel}\frac{B_{0}^{2}}{D}\frac{Z}{m}\left(-\hat{\delta}\frac{\partial\lambda}{\partial\theta}+I\frac{\partial\lambda}{\partial\psi}\right)\\
 & +\frac{1}{D}\frac{1}{Z}\frac{\partial\varepsilon}{\partial B_0}\left(\hat{\delta}\frac{\partial B_{0}}{\partial\theta}-I\frac{\partial B_{0}}{\partial\psi}\right)+\frac{1}{D}\left(\hat{\delta}\partial_{\theta}\phi_{b\parallel}-I\partial_{\psi}\phi_{b\parallel}\right)\label{eq:zetadot}
\end{split}
\end{align}
\begin{align}
\begin{split}
    \dot{\psi} =& \frac{1}{Z}\frac{\partial\varepsilon}{\partial B_0}\frac{1}{D}\left(I\frac{\partial B_{0}}{\partial\zeta}-g\frac{\partial B_{0}}{\partial\theta}\right)+\frac{1}{D}\left[I\partial_{\zeta}\phi_{b\parallel}-g\partial_{\theta}\phi_{b\parallel}\right]\\
 & +\frac{\rho_{\parallel}B_{0}^{2}}{D}\frac{Z}{m}\left(g\frac{\partial\lambda}{\partial\theta}-I\frac{\partial\lambda}{\partial\zeta}\right)
\end{split}
\end{align}
\begin{align}
\begin{split}
    \dot{\theta} =& \rho_{\parallel}B_{0}^{2}\frac{Z}{m}\frac{1}{D}-\frac{\rho_{\parallel}B_{0}^{2}}{D}\frac{Z}{m}\left(g'-\partial_{\zeta}\hat{\delta}\right)\left(\rho_{\parallel}+\lambda\right)+\frac{1}{D}\frac{1}{Z}\frac{\partial\varepsilon}{\partial B_0}\left(g\frac{\partial B_{0}}{\partial\psi}-\hat{\delta}\frac{\partial B_{0}}{\partial\zeta}\right)\\
 & +\frac{1}{D}\left(g\partial_{\psi}\phi_{b\parallel}-\hat{\delta}\partial_{\zeta}\phi_{b\parallel}\right)+\frac{\rho_{\parallel}B_{0}^{2}}{D}\frac{Z}{m}\left(\hat{\delta}\frac{\partial\lambda}{\partial\zeta}-g\frac{\partial\lambda}{\partial\psi}\right)
\end{split}
\end{align}
\begin{align}
\begin{split}
    \dot{\rho}_{\parallel} =& \frac{1}{D}\left(-1+\left(\rho_{\parallel}+\lambda\right)\left(g'-\partial_{\zeta}\hat{\delta}\right)-\left(\hat{\delta}\frac{\partial\lambda}{\partial\zeta}-g\frac{\partial\lambda}{\partial\psi}\right)\right)\left(\frac{1}{Z}\frac{\partial\varepsilon}{\partial B_0}\frac{\partial B_{0}}{\partial\theta}+\partial_{\theta}\phi_{b\parallel}\right)\\
 & -\frac{1}{D}\left(q+\left(\rho_{\parallel}+\lambda\right)\left(I'-\partial_{\theta}\hat{\delta}\right)+\left(I\frac{\partial\lambda}{\partial\psi}-\hat{\delta}\frac{\partial\lambda}{\partial\theta}\right)\right)\left(\frac{1}{Z}\frac{\partial\varepsilon}{\partial B_0}\frac{\partial B_{0}}{\partial\zeta}+\partial_{\zeta}\phi_{b\parallel}\right)\\
 & -\frac{1}{D}\left(g\frac{\partial\lambda}{\partial\theta}-I\frac{\partial\lambda}{\partial\zeta}\right)\left(\frac{1}{Z}\frac{\partial\varepsilon}{\partial B_0}\frac{\partial B_{0}}{\partial\psi}+\partial_{\psi}\phi_{b\parallel}\right)-\frac{\partial\lambda}{\partial t}\label{eq:rhoparadot}
\end{split}
\end{align}
where $I'=\partial_{\psi}I$, $g'=\partial_{\psi}g$, $\lambda=\left\langle\delta A_{\parallel}^L + \delta A_\parallel^{NL}\right\rangle/B_{0}$,
$D=gq+I+\rho_{\parallel}\left[\left(I'-\partial_{\theta}\hat{\delta}\right)g-I\left(g'-\partial_{\zeta}\hat{\delta}\right)\right]$,
$\partial\varepsilon/\partial B_0=\left(\mu+\rho_{\parallel}^{2}\Omega\right)$,
$\phi_{b\parallel}=\left\langle\phi\right\rangle+\frac{\mu}{Z}\left\langle\left\langle\delta B_{\parallel}\right\rangle\right\rangle$. In the simulation with the static RMP island field, the equilibrium $\mathbf{B}$ field should include both the original magnetic field and the RMP field. But for simplicity, we treat the RMP island field $\delta \mathbf{B}_{eq}$ as part of the perturbation, such that the definition of $B_0$, $\psi$, $\theta$, $\zeta$, $\rho_\parallel$ keep the original definition. In this way, the $\lambda$ should be replaced to include an island term $\lambda'=\lambda+\lambda_{eq} = \langle\delta A_\parallel^L + \delta A_\parallel^{NL} + \delta A_{\parallel eq}\rangle/B_0$ in Eqs \eqref{eq:zetadot}-\eqref{eq:rhoparadot}.

We should notice that in Eq \eqref{eq:rhoparadot}, the zonal potential should be removed from $\left\langle\phi\right\rangle$ in the simulations. Although we should define the zonal part such that there is no parallel acceleration from $\nabla\left\langle\phi\right\rangle_{00}$, in the simulation, we often use the averaged value along the unperturbed flux surface to act as the zonal fields for simplicity. This approximation, however, introduces the unphysical acceleration $\delta\mathbf{B}_\perp\cdot\nabla\left\langle\phi\right\rangle_{00}$ in Eq \eqref{eq:rhoparadot}. Therefore, we need to subtract the error explicitly.

Sometimes, the nonlinearity caused by parallel acceleration from the perturbed potential in $\dot{\rho}_{\parallel}$ is not important and often induces numerical noise. We can choose to neglect these nonlinearities from Eq \eqref{eq:rhoparadot},
\begin{align}
\begin{split}
    \dot{\rho}_{\parallel} =& \frac{1}{D}\left(-1+\rho_{\parallel}\left(g'-\partial_{\zeta}\hat{\delta}\right)\right)\left(\frac{1}{Z}\frac{\partial\varepsilon}{\partial B_0}\frac{\partial B_{0}}{\partial\theta}\right)\\
 & +\frac{1}{D}\left(-q-\rho_{\parallel}\left(I'-\partial_{\theta}\hat{\delta}\right)\right)\left(\frac{1}{Z}\frac{\partial\varepsilon}{\partial B_0}\frac{\partial B_{0}}{\partial\zeta}\right)\\
 & +\frac{\rho_{\parallel}^{2}B_{0}}{D}\frac{Z}{m}\left[g'\lambda+g\partial_{\psi}\lambda-\lambda\partial_{\zeta}\hat{\delta}-\hat{\delta}\partial_{\zeta}\lambda\right]\partial_{\theta}B_{0}\\
 & -\frac{\rho_{\parallel}^{2}B_{0}}{D}\frac{Z}{m}\left[I'\lambda+I\partial_{\psi}\lambda-\lambda\partial_{\theta}\hat{\delta}-\hat{\delta}\partial_{\theta}\lambda\right]\partial_{\zeta}B_{0}\\
 & +\frac{\rho_{\parallel}^{2}B_{0}}{D}\frac{Z}{m}\left(I\partial_{\zeta}\lambda-g\partial_{\theta}\lambda\right)\partial_{\psi}B_{0}\\
 & +\frac{\rho_{\parallel}}{B_{0}}\frac{1}{D}\left[\left(\hat{\delta}\frac{\partial B_{0}}{\partial\theta}-I\frac{\partial B_{0}}{\partial\psi}\right)\frac{\partial\left\langle\delta\phi\right\rangle_{nz}}{\partial\zeta}+\left(I\frac{\partial B_{0}}{\partial\zeta}-g\frac{\partial B_{0}}{\partial\theta}\right)\frac{\partial\left\langle\delta\phi\right\rangle_{nz}}{\partial\psi}\right.\\
 &\left.+\left(g\frac{\partial B_{0}}{\partial\psi}-\hat{\delta}\frac{\partial B_{0}}{\partial\zeta}\right)\frac{\partial\left\langle\delta\phi\right\rangle_{nz}}{\partial\theta}\right]
\end{split}
\end{align}
Eqs \eqref{eq:zetadot}-\eqref{eq:rhoparadot} are derived from the Euler-Lagrangian equation in gyrokinetic phase space. White et al \cite{White_1984} made an additional transformation to the parallel gyrocenter velocity to get rid of $\hat{\delta}$ term, and derived the equation of motion from the Hamiltonian method. In fact, by dropping all $\hat{\delta}$ terms in Eqs \eqref{eq:zetadot}-\eqref{eq:rhoparadot}, one obtains the identical equations of motion derived in \cite{White_1984} formally, although the definition of the parallel coordinate is different by a distance $\sim\rho_\parallel$. We choose not to do the transformation here in order to keep the Maxwell equations as the standard gyrokinetic form. In practice, dropping $\hat{\delta}$ while not doing the transformation will cause an error of $\mathcal{O}(k_\parallel\rho_\parallel)$ in the particle motion, which is negligible in the gyrokinetic framework.

In the delta-f simulation, we define the ion particle weight $w_s \equiv \delta f_s / f_s$. The ion weight equation, Eq \eqref{eq:Ldfs} can be expressed as
\begin{align}
\begin{split}
    \frac{dw_s}{dt} = & (1-w_s) \times\left\{ -\left[v_{\parallel}\frac{-\mathbf{b}_{0}\times\nabla\left\langle\delta A_{\parallel}\right\rangle}{B_{\parallel}^{*}}+\frac{\delta\mathbf{B}_{eq}}{B_\parallel^*}+\frac{(\mathbf{b}_{0}+\delta\mathbf{B}_{eq}/B_0)\times\nabla \phibparaanglept}{B_{\parallel}^{*}}\right]\cdot\nabla|_{v\perp}F_{0}\frac{1}{F_{0}}\right.\\
	&-\frac{\mu(\left\langle\delta\mathbf{B}_{\perp}\right\rangle + \delta\mathbf{B}_{eq})\cdot\nabla B_{0}}{B_{\parallel}^{*}}v_{\parallel}\frac{1}{F_{0}}\left(\frac{\partial F_{0}}{\partial\left(\mu B_{0}\right)}-\frac{1}{m_s v_\parallel}\frac{\partial F_{0}}{\partial v_{\parallel}}\right)\\
	&+\frac{\mu(\mathbf{b}_{0}+\delta\mathbf{B}_{eq}/B_0)\times\nabla B_{0}}{B_{\parallel}^{*}}\cdot\nabla\phibparaanglept \frac{1}{F_{0}}\left(\frac{\partial F_{0}}{\partial\left(\mu B_{0}\right)}-\frac{1}{m_sv_{\parallel}}\frac{\partial F_{0}}{\partial v_{\parallel}}\right)\\
	&+\frac{\mu(\mathbf{b}_{0}+\delta\mathbf{B}_{eq}/B_0)\times\nabla B_{0}}{B_{\parallel}^{*}}\cdot\nabla\phibparaanglept \frac{1}{m_sv_{\parallel}}\frac{\partial F_{0}}{F_0\partial v_{\parallel}}\\
	&+\left(-Z_s\left\langle E_{\parallel}\right\rangle+\frac{\mu\mathbf{B}_{0}}{B_{\parallel}^{*}}\cdot\nabla\left\langle\left\langle\delta B_{\parallel}\right\rangle\right\rangle \right)\frac{1}{m_s}\frac{\partial F_{0}}{F_0\partial v_{\parallel}}\\
	&+\left[Z_s\frac{mv_{\parallel}}{Z_sB_{\parallel}^{*}}\nabla\times\left(\mathbf{b}_{0}+\frac{\delta \mathbf{B}_{eq}}{B_0}\right)\cdot\nabla \phibparaanglept \right]\frac{1}{m_s}\frac{\partial F_{0}}{F_0\partial v_{\parallel}}\\
	&+\left[Z_s\frac{(\langle\delta\mathbf{B}_{\perp}\rangle+\delta\mathbf{B}_{eq})}{B_{\parallel}^{*}}\cdot\nabla \left(\phibparaangle[nz]+\langle\phi\rangle_{eq}\right) \right]\frac{1}{m_s}\frac{\partial F_{0}}{F_0\partial v_{\parallel}}\\
    &\left.-Z_s\frac{\delta\mathbf{B}_{eq}}{B_\parallel^*}\cdot\nabla\langle\phi\rangle_{eq}\frac{1}{m_s}\frac{\partial F_0}{F_0\partial v_\parallel}\right\},\label{eq:ion_weight_equation}
\end{split}
\end{align}
where $\langle E_\parallel\rangle = -\bunit\cdot\nabla\langle\phi\rangle_{eff}$, the symbol $\left.\nabla\right|_{v_{\perp}}f_{0}$ means the derivative keeping $v_{\perp}$ constant, $\left.\nabla\right|_{v_{\perp}}f_{0}=\left.\nabla\right|_{\mu}f_{0}+\mu\nabla B_0f_0/T$. The last five rows stand for the contribution of time-static perturbations like the RMP field. Note that in the 5th line and the last line, the zonal potential is artificially removed for the same reason as explained in Eq \eqref{eq:rhoparadot}. And we can write out the expressions in Boozer coordinates of some terms above,
\begin{align*}
\begin{split}
    &-\left(v_{\parallel}\frac{- \mathbf{b}_{0}\times\nabla\left\langle\delta A_{\parallel}\right\rangle}{B_{\parallel}^{*}} + v_\parallel \frac{\delta\mathbf{B}_{eq}}{B_\parallel^*}+\frac{(\mathbf{b}_{0}+\delta\mathbf{B}_{eq}/B_0)\times\nabla\phibparaanglept}{B_{\parallel}^{*}}\right)\cdot\nabla|_{v\perp}F_{0}\frac{1}{F_{0}}\\
    =&\frac{v_{\parallel}B_0}{D}\left[g\partial_{\theta}\lambda'-I\partial_{\zeta}\lambda'\right]\kappa_s\\
    &+\frac{1}{D}\left[(I+\dbeq{\theta})\partial_{\zeta}\phibparaanglept-(g+\dbeq{\zeta})\partial_{\theta}\phibparaanglept\right]\kappa_s,
\end{split}
\end{align*}
where $\lambda' = \lambda + \lambda_{eq}$.
\begin{align*}
\begin{split}
    &-\frac{\mu(\langle\delta\mathbf{B}_{\perp}\rangle+\delta\mathbf{B}_{eq})\cdot\nabla B_{0}}{B_{\parallel}^{*}}\\
    =&-\mu\frac{B_{0}}{D}\left(\partial_{\psi}\lambda'\left(I\partial_{\zeta}B_{0}-g\partial_{\theta}B_{0}\right)+\partial_{\theta}\lambda'\left(g\partial_{\psi}B_{0}-\delta\partial_{\zeta}B_{0}\right)+\partial_{\zeta}\lambda'\left(\delta\partial_{\theta}B_{0}-I\partial_{\psi}B_{0}\right)\right)\\
    &-\frac{\mu B_0}{D}\lambda'\mathcal{J}\left[\curlBu{\theta}\partial_\theta B_0+\curlBu{\zeta}\partial_\zeta B_0\right],
\end{split}
\end{align*}
\begin{align*}
\begin{split}
    &\frac{\mu(\mathbf{b}_{0}+\delta\mathbf{B}_{eq}/B_0)\times\nabla B_{0}}{B_{\parallel}^{*}}\cdot\nabla\phibparaanglept\\
    =&\frac{\mu}{D}\left[(\hdelta+\dbeq{\psi})\left(\partial_{\zeta}\phibparaanglept\partial_{\theta}B_{0}-\partial_{\theta}\phibparaanglept\partial_{\zeta}B_{0}\right)\right.\\
    &+(I+\dbeq{\theta})\left(\partial_{\psi}\phibparaanglept\partial_{\zeta}B_{0}-\partial_{\zeta}\phibparaanglept\partial_{\psi}B_{0}\right)\\
    &\left.+(g+\dbeq{\zeta})\left(\partial_{\theta}\phibparaanglept\partial_{\psi}B_{0}-\partial_{\psi}\phibparaanglept\partial_{\theta}B_{0}\right)\right],
\end{split}
\end{align*}
\begin{align*}
\begin{split}
    &-Z_sE_{\parallel}+\frac{\mu\mathbf{B}_{0}}{B_{\parallel}^{*}}\cdot\nabla\delta B_{\parallel}\\
    =&Z_s\frac{B_0}{gq+I}\left(\partial_\theta\phi_{eff}+q\partial_\zeta\phi_{eff}\right)+\frac{\mu B_{0}}{D}\left(q\partial_{\zeta}\delta B_{\parallel}+\partial_{\theta}\delta B_{\parallel}\right),
\end{split}
\end{align*}
\begin{align*}
\begin{split}
    &Z_s\frac{m_sv_{\parallel}}{Z_sB_{\parallel}^{*}}\nabla\times(\mathbf{b}_{0}+\delta\mathbf{B}_{eq}/B_0)\cdot\nabla\phibparaanglept\\
    =&\frac{mv_{\parallel}}{B_{0}D}\left\{\partial_{\psi}\phibparaanglept\left[(I-\dbeq{\psi})\partial_{\zeta}B_{0}-(g-\dbeq{\zeta})\partial_{\theta}B_{0}\right]\right.\\
    &+\partial_{\theta}\phibparaanglept\left[(g+\dbeq{\zeta})\partial_{\psi}B_{0}-(\hdelta+\dbeq{\psi})\partial_{\zeta}B_{0}\right]\\
    &\left.+\partial_{\zeta}\phibparaanglept\left[(\hdelta+\dbeq{\psi})\partial_{\theta}B_{0}-(I+\dbeq{\theta})\partial_{\psi}B_{0}\right]\right\} \\
    &+\frac{mv_\parallel}{D}\mathcal{J}\left[\curlBu{\theta}\cdot\partial_\theta\phibparaanglept\right]\\
    &\left.+\curlBu{\zeta}\partial_\zeta\phibparaanglept\right]\\
    +&\frac{mv_\parallel}{D}\left[(\partial_\theta\dbeq{\zeta}-\partial_\zeta\dbeq{\theta})\partial_\psi\phibparaanglept\right.\\
    &+(\partial_\zeta\dbeq{\psi}-\partial_\psi\dbeq{\zeta})\partial_\theta\phibparaanglept\\
    &+\left.(\partial_\psi\dbeq{\theta}-\partial_\theta\dbeq{\psi})\partial_\zeta\phibparaanglept\right],
\end{split}
\end{align*}

\begin{align*}
\begin{split}
    &Z_s\frac{(\langle\delta\mathbf{B}_{\perp}\rangle+\delta\mathbf{B}_{eq}/B_0)}{B_{\parallel}^{*}}\cdot\nabla \left(\phibparaangle[nz]+\langle\phi\rangle_{eq}\right)\\
    =&Z_s\frac{B_{0}}{D}\left[\partial_{\psi}\lambda'\left(I\partial_{\zeta}\phibparaangle[nz]-g\partial_{\theta}\phibparaangle[nz]\right)\right.\\
    &+\partial_{\theta}\lambda'\left(g\partial_{\psi}\phibparaangle[nz]-\hdelta\partial_{\zeta}\phibparaangle[nz]\right)\\
    &\left.+\partial_{\zeta}\lambda'\left(\hdelta\partial_{\theta}\phibparaangle[nz]-I\partial_{\psi}\phibparaangle[nz]\right)\right]\\
    &+Z_s\frac{B_0}{D}\partial_{\psi}\langle\phi_{eq}\rangle\left(g\partial_\theta\lambda' - I\partial_\zeta\lambda'\right)\\
    &+\frac{Z_s B_0}{D}\lambda'\mathcal{J}\left[\curlBu{\zeta}\partial_{\zeta}\phibpara{nz}-\curlBu{\theta}\partial_{\theta}\phibpara{nz}\right].
\end{split}
\end{align*}
Where
\begin{equation}
\begin{split}
    \dbeq{\psi} &= \sum_{i=1}^{3}\left(\partial_{\alpha_{i}}\lambda_{eq}(\mathsf{g}^{{\alpha_{i}}\zeta}-q\mathsf{g}^{{\alpha_{i}}\theta})\right) + \lambda_{eq} \curlBd{\psi},\\
    \dbeq{\theta} &= -\sum_{i=1}^{3}\left(\partial_{\alpha_{i}}\lambda_{eq}\mathsf{g}^{{\alpha_{i}}\psi}\right) + \lambda_{eq}\curlBd{\theta},\\
    \dbeq{\zeta} &= q\sum_{i=1}^{3}\left(\partial_{\alpha_{i}}\lambda_{eq} \mathsf{g}^{{\alpha_{i}}\psi}\right) + \lambda_{eq}\curlBd{\zeta},
\end{split}
\end{equation}
where $(\alpha_1,\alpha_2,\alpha_3)=(\psi,\theta,\zeta)$
\begin{equation}
\begin{split}
    \curlBu{\psi}=&0,\\
    \curlBu{\theta}=&\frac{1}{\mathcal{J}}\left(\partial_\zeta\hdelta - \partial_\psi g\right),\\
    \curlBu{\zeta}=&\frac{1}{\mathcal{J}}\left(\partial_\psi I-\partial_\theta\hdelta\right),\\
    \curlBd{\alpha_i}=&\sum_{j=1}^{3}\curlBu{\alpha_j}\mathsf{g}_{\alpha_i\alpha_j}.
\end{split}
\end{equation}

And for the shifted Maxwellian distribution function, we have
\begin{align*}
\begin{split}
    \frac{1}{F_0}\left(\frac{\partial F_0}{\partial(\mu B_0)}-\frac{1}{m_sv_\parallel}\frac{\partial F_0}{\partial v_\parallel}\right) &= -\frac{1}{T_s}\frac{u_{\parallel0,s}}{v_{\parallel}},\\
    \frac{1}{m_sF_0}\frac{\partial F_0}{\partial v_\parallel} &= -\frac{1}{T}\left(v_{\parallel}-u_{\parallel0}\right).
\end{split}
\end{align*}
For slowing down distribution as expressed in Eq \eqref{eq:sd_dist}, we have
\begin{align*}
\begin{split}
    \frac{1}{F_0}\left(\frac{\partial F_0}{\partial(\mu B_0)}-\frac{1}{m_sv_\parallel}\frac{\partial F_0}{\partial v_\parallel}\right) &=\frac{4}{m_sv^{2}}\frac{-\left(\Lambda-\Lambda_{0}\right)\left(B_{a}/B_{0}-2\Lambda\right)}{\Delta\Lambda^{2}},\\
    \frac{1}{m_sF_0}\frac{\partial F_0}{\partial v_\parallel} &= -\frac{1}{m_s}\frac{3vv_\parallel}{v^{3}+v_{c}^{3}}-\frac{4v_\parallel}{m_sv^{2}}\frac{\left(\Lambda-\Lambda_{0}\right)\Lambda}{\Delta\Lambda^{2}}.
\end{split}
\end{align*}

In the above terms, $\kappa_s = -\partial_\psi\ln f_{0s} = \kappa_{0s} + \kappa_{v,s}$, $\kappa_{0s}=-\partial_\psi\ln n_{0s} - \partial_{\psi}\ln T_s\left[\frac{m_s(v_\parallel-u_{\parallel0s})^2}{2T_s}+\frac{\mu B_0}{T_s}-1.5\right]$, $\kappa_{v,s}=-\frac{m_s\left(v_{\parallel}-u_{\parallel0s}\right)}{T_s}\partial_\psi u_{\parallel0s}$. The ion equilibrium parallel flow is assumed as a function of $\psi$. For the slowing down distribution function, $\kappa_{v}$ is assumed to be 0. $\phi_{nz}$ and $\phi_{00}$ are the non-zonal and zonal parts of the perturbed electrostatic potential, respectively.

Define the electron particle weight as $w_e=\delta h_e/f_e$. If we take the local Maxwellian distribution as the equilibrium electron distribution, the electron weight equation is given by
\begin{equation}
\begin{split}
    \frac{d w_e}{dt} =& -\left(1-\frac{\delta f_e^{ad}}{f_e}-w_e\right)\left(\frac{1}{f_{0e}}\delta Lf_{0e}+L\frac{\delta f_{e}^{ad}}{f_{0e}} + \frac{\delta f_{e}^{ad}/f_{0e}}{f_{0e}}\delta Lf_{0e}\right)\\
    =&\left(1-\frac{\delta f_e^{ad}}{f_e}-w_e\right)\left\{-\frac{1}{f_{0e}}\left(\delta\mathbf{v}_{E}+\mathbf{v}_{b\parallel}\right)\cdot\nabla|_{v\perp}f_{0e}\right.\\
    &-\frac{\partial}{\partial t}\frac{\delta f_{e}^{ad}}{f_{0e}}+\frac{1}{f_{0e}}\frac{e}{m_e}\frac{\delta\mathbf{B}_{\perp}}{B_{\parallel}^{*}}\cdot\nabla\left(\delta\phi-\frac{\mu}{e}\delta B_{\parallel}\right)\frac{\partial f_{0e}}{\partial v_{\parallel}}+v_{\parallel}\frac{\delta\mathbf{B}_{\perp}}{B_{\parallel}^{*}}\cdot\nabla|_{\mu}\frac{\delta f_{0e}^{ad}}{f_{0e}}\\
    &-\frac{1}{f_{0e}}\left(\delta\mathbf{v}_{E}+\mathbf{v}_{b\parallel}\right)\cdot\mu\nabla B_{0}\frac{\partial f_{0e}}{\partial(\mu B_{0})}-\left(-v_{\parallel}\frac{mv_{\parallel}\nabla\times\bunit}{eB}+\mathbf{v}_{g}\right)\cdot\nabla\frac{\delta f_{e}^{ad}}{f_{0e}}\\
    &+\frac{1}{f_{0e}}\frac{v_{\parallel}\nabla\times\bunit}{B_{\parallel}^{*}}\nabla\left(\delta\phi-\frac{\mu}{e}\delta B_{\parallel}\right)\frac{\partial f_{0}}{\partial v_{\parallel}}-\left(\mathbf{v}_{E}+\mathbf{v}_{b\parallel}\right)\cdot\nabla\frac{\delta f_{e}^{ad}}{f_{0}}\\
    &+\frac{1}{m_{e}}\frac{\mathbf{B}^{*}}{B_{\parallel}}\cdot\left(\mu\nabla B_{0}-e\nabla\phi+\mu\nabla\delta B_{\parallel}\right)\frac{\partial}{\partial v_{\parallel}}\left(\frac{\partial\ln f_{0}}{\partial\psi_{0}}\delta\psi+\frac{\partial\ln f_{0}}{\partial\alpha}\delta\alpha\right)\\
    &-\frac{\delta f_{e}^{ad}}{f_{0}}\frac{1}{f_{0e}}\delta Lf_{0e}\\
    &=\left(1-\frac{\delta f_{e}^{ad}}{f_e}-w_e\right)\times\left(w_{drive}+w_{para}+w_{drift,ind}+w_{drift}+w_{dv}+w_{highorder}\right),\label{eq:wedot_vector}
\end{split}
\end{equation}
where the nonlinear part $\delta L \delta f_e^{ad}$ has been neglected, $\mathbf{v}_{d}=\frac{\mu}{Z_{s}B_{0}}\mathbf{b}_{0}\times\nabla B_{0}+\frac{m_{s}v_\parallel^2}{Z_{s}B_{0}^{2}}\nabla\times\mathbf{b}_{0}$, $\delta\phi_{ind}^{ad}=\phi_{eff}^{ad}-\phi$ is the adiabatic part of inductive potential. In the Boozer coordinate system, the detailed implementation can be separated into several parts.

The terms in the last line of Eq \eqref{eq:wedot_vector} are defined as
\begin{equation*}
\begin{aligned}
    w_{drive}
    =&\frac{\mathbf{b}_{0}\times\nabla(\delta\phi-\frac{\mu}{e}\delta B_\parallel)}{B_{\parallel}^{*}}\cdot\nabla\psi\frac{\partial \ln f_{0e}}{\partial \psi_{0}}|_{v\perp}\\
    =&\frac{1}{D}\left[I\partial_{\zeta}(\delta\phi-\frac{\mu}{e}\delta B_\parallel)-g\partial_{\theta}(\delta\phi-\frac{\mu}{e}\delta B_\parallel)\right]\kappa_e.
\end{aligned}
\end{equation*}
\begin{equation*}
\begin{aligned}
w_{para} =& -\frac{\partial}{\partial t}\frac{\delta f_e^{ad}}{f_{0e}} -\frac{1}{f_{0e}}\frac{e}{m}\frac{\delta\mathbf{B}_{\perp}}{B_{\parallel}^{*}}\cdot\nabla\left(\delta\phi-\frac{\mu}{e}\delta B_{\parallel}\right)\frac{\partial f_{0e}}{\partial v_{\parallel}} - v_\parallel\frac{\delta\mathbf{B}_\perp}{B_\parallel^*}\cdot\nabla\frac{\delta f_{e}^{ad}}{f_{0e}}\\
=&-\frac{\partial}{\partial t}\frac{\delta f_e^{ad}}{f_{0e}}-\left(1-\frac{\mu B_{0}}{T_{e}}\right)\frac{\partial_{t}\delta B_{\parallel}}{B_{0}}-\frac{1}{q}\frac{\partial\phi_{ind}^{ad}}{\partial\theta}\left(\kappa_e-\kappa_{n,e}\right)\\
 &-\frac{eu_{\parallel0e}}{T_{e}}\frac{B_{0}}{D}\left[\partial_{\psi}\lambda\left(I\partial_{\zeta}(\phi_{nz}-\frac{\mu}{e}\delta B_\parallel)-g\partial_{\theta}(\phi_{nz}-\frac{\mu}{e}\delta B_\parallel)\right)\right.\\
 &\left.+\partial_{\theta}\lambda\left(g\partial_{\psi}(\phi_{nz}-\frac{\mu}{e}\delta B_\parallel)-\hat{\delta}\partial_{\zeta}(\phi_{nz}-\frac{\mu}{e}\delta B_\parallel)\right)\right.\\
 &\left.+\partial_{\zeta}\lambda\left(\hat{\delta}\partial_{\theta}(\phi_{nz}-\frac{\mu}{e}\delta B_\parallel)-I\partial_{\psi}(\phi_{nz}-\frac{\mu}{e}\delta B_\parallel)\right)\right]\\
 &- \frac{eu_{\parallel0e}}{T_{e}}\frac{\lambda\mathcal{J}B_{0}}{D}\sum_{j=1}^{3}\left(\nabla\times\mathbf{B}_{0}\right)^{\alpha_{j}}\cdot\partial_{\alpha_{j}}\left(\delta\phi_{nz}-\frac{\mu}{e}\delta B_{\parallel}\right)\\
 &-\frac{e v_\parallel}{T_{e}}\frac{B_{0}}{D}\left[\partial_{\psi}\lambda\left(I\partial_{\zeta}\delta\phi_{ind}-g\partial_{\theta}\delta\phi_{ind}\right)+\partial_{\theta}\lambda\left(g\partial_{\psi}\delta\phi_{ind}-\hat{\delta}\partial_{\zeta}\delta\phi_{ind}\right)\right.\\
 &\left.+\partial_{\zeta}\lambda\left(\hat{\delta}\partial_{\theta}\delta\phi_{ind}-I\partial_{\psi}\delta\phi_{ind}\right)\right] - \frac{ev_\parallel}{T_{e}}\frac{\lambda\mathcal{J}B_{0}}{D}\sum_{j=1}^{3}\left(\nabla\times\mathbf{B}_{0}\right)^{\alpha_{j}}\cdot\partial_{\alpha_{j}}\delta\phi_{ind}\\
 &-\frac{v_{\parallel}B_{0}}{D}\left[\left(g\partial_{\theta}\lambda-I\partial_{\zeta}\lambda\right)\mathcal{R}^{ad}_{\psi}+\left(\hdelta\partial_{\zeta}\lambda-g\partial_{\psi}\lambda\right)\mathcal{R}^{ad}_{\theta}+\left(I\partial_{\psi}\lambda-\hdelta\partial_{\theta}\lambda\right)\partial_{\zeta}\mathcal{R}^{ad}_{\zeta}\right]\\
 &-\frac{v_{\parallel}\mathcal{J}B_{0}}{D}\lambda\sum_{j=1}^{3}\left(\nabla\times\mathbf{B}_{0}\right)^{\alpha_{j}} \mathcal{R}^{ad}_{\alpha_j}.
\end{aligned}
\end{equation*}
where we have used the governing equation for $\delta \psi^a$ and $\delta\alpha ^a$, only the $\alpha$ derivative of $f_{0e}$ and $u_{\parallel0e}$ are assumed to be 0. The time derivative of $\delta f_{e}^{ad} / f_{0e}$ can be obtained from Eq \eqref{eq:dfad_dt_cons} or \eqref{eq:dfad_dt_hybr}. The expression of $\mathcal{R}^{ad}$ in the direction of $\alpha_{j}$ is given by
\begin{equation*}
\begin{split}
    \mathcal{R}^{ad}_{\alpha_j} = \left(e\phi_{eff} + \mu\delta B_\parallel\right)\frac{\partial}{\partial \alpha_j}\frac{1}{T_e}-\frac{\partial}{\partial\alpha_j}\left(\kappa_{e}\delta\psi\right) + \frac{\partial}{\partial\alpha_j}\left(\frac{e}{T_e}\frac{\partial\phi_{eq}}{\partial\psi}\delta\psi\right)
\end{split}
\end{equation*}
\begin{equation}
\begin{split}
w_{drift,ind} =& -\frac{1}{f_{0e}}\left(\delta\mathbf{v}_{E}+\mathbf{v}_{b\parallel}\right)\cdot\mu\nabla B_{0}\frac{\partial f_{0}}{\partial(\mu B_{0})}-\left(v_{\parallel}\frac{\frac{mv_{\parallel}}{Z_{e}}\nabla\times\bunit}{B}+\mathbf{v}_{g}\right)\cdot\nabla\frac{\delta f_{e}^{ad}}{f_{0}}\\
&-\frac{1}{f_{0e}}\frac{e}{m}\frac{mv_{\parallel}/Z_e\nabla\times\bunit}{B_{\parallel}^{*}}\nabla\left(\delta\phi-\frac{\mu}{e}\delta B_{\parallel}\right)\frac{\partial f_{0}}{\partial v_{\parallel}} \\
=&\frac{1}{T_{e}}\frac{1}{D}\frac{\partial\mathcal{E}_{0}}{\partial B}\left[\left(g\partial_{\theta}B_{0}-I\partial_{\zeta}B_{0}\right)\partial_{\psi}\left(\delta\phi-\frac{\mu}{e}\delta B_{\parallel}\right)+\left(\hat{\delta}\partial_{\zeta}B_{0}-g\partial_{\psi}B_{0}\right)\partial_{\theta}\left(\delta\phi-\frac{\mu}{e}\delta B_{\parallel}\right)\right.\\
&\left.+\left(I\partial_{\psi}B_{0}-\hat{\delta}\partial_{\theta}B_{0}\right)\partial_{\zeta}\left(\delta\phi-\frac{\mu}{e}\delta B_{\parallel}\right)\right]\\
&+ \frac{m_{e}v_{\parallel}u_{\parallel0}}{T_{e}}\frac{\mathcal{J}}{D}\sum_{j=1}^{3}\left(\nabla\times\mathbf{B}_{0}\right)^{\alpha_{j}}\partial_{\alpha_{j}}\phibparaele{}\\
& -\frac{1}{T_{e}}\frac{1}{D}\frac{\partial\mathcal{E}}{\partial B}\left[\left(g\partial_{\theta}B_{0}-I\partial_{\zeta}B_{0}\right)\partial_{\psi}\phi_{ind}+\left(\hat{\delta}\partial_{\zeta}B_{0}-g\partial_{\psi}B_{0}\right)\partial_{\theta}\phi_{ind}+\left(I\partial_{\psi}B_{0}-\hat{\delta}\partial_{\theta}B_{0}\right)\partial_{\zeta}\phi_{ind}\right]\\
&+\frac{m_{e}v_{\parallel}^2}{T_{e}}\frac{\mathcal{J}}{D}\sum_{j=1}^{3}\left(\nabla\times\mathbf{B}_{0}\right)^{\alpha_{j}}\partial_{\alpha_{j}}\phi_{ind}\\
& -\frac{1}{T_{e}}\frac{1}{D}\frac{\partial\mathcal{E}}{\partial B}\left[\left(g\partial_{\theta}B_{0}-I\partial_{\zeta}B_{0}\right)\mathcal{R}_{\psi}^{ad}+\left(\hat{\delta}\partial_{\zeta}B_{0}-g\partial_{\psi}B_{0}\right)\mathcal{R}_{\theta}^{ad}+\left(I\partial_{\psi}B_{0}-\hat{\delta}\partial_{\theta}B_{0}\right)\mathcal{R}_{\zeta}^{ad}\right]\\
& + \frac{mv_{\parallel}^2}{T_{e}}\frac{\mathcal{J}}{D}\sum_{j=1}^{3}\left(\nabla\times\mathbf{B}_{0}\right)^{\alpha_{j}}\mathcal{R}_{\alpha_{j}}^{ad}
\end{split}
\end{equation}
where we have kept the terms of the order $O(1/k_{\perp} L_p)$, and dropped the terms of the order $O(1/k_\perp L_B)$.
\begin{equation*}
\begin{aligned}
    w_{drift}
    =&-\left(\mathbf{v}_{E}+\mathbf{v}_{b\parallel}\right)\cdot\nabla\frac{\delta f_{e}^{ad}}{f_{0e}}\\
    =&\frac{1}{D}\left[\left(I\partial_{\zeta}\left(\phi-\frac{\mu}{e}\delta B_{\parallel}\right)-g\partial_{\theta}\left(\phi-\frac{\mu}{e}\delta B_{\parallel}\right)\right)\partial_{\psi}\frac{\delta f_{e}^{ad}}{f_{0e}}\right.\\
    &\left.+\left(g\partial_{\psi}\left(\phi-\frac{\mu}{e}\delta B_{\parallel}\right)-\hat{\delta}\partial_{\zeta}\left(\phi-\frac{\mu}{e}\delta B_{\parallel}\right)\right)\partial_{\theta}\frac{\delta f_{e}^{ad}}{f_{0e}}\right.\\
    &\left.+\left(\hat{\delta}\partial_{\theta}\left(\phi-\frac{\mu}{e}\delta B_{\parallel}\right)-I\partial_{\psi}\left(\phi-\frac{\mu}{e}\delta B_{\parallel}\right)\right)\partial_{\zeta}\frac{\delta f_{e}^{ad}}{f_{0e}}\right].
\end{aligned}
\end{equation*}
Note that $\phi=\delta\phi + \phi_{eq}$.
\begin{equation*}
\begin{aligned}
    w_{dv}
    =&\frac{1}{m_{e}}\frac{\mathbf{B}^{*}}{B_{\parallel}}\cdot\left(\mu\nabla B_{0}-e\nabla\phi+\mu\nabla\delta B_{\parallel}\right)\frac{\partial}{\partial v_{\parallel}}\frac{\delta f_{e}^{ad}}{f_{e0}}\\
    =&\left[\frac{1}{m_e}\frac{B_{0}}{D}\left(q\partial_{\zeta}\mathcal{F}+\partial_{\theta}\mathcal{F}\right)-\frac{v_\parallel}{e}\frac{1}{D}\left(\left(I'-\partial_{\theta}\hat{\delta}\right)\partial_{\zeta}\mathcal{F}-\left(g'-\partial_{\zeta}\hat{\delta}\right)\partial_{\theta}\mathcal{F}\right)\right.\\
    &-\frac{v_\parallel}{eB_0D}\left((I\partial_\zeta B_0 - g\partial_\theta B_0)\partial_\psi\mathcal{F} + (g\partial_\psi B_0 - \hdelta\partial_\zeta B_0)\partial_\theta\mathcal{F}\right.\\
    &\left.\left.(\hdelta\partial_\theta B_0 - I\partial_\psi B_0)\right)\partial_\zeta\mathcal{F}\right]\delta\psi\left(\partial_\psi\ln T_{e}\frac{m_e(v_{\parallel}-u_{\parallel0e})}{T_e}+\partial_\psi u_{\parallel0e}\frac{m_{e}}{T_{e}}\right),
\end{aligned}
\end{equation*}
where $\mathcal{F}$ is defined as $\mu B_0 + \mu\delta B_\parallel - e\phi$.
\begin{equation*}
\begin{split}
 w_{highorder} = -\frac{\delta f_{e}^{ad}}{f_{0}}\frac{1}{f_{0e}}\delta Lf_{0e},
\end{split}
\end{equation*}
and the expression of $(\delta Lf_{0e})/f_{0e}$ can be found by dividing Eq \eqref{eq:ion_weight_equation} by $-(1-w)$.

We need to store $\delta B_\parallel$ and $\delta n_e$ from the last time step to take the
time-centered derivative for a better numerical property. In deriving the weight equation, we
assume the equilibrium distribution is shifted Maxwellian and the spatial
variation of $n_{0}$, $T$, and $P_{0}$ is only in $\psi$ direction.

The electron continuity equation is given by
\begin{equation}
    \frac{\partial\delta n_{e}}{\partial t}=-(w_{\parallel}+w_{drive}+w_{drift}+w_{eqc}+w_{*}+w_{b\parallel}+w_{\perp nl}+w_{*nl}),
\end{equation}
where the $w$ terms represent the contribution from various flows. $w_{\parallel}$ denotes the parallel perturbed flow,
\begin{equation}
\begin{aligned}
    w_{\parallel}=&\mathbf{B}_{0}\cdot\nabla\left(\frac{n_{0e}\delta u_{\parallel e}}{B_{0}}\right)
    = \frac{n_{0e}}{\mathcal{J}B_{0}}\left(q\partial_{\zeta}\delta u_{e\parallel}+\partial_{\theta}\delta u_{e\parallel}\right).
\end{aligned}
\end{equation}
where $\mathcal{J}$ is the Jacobian for Boozer coordinates $\mathcal{J}=(gq+I)/B^{2}$.

$w_{drive}$ denotes the diamagnetic flow,
\begin{equation}
\begin{aligned}
w_{drive}=&-n_{0}\mathbf{v}_{*}\cdot\frac{\nabla B_{0}}{B_{0}}\\
=&\frac{1}{\mathcal{J}eB_{0}^{3}}\left[\frac{\partial B}{\partial\psi}\left(I\partial_{\zeta}\delta P_{e}-g\partial_{\theta}\delta P_{e}\right)+\frac{\partial B}{\partial\theta}\left(g\partial_{\psi}\delta P_{e}-\hat{\delta}\partial_{\zeta}\delta P_{e}\right)+\frac{\partial B}{\partial\zeta}\left(\hat{\delta}\partial_{\theta}\delta P_{e}-I\partial_{\psi}\delta P_{e}\right)\right],
\end{aligned}
\end{equation}
where $\delta P_{e}$ is the total perturbed electron pressure.
\begin{equation}
\begin{aligned}
\delta P_{e}=&\delta P_{\perp e}^{ad}+\delta P_{\parallel e}^{ad}+\delta P_{\perp e}^{na}+\delta P_{\parallel e}^{na}.
\end{aligned}
\end{equation}
$w_{drift}$, $w_{*}$, $w_{*nl}$, $w_{drift0}$, and $w_{eb0}$ denotes the flow due to $E\times B$ drift velocity,
\begin{align}
\begin{split}
& w_{*} + w_{drift}+ w_{*nl}+w_{drift0}+w_{eb0}\\
=&B_{0}\mathbf{v}_{E}\cdot\nabla\left(\frac{n_{0e}+\delta n_e}{B_{0}}\right)-(n_{0e}+\delta n_e)\mathbf{v}_{E}\cdot\frac{\nabla B_{0}}{B_{0}}\\
=& \frac{1}{\mathcal{J}B_{0}^{2}}\frac{\partial n_{0e}}{\partial\psi}\left(I\partial_{\zeta}\phi_{pt}-g\partial_{\theta}\phi_{pt}\right)\\
&+\frac{2(n_{0e}+\delta n_e)}{\mathcal{J}B_{0}^{3}}\left[\frac{\partial B}{\partial\psi}\left(g\partial_{\theta}\phi_{pt}-I\partial_{\zeta}\phi_{pt}\right)+\frac{\partial B}{\partial\theta}\left(\hat{\delta}\partial_{\zeta}\phi_{pt}-g\partial_{\psi}\phi_{pt}\right)+\frac{\partial B}{\partial\zeta}\left(I\partial_{\psi}\phi_{pt}-\hat{\delta}\partial_{\theta}\phi_{pt}\right)\right]\\
&+\frac{1}{\mathcal{J}B_{0}^{2}}\left[\frac{\partial\delta n_{e}}{\partial\psi}\left(I\partial_{\zeta}\phi_{pt}-g\partial_{\theta}\phi_{pt}\right)+\frac{\partial\delta n_{e}}{\partial\theta}\left(g\partial_{\psi}\phi_{pt}-\hat{\delta}\partial_{\zeta}\phi_{pt}\right)\right.\\
&\left.+\frac{\partial\delta n_{e}}{\partial\zeta}\left(\hat{\delta}\partial_{\theta}\phi_{pt}-I\partial_{\psi}\phi_{pt}\right)\right]\\
&+\frac{2\delta n_e}{\mathcal{J}B_0^3}\left(I\frac{\partial B_0}{\partial\zeta}-g\frac{\partial B_0}{\partial\theta}\right)\partial_\psi\phi_{eq}\\
&+\frac{1}{\mathcal{J}B_0^2}\left(g\frac{\partial\delta n_e}{\partial\theta}-I\frac{\partial\delta n_e}{\partial\zeta}\right)\partial_\psi\phi_{eq}.
\end{split}
\end{align}
$w_{eqc}$ denotes the equilibrium flow,

\begin{align}
\begin{split}
w_{eqc}=	&\delta\mathbf{B}_{\perp}\cdot\nabla\left(\frac{n_{0e}u_{\parallel0e}}{B_{0}}\right)-\frac{\nabla\times\mathbf{B_{0}}}{eB_{0}^{2}}\cdot\left(\nabla\delta P_{\parallel e}+\frac{\left(\delta P_{\perp e}-\delta P_{\parallel e}\right)\nabla B_{0}}{B_{0}}-n_{0e}e\nabla\phi\right)+\frac{\delta n_{e}}{B_{0}^{2}}\nabla\times\mathbf{B}_{0}\cdot\nabla\phi\\
=&-\frac{1}{e\mu_{0}\mathcal{J}}\left[\left(\partial_{\theta}\lambda g-\partial_{\zeta}\lambda I\right)\partial_{\psi}S+\left(\partial_{\zeta}\lambda\hat{\delta}-\partial_{\psi}\lambda g\right)\partial_{\theta}S+\left(\partial_{\psi}\lambda I-\partial_{\theta}\lambda\hat{\delta}\right)\partial_{\zeta}S\right]\\
&+\frac{1}{\mathcal{J}eB_{0}}\left(\partial_{\theta}\lambda g-\partial_{\zeta}\lambda I\right)\sum_{s\neq e}n_{0s}Z_{s}\partial_{\psi}u_{\parallel0}\\
&-\frac{1}{\mathcal{J}eB_{0}^{2}}\left[\left(\partial_{\theta}\lambda g-\partial_{\zeta}\lambda I\right)\partial_{\psi}B_{0}+\left(\partial_{\zeta}\lambda\hat{\delta}-\partial_{\psi}\lambda g\right)\partial_{\theta}B_{0}+\left(\partial_{\psi}\lambda I-\partial_{\theta}\lambda\hat{\delta}\right)\partial_{\zeta}B_{0}\right]\sum_{s\neq e}n_{0s}Z_{s}u_{\parallel0}\\
&+\frac{1}{eB_{0}^{2}\mathcal{J}}\left(g'-\frac{\partial\hat{\delta}}{\partial\zeta}\right)\left[\frac{B_{0}^{2}}{\mu_{0}}\lambda\partial_{\theta}S+\lambda\partial_{\theta}B_{0}\sum_{s\neq e}n_{0}Z_{s}u_{\parallel0}+\partial_{\theta}\delta P_{\parallel e}\right.\\
&\left.+\frac{1}{B_{0}}\partial_{\theta}B_{0}\left(\delta P_{\perp e}-\delta P_{\parallel e}\right)-\left(n_{0e}+\delta n_{e}\right)e\partial_{\theta}\phi_{pt}\right]\\
&+\frac{1}{eB_{0}^{2}\mathcal{J}}\left(\frac{\partial\hat{\delta}}{\partial\theta}-I'\right)\left[\frac{B_{0}^{2}}{\mu_{0}}\lambda\partial_{\zeta}S+\lambda\partial_{\zeta}B_{0}\sum_{s\neq e}n_{0}Z_{s}u_{\parallel0}+\partial_{\zeta}\delta P_{\parallel e}\right.\\
&\left.+\frac{1}{B_{0}}\partial_{\zeta}B_{0}\left(\delta P_{\perp e}-\delta P_{\parallel e}\right)-\left(n_{0e}+\delta n_{e}\right)e\partial_{\zeta}\phi_{pt}\right]
\end{split}
\end{align}

where $\delta P_{(\perp,\parallel) e} = \delta P_{(\perp,\parallel) e}^{ad} + \delta P_{(\perp,\parallel) e} ^{na}$ can be obtained from Eqs \eqref{eq:delpperp_orig} and \eqref{eq:delppara_orig}.
In the limit neglecting $u_{\parallel e0}$, $\delta P_{\perp e}-\delta P_{\parallel e}
=\delta P_{\perp e}^{na}-\delta P_{\parallel e}^{na}-\delta B_{\parallel}P_{\perp0e}/B_{0}$ can be used. All ion species are assumed to have the same parallel flow velocity $u_{\parallel0}\left(\psi\right)$, and the equilibrium potential is assumed to only depend on $\psi$.
$S=\frac{\mu_0J_{\parallel0}}{B_{0}}$ is proportional to the parallel current.

$w_{b\parallel}$ denotes the flow from the magnetic compressional perturbation
\begin{align}
\begin{split}
w_{b\parallel}
=&-\frac{\mathbf{b}_{0}\times\nabla\delta B_{\parallel}}{e}\cdot\nabla\left(\frac{\delta P_{\perp e}+P_{\perp0e}}{B_{0}^{2}}\right)-\frac{\nabla\times\mathbf{b}_{0}\cdot\nabla\delta B_{\parallel}}{eB_{0}^{2}}\left(\delta P_{\perp e}+P_{\perp0e}\right)\\
=& \frac{3P_{\perp e}^{tot}}{\mathcal{J}B_{0}^{4}e}\left[\frac{\partial\delta B_{\parallel}}{\partial\psi}\left(g\frac{\partial B_{0}}{\partial\theta}-I\frac{\partial B_{0}}{\partial\zeta}\right)+\frac{\partial\delta B_{\parallel}}{\partial\theta}\left(\hat{\delta}\frac{\partial B_{0}}{\partial\zeta}-g\frac{\partial B_{0}}{\partial\psi}\right)\right.\\
 &\left.+\frac{\partial\delta B_{\parallel}}{\partial\zeta}\left(I\frac{\partial B_{0}}{\partial\psi}-\hat{\delta}\frac{\partial B_{0}}{\partial\theta}\right)\right]\\
 &-\frac{1}{e\mathcal{J}B_{0}^{3}}\left[\hat{\delta}\left(\partial_{\theta}\delta B_{\parallel}\partial_{\zeta}\delta P_{\perp e}-\partial_{\zeta}\delta B_{\parallel}\partial_{\theta}\delta P_{\perp e}\right)\right.\\
 &\left.I\left(\partial_{\zeta}\delta B_{\parallel}\partial_{\psi}P_{\perp e}^{tot}-\partial_{\psi}\delta B_{\parallel}\partial_{\zeta}\delta P_{\perp e}\right)+g\left(\partial_{\psi}\delta B_{\parallel}\partial_{\theta}\delta P_{\perp e}-\partial_{\theta}\delta B_{\parallel}\partial_{\psi}P_{\perp e}^{tot}\right)\right]\\
 &+\frac{P_{\perp e}^{tot}}{eB_{0}^{3}\mathcal{J}}\left[\left(g'-\partial_{\zeta}\hat{\delta}\right)\partial_{\theta}\delta B_{\parallel}-\left(I'-\partial_{\theta}\hat{\delta}\right)\partial_{\zeta}\delta B_{\parallel}\right].
\end{split}
\end{align}
Here $P_{\perp e}^{tot}=P_{\perp0e}+\delta P_{\perp e}$ is the total
perpendicular electron pressure, $P_{\perp e}^{tot}=n_{0e}T_{e}+\delta P_{\perp e}^{na}+(\delta n_{e}-\int{\delta h_e}d\mathbf{v})T_{e}+n_{0e}\partial_{\psi}T_{e}\delta\psi-n_{0e}T_{e}\delta B_{\parallel}/B_{0}$.

$w_\perp^{nl}$ represents the nonlinear magnetic flutter flow,
\begin{align}
\begin{split}
w_{\perp nl} & =\delta\mathbf{B}_{\perp}\cdot\nabla\left(\frac{n_{0e}\delta u_{\parallel e}}{B_{0}}\right)\\
 & =\frac{1}{\mathcal{J}}\left[\hat{\delta}\left(\partial_{\zeta}\lambda\partial_{\theta}\delta S-\partial_{\theta}\lambda\partial_{\zeta}\delta S\right)\right.\\
 & \left.+I\left(\partial_{\psi}\lambda\partial_{\zeta}\delta S-\partial_{\zeta}\lambda\partial_{\psi}\delta S\right)+g\left(\partial_{\theta}\lambda\partial_{\psi}\delta S-\partial_{\psi}\lambda\partial_{\theta}\delta S\right)\right]
\end{split}
\end{align}
where $\delta S$ is the perturbed electron parallel flow, $\delta S=n_{0e}\delta u_{\parallel e}/B_{0}$.
In the continuity equation, we have dropped the terms at order higher than $O(k_\perp L_B)$, but we keep the second order derivative of the equilibrium magnetic field $\nabla S$, since it is the driving term for kink instability.

\section{The construction of the Boozer coordinate system in the axisymmetric system}\label{sec:XMAP}
In the benchmark of the internal kink modes, it is proven that the self-consistent equilibrium, especially the consistency between the magnetic field and the parallel current, is important for the simulation accuracy. Here, we describe a numerical algorithm to use the GEQDSK file to construct the Boozer coordinate system in the axisymmetric system. This algorithm is implemented using the Python language and is used to generate the equilibrium data file for GTC simulations for the benchmark.
The information from the GEQDSK file used to construct the Boozer coordinate system contains $\psi(R,Z),g(\psi),q(\psi)$, the last closed flux surface as a function of $(R,Z)$, and the position of the magnetic axis $(R_0,Z_0)$. An open-source GEQDSK file parser using Python language can be found at \cite{Greader}.

First of all, we can use the $\psi(R,Z)$ data to get the function $R(\psi,\theta_0)$ and $Z(\psi,\theta_0)$ on $(\psi,\theta_0)$ grids. Here $\theta_0$ is the geometrical poloidal angle, $\cos\theta_0=(R-R_0)/\sqrt{\left(R-R_0\right)^2+\left(Z-Z_0\right)^2}$, where $(R_0,Z_0)$ denotes the position of magnetic axis. This can be done using the RBF(radial basis functions) inverse interpolation capability in the Scipy module\cite{Scipy}. Then, the poloidal magnetic field can be calculated from
\begin{equation}
\begin{aligned}
    B_R(\psi,\theta_0)&=-\frac{1}{R}\frac{\partial\psi}{\partial Z},\\
    B_Z(\psi,\theta_0)&=\frac{1}{R}\frac{\partial\psi}{\partial R}.
\end{aligned}
\end{equation}
The toroidal magnetic field can simply be written as $B_t(\psi,\theta_0)=g/R$.

The next step is to transform $\theta_0$ to a new $\theta'$ angle to get the straight fieldline coordinate $(\theta',\phi)$ for a given $\psi$. A relation between $\theta_0$ and $\phi$ on a field line is needed for this purpose. Without losing the generality, we choose to trace the magnetic field line that crosses $(\theta_0=0,\phi=0)$, using the expressions of $(B_R,B_Z,B_t)$. For each flux surface, we can integrate numerically to get the relation between $\theta_0$ and the toroidal angle $\phi$.
\begin{equation}
\begin{aligned}
\theta_0 = \int_0{d\phi \frac{R B_p}{r B_t} \frac{B_Z \cos{\vartheta}-B_R\sin{\vartheta}}{B_p}},
\end{aligned}
\end{equation}
where $B_p=\sqrt{B_R^2+B_Z^2}$, $r=\sqrt{\left(R-R_0\right)^2+\left(Z-Z_0\right)^2}$, $\vartheta$ is the local $\theta_0$ value during the integration. Note that the integration is done following the field line, so the integration kernel depends on $\phi$. We have the freedom to choose $\theta'=\theta_0$ at the point $\theta_0=0$, and the integration is done from $\theta_0=0$ to $\theta=2\pi$. Now we have the function of $\theta_0(\phi)$ for arbitrary $\psi$. The $q$ value is also obtained, $q=\frac{\phi(\theta_0=2\pi)}{2\pi}$. ($q$ is also given in the GEQDSK file, which should be consistent with the field line tracing result.) Following the definition of straight field line coordinate, $\theta'$ should be proportional to $\phi$ on a certain field line. On the field line we just have traced, $\theta'=\phi/q$. Using this relation, it's easy to get the function of $\theta_0(\theta')$ on this field line, and for every other field line due to toroidal axisymmetry. Accordingly, the functions $R(\psi,\theta'), Z(\psi,\theta'), B_R(\psi,\theta'), B_Z(\psi,\theta')$, and $B_t(\psi,\theta')$ can be obtained. 

Now the magnetic field satisfies the contra-variant form of the straight field line coordinates representation, $\mathbf{B}=q(\psi)\nabla\psi\times\nabla\theta'-\nabla\psi\times\nabla\zeta_0$, where $\zeta_0=-\phi$. We can also express the magnetic field following the co-variant form $\mathbf{B}=\delta'\nabla\psi+I'\nabla\theta'+g'\nabla\zeta_0$. It is proven that $\mathbf{B}$ can be written as $\mathbf{B}=\delta'\nabla\psi+I(\psi)\nabla\theta'+g(\psi)\nabla\zeta_0+\nabla H(\psi,\theta')$, where $I(\psi)$ and $g(\psi)$ is exactly the co-variant component of $\mathbf{B}$ in Boozer coordinates.\cite{Helander2014} This transformation will be performed on $\theta'$ and $\zeta_0$ simultaneously, it can starts with getting $I(\psi)$ from $I'$. From the co-variant form, we have 
\begin{equation}
    I'=\mathbf{B}\cdot\frac{\partial\mathbf{x}}{\partial \theta'}=B_R\frac{\partial R}{\partial \theta'} + B_Z\frac{\partial Z}{\partial \theta'},
\end{equation}
and by flux averaging, we have $I=\oint{I'd\theta}'$, and $H=\int(I'-I)d\theta'$. The transform on poloidal and toroidal angles will be $\theta_b=\theta'+\nu$, and $\zeta=\zeta_0+q\nu$, where $\nu=H/(gq+I)$. The three coordinates $(\psi,\theta_b,\zeta)$ are the Boozer coordinates. And we can get the equilibrium field quantities as functions of $(\psi,\theta_b)$ since we know the relation $\theta'(\theta_b)$. We can demonstrate that $\mathbf{B}=q\nabla\psi\times\nabla\theta_b-\nabla\psi\times\nabla\zeta$, and $\mathbf{B}=\hat{\delta}\nabla\psi+I\nabla\theta_b+g\nabla\zeta$, where $\hat\delta$ can be simply calculated from $$\hat{\delta} = B_R\frac{\partial R}{\partial\psi}+B_Z\frac{\partial Z}{\partial \psi}.$$

\begin{figure}[H]
    \includegraphics[width=0.9\textwidth]{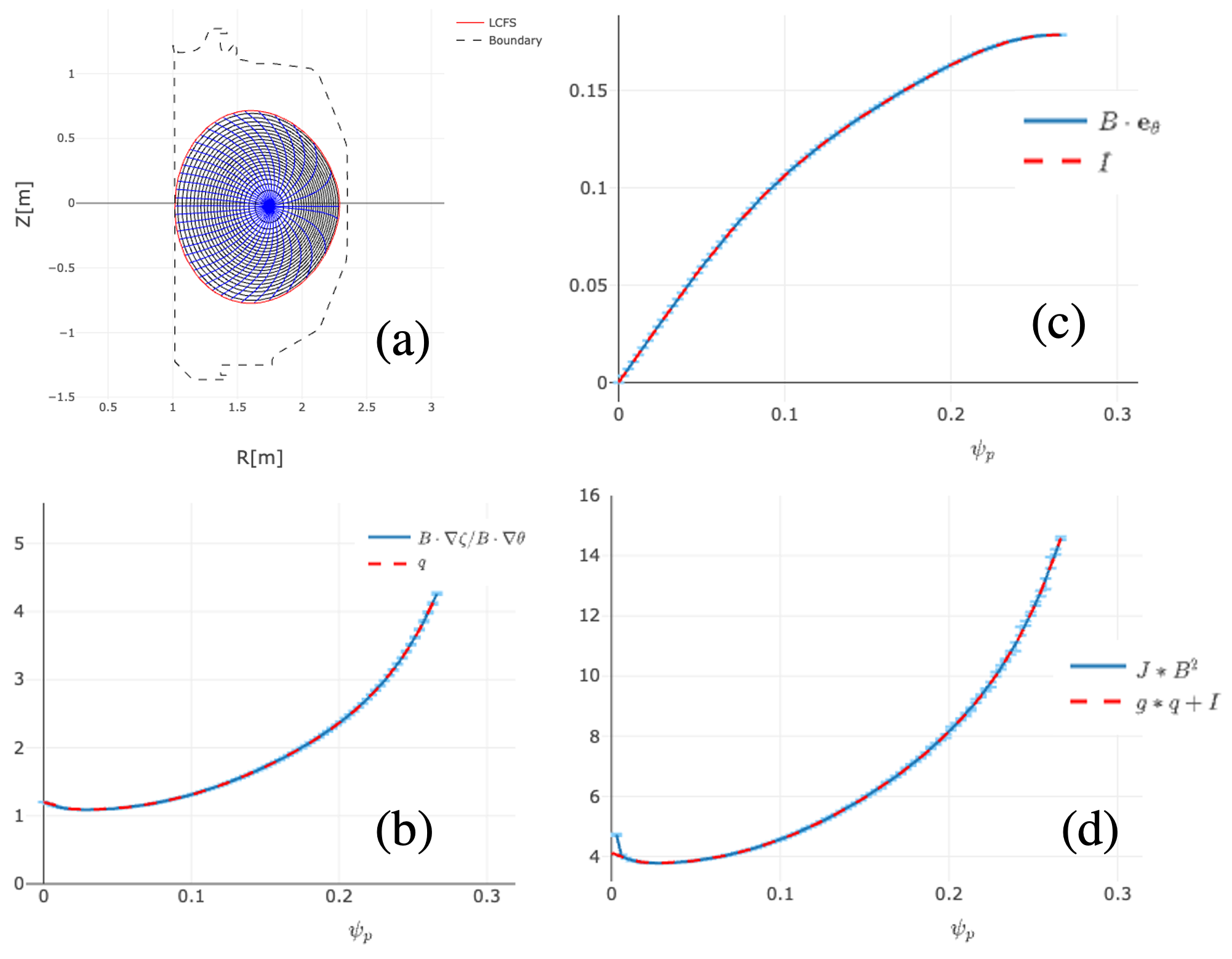}
    \centering
    \caption{The generated Boozer coordinate of DIII-D shot\#178631. (a) Boozer coordinates on the poloidal plane. (b) The consistency check between $q(\psi)$ and $\mathbf{B}\cdot\nabla\zeta/\mathbf{B}\cdot\nabla\theta$. (c) The consistency check between $I(\psi)$ and $\mathbf{B}\cdot\mathbf{e}_\theta$. (d) The consistency check between $\mathbf{e}_\psi\times\mathbf{e}_\theta\cdot\mathbf{e}_\zeta$ and $(gq+I)/B^2$. }
    \label{fig:xmap_check}
\end{figure}

An example of a constructed Boozer coordinate system is shown in Figure \ref{fig:xmap_check}. The blue error bars in subplots (b), (c), (d) of Figure \ref{fig:xmap_check} show the variation of quantities on one flux surface. These three subplots show the consistency between the covariant and contravariant form of $B$ field and also the invariance of $g,q,I$ on flux surfaces. So the requirements of Boozer coordinate are satisfied.

\section{The implementation of Laplacian operator}\label{sec:LaplacionOp}
In a general curvilinear coordinate system, the Laplacian operator is expressed as in
\begin{equation}
    \nabla^{2}f=\sum_{\alpha=1,2,3}\sum_{\beta=1,2,3}\frac{1}{\mathcal{J}}\frac{\partial}{\partial\xi^{\alpha}}\left(\mathcal{J}\nabla\xi^{\alpha}\cdot\nabla\xi^{\beta}\frac{\partial}{\partial\xi^{\beta}}f\right),
\end{equation}
where $\left(\xi^{1},\xi^{2},\xi^{3}\right)=\left(\psi,\theta,\zeta\right)$, and $\mathcal{J}$ is the Jacobian, $f$ is a perturbed field. Then, for the Boozer coordinate system, we have
\begin{equation}
\begin{aligned}
    \nabla^{2}f=&\mathsf{g}^{\psi\psi}\frac{\partial^{2}f}{\partial\psi^{2}}+\frac{1}{\mathcal{J}}\frac{\partial f}{\partial\psi}\left(\frac{\partial\mathcal{J}\mathsf{g}^{\psi\psi}}{\partial\psi}+\frac{\partial\mathcal{J}\mathsf{g}^{\psi\theta}}{\partial\theta}+\frac{\partial\mathcal{J}\mathsf{g}^{\psi\zeta}}{\partial\zeta}\right)\\
	&+\mathsf{g}^{\theta\theta}\frac{\partial^{2}f}{\partial\theta^{2}}+\frac{1}{\mathcal{J}}\frac{\partial f}{\partial\theta}\left(\frac{\partial\mathcal{J}\mathsf{g}^{\psi\theta}}{\partial\psi}+\frac{\partial\mathcal{J}\mathsf{g}^{\theta\theta}}{\partial\theta}+\frac{\partial\mathcal{J}\mathsf{g}^{\theta\zeta}}{\partial\zeta}\right)\\
	&+\mathsf{g}^{\zeta\zeta}\frac{\partial^{2}f}{\partial\zeta^{2}}+\frac{1}{\mathcal{J}}\frac{\partial f}{\partial\zeta}\left(\frac{\partial\mathcal{J}\mathsf{g}^{\psi\zeta}}{\partial\psi}+\frac{\partial\mathcal{J}\mathsf{g}^{\theta\zeta}}{\partial\theta}+\frac{\partial\mathcal{J}\mathsf{g}^{\zeta\zeta}}{\partial\zeta}\right)\\
	&+2\mathsf{g}^{\psi\theta}\frac{\partial^{2}f}{\partial\psi\partial\theta}+2\mathsf{g}^{\theta\zeta}\frac{\partial^{2}f}{\partial\theta\partial\zeta}+2\mathsf{g}^{\psi\zeta}\frac{\partial^{2}f}{\partial\psi\partial\zeta}
\end{aligned}
\end{equation}
Now we define the flux aligned coordinate $\left(\psi,\theta_{0},\zeta_{0}\right)$, with $\theta_{0}=\theta-\zeta/q$, $\zeta_{0}=\zeta$. The expression of the Laplacian operator in the flux-aligned coordinates is given by
\begin{align}
\begin{split}
    \nabla^{2}f=&\mathsf{g}^{\psi\psi}\frac{\partial^{2}f}{\partial\psi^{2}}+\frac{1}{\mathcal{J}}\frac{\partial f}{\partial\psi}\left[\frac{\partial \mathcal{J}\mathsf{g}^{\psi\psi}}{\partial\psi}+\frac{\partial \mathcal{J}\mathsf{g}^{\psi\theta}}{\partial\theta_{0}}+\left(\frac{\partial}{\partial\zeta_{0}}-\frac{1}{q}\frac{\partial}{\partial\theta_{0}}\right)\left(\mathcal{J}\mathsf{g}^{\psi\zeta}\right)\right]\\
	&+\mathsf{g}^{\theta\theta}\frac{\partial^{2}f}{\partial\theta_{0}^{2}}+\frac{1}{\mathcal{J}}\frac{\partial f}{\partial\theta_{0}}\left[\frac{\partial \mathcal{J}\mathsf{g}^{\psi\theta}}{\partial\psi}+\frac{\partial \mathcal{J}\mathsf{g}^{\theta\theta}}{\partial\theta_{0}}+\left(\frac{\partial}{\partial\zeta_{0}}-\frac{1}{q}\frac{\partial}{\partial\theta_{0}}\right)\left(\mathcal{J}\mathsf{g}^{\theta\zeta}\right)\right]\\
	&+\mathsf{g}^{\zeta\zeta}\left(\frac{\partial}{\partial\zeta_{0}}-\frac{1}{q}\frac{\partial}{\partial\theta_{0}}\right)^{2}f+\frac{1}{\mathcal{J}}\left(\frac{\partial}{\partial\zeta_{0}}-\frac{1}{q}\frac{\partial}{\partial\theta_{0}}\right)f\times\left[\frac{\partial \mathcal{J}\mathsf{g}^{\psi\zeta}}{\partial\psi}+\frac{\partial \mathcal{J}\mathsf{g}^{\theta\zeta}}{\partial\theta_{0}}+\left(\frac{\partial}{\partial\zeta_{0}}-\frac{1}{q}\frac{\partial}{\partial\theta_{0}}\right)\left(\mathcal{J}\mathsf{g}^{\zeta\zeta}\right)\right]\\
	&+2\mathsf{g}^{\psi\theta}\frac{\partial^{2}f}{\partial\psi\partial\theta_{0}}+2\mathsf{g}^{\theta\zeta}\left(\frac{\partial}{\partial\zeta_{0}}-\frac{1}{q}\frac{\partial}{\partial\theta_{0}}\right)\frac{\partial f}{\partial\theta_{0}}+2\mathsf{g}^{\psi\zeta}\left(\frac{\partial}{\partial\zeta_{0}}-\frac{1}{q}\frac{\partial}{\partial\theta_{0}}\right)\frac{\partial f}{\partial\psi}.
\end{split}
\end{align}
On the other hand,
\begin{align}
\begin{split}
    \nabla_\perp^2 f =& (\nabla - \bunit\bunit\cdot\nabla)\cdot(\nabla - \bunit\bunit\cdot\nabla) f\\
    =& \nabla^2 f - \bunit\bunit:\nabla\nabla f + \frac{\bunit}{B_0}\cdot[\bunit\cdot\nabla\mathbf{B}_0](\bunit\cdot\nabla)f.
\end{split}
\end{align}
It turns out $\nabla_\perp^2 = \nabla^2 + \mathcal{O}(\epsilon^2)$, where $\epsilon\sim r/R\sim k_\parallel/k_\perp\sim\mathsf{g}^{\psi\zeta}\partial_{\zeta}\partial_{\psi} / k_\perp^2 \sim \mathsf{g}^{\theta\zeta}\partial_{\zeta}\partial_{\theta} / k_\perp^2$. Here, we only solve the gyrokinetic Poisson equation to the first order, and the perpendicular Laplacian operator can be approximately obtained from the full Laplacian operator by dropping all the second-order terms,
\begin{align}
\begin{split}
    \nabla_{\perp}^{2}f=&\mathsf{g}^{\psi\psi}\frac{\partial^{2}f}{\partial\psi^{2}}+\frac{1}{\mathcal{J}}\frac{\partial f}{\partial\psi}\left[\frac{\partial\mathcal{J}\mathsf{g}^{\psi\psi}}{\partial\psi}+\frac{\partial\mathcal{J}\mathsf{g}^{\psi\theta}}{\partial\theta_{0}}-\frac{1}{q}\frac{\partial}{\partial\theta_{0}}\left(\mathcal{J}\mathsf{g}^{\psi\zeta}\right)\right]\\
	&+\left(\mathsf{g}^{\theta\theta}+\mathsf{g}^{\zeta\zeta}\frac{1}{q^{2}}\right)\frac{\partial^{2}f}{\partial\theta_{0}^{2}}+\frac{1}{\mathcal{J}}\frac{\partial f}{\partial\theta_{0}}\left[\frac{\partial\mathcal{J}\mathsf{g}^{\psi\theta}}{\partial\psi}+\frac{\partial\mathcal{J}\mathsf{g}^{\theta\theta}}{\partial\theta_{0}}-\frac{1}{q}\frac{\partial}{\partial\theta_{0}}\left(\mathcal{J}\mathsf{g}^{\theta\zeta}\right)\right]\\
	&-\frac{1}{J}\frac{1}{q}\frac{\partial f}{\partial\theta_{0}}\times\left[\frac{\partial\mathcal{J}\mathsf{g}^{\psi\zeta}}{\partial\psi}+\frac{\partial\mathcal{J}\mathsf{g}^{\theta\zeta}}{\partial\theta_{0}}-\frac{1}{q}\frac{\partial}{\partial\theta_{0}}\left(\mathcal{J}\mathsf{g}^{\zeta\zeta}\right)\right]\\
	&+2\mathsf{g}^{\psi\theta}\frac{\partial^{2}f}{\partial\psi\partial\theta_{0}}-2\mathsf{g}^{\theta\zeta}\frac{1}{q}\frac{\partial}{\partial\theta_{0}}\frac{\partial f}{\partial\theta_{0}}-2\mathsf{g}^{\psi\zeta}\frac{1}{q}\frac{\partial}{\partial\theta_{0}}\frac{\partial f}{\partial\psi}
\end{split}
\end{align}

\section{Energy conservation and energy transferring rate}\label{sec:GKenergy}
Following \cite{Brizard_2017}, we can write out the nonlinear gyrokinetic electromagnetic energy invariant in the parallel-symplectic representation,
\begin{equation}
\begin{split}
    \mathcal{E}_{GY} =& \int d\mathbf{x}\int d\mathbf{v} f_{s} H_{GY} + \int d\mathbf{x}\left[-\frac{\epsilon_0}{2}\left|\nabla\phi\right|^2 + \frac{1}{2\mu_0}\left|\mathbf{B}_0+\delta\mathbf{A}\right|^2\right],
\end{split}
\end{equation}
where $H_{GY}$ is the gyrokinetic Hamiltonian expanded to the 2nd order of $e\delta\phi/T_e$,
\begin{equation}
\begin{split}
    H_{GY} = &\mu B_0 + \frac{1}{2}mv_\parallel^2 + Z_s\gpgc - Z_s\left\langle\delta\mathbf{A}_{\perp}\cdot\mathbf{v}_\perp\right\rangle + \frac{Z_s^2}{2m_s}\left(\left\langle\left|\delta\mathbf{A}_{\perp}\right|^2\right\rangle + \left\langle\delta\tilde{A}_{\parallel}^2\right\rangle\right)\\
    &-\frac{1}{2}\savg{Z_s\tilde{\psi}}.
\end{split}
\end{equation}
$\{,\}$ is the unperturbed Poisson bracket in the guiding-center space, and 
\begin{equation}
\begin{split}
    S_1 &\approx\frac{q_s}{\Omega_s}\int\tilde{\psi}d\alpha \equiv \frac{q_s}{\Omega_s}\tilde{\Psi},\\
    \psi &\equiv \phi - v_\parallel\delta A_{\parallel} - \delta\mathbf{A}_{\perp}\cdot\mathbf{v}_\perp,\\
    \tilde{\psi} &\equiv\psi - \langle\psi\rangle,\\
    \delta \tilde{A}_{\parallel} &\equiv\delta A_\parallel - \left\langle\delta A_\parallel\right\rangle.
\end{split}
\end{equation}
The approximate definition of $S_1$ stays valid as long as $\omega/\Omega_c\ll1$. The Poisson bracket $\savg{Z_s\tilde{\psi}}$ can be written as
\begin{equation}
\begin{split}
    \savg{Z_s\tilde{\psi}} & = \frac{Z_s^2}{\Omega_s}\left\langle\left\{\tilde{\Psi},\partial_\alpha\tilde{\Psi}\right\}\right\rangle\\
    &= \frac{Z_s^3}{m_s\Omega_s}\left\langle\frac{\partial\tilde{\Psi}}{\partial\alpha}\frac{\partial^2\tilde{\Psi}}{\partial\alpha\partial\mu} - \frac{\partial\tilde{\Psi}}{\partial\mu}\frac{\partial^2\tilde{\Psi}}{\partial\alpha^2}\right\rangle\\
    &= \frac{Z_s^2}{B_0}\frac{\partial}{\partial\mu}\left\langle\tilde{\psi}^2\right\rangle
\end{split}
\end{equation}
In the GTC simulation, we define kinetic energy and field energy as
\begin{equation}
\begin{split}
    \mathcal{E}_{k,s} = &\int d\mathbf{x}\int d\mathbf{v} f_s \left(\mu B_0 + \frac{1}{2}m_s v_\parallel^2\right),\\
    \mathcal{E}_{f} = &\int d\mathbf{x}\left[-\frac{\epsilon_0}{2}\left|\nabla\phi\right|^2 + \frac{1}{2\mu_0}\left|\mathbf{B}_0+\delta\mathbf{A}\right|^2\right] \\
    &+ \sum_{s}\int d\mathbf{x}\int d\mathbf{v} f_s \left[Z_s\left(\gpgc - \avperpgc\right) + \frac{Z_s}{2m_s^2} \left(\left\langle\left|\delta\mathbf{A}_{\perp}\right|^2\right\rangle + \left\langle\delta\tilde{A}_{\parallel}^2\right\rangle\right) \right.\\
    & \left. -\frac{1}{2} \frac{Z_s^2}{B_0}\frac{\partial}{\partial\mu} \left\langle\tilde{\psi}^2\right\rangle \right].
\end{split}
\end{equation}
The last term in $\mathcal{E}_f$ can be approximately calculated using the feature of the equilibrium distribution function
\begin{equation}
\begin{split}
    -\frac{1}{2}\int d\mathbf{x}\int d\mathbf{v} f_s\frac{Z_s^2}{B_0}\frac{\partial}{\partial\mu}\left\langle\tilde{\psi}^2\right\rangle \approx \frac{1}{2} \int d\mathbf{x}\int d\mathbf{v} \frac{Z_s^2}{B_0}\frac{\partial f_{0s}}{\partial\mu} \left\langle\tilde{\psi}^2\right\rangle.
\end{split}
\end{equation}
For the Maxwellian distribution
$$\frac{\partial f_{0s}}{\partial\mu} = -\frac{B_0}{T_s} f_{0s},$$
and for anisotropic slowing down distribution,
$$\frac{\partial f_{0s}}{\partial\mu} = -\left(\frac{3v}{v^{3}+v_{c}^{3}}\frac{B_{0}}{m_s}+4\frac{B_{a}}{m_sv^{2}}\frac{\left(\Lambda-\Lambda_{0}\right)\left(1-B_{0}\Lambda/B_{a}\right)}{\Delta\Lambda^{2}}\right)f_{0s}.$$
The energy conservation of the system can be verified by checking that $\sum_{s}\mathcal{E}_{k,s} + \mathcal{E}_f = const$.
When the particle moves in the phase space from $(\mathbf{R},\mu,v_\parallel)$ to $(\mathbf{R}',\mu,v_\parallel')$ in a short time period $\Delta t$, we can write the change of the kinetic energy as
\begin{equation}
    \begin{split}
        \Delta\mathcal{E}_{k,s}=&\int d\mathbf{x} \int d\mathbf{v} f_s (\mathbf{R}',\mu,v_\parallel',t+\Delta t)\times\left(\mu B_0(\mathbf{R}')+\frac{1}{2}m_sv_\parallel'^2\right)\\
        &- \int d\mathbf{x} \int d\mathbf{v} f_s(\mathbf{R},\mu,v_\parallel,t)\times\left(\mu B_0(\mathbf{R})+\frac{1}{2}m_sv_\parallel^2\right).
    \end{split}
\end{equation}
Since the evolution of particle distribution function conserves phase shape density and volume, we have $d\mathbf{x}d\mathbf{v} f_s (\mathbf{R}',\mu,v_\parallel',t+\Delta t) = d\mathbf{x}d\mathbf{v} f_s(\mathbf{R},\mu,v_\parallel,t)$, and
\begin{equation}
    \Delta\mathcal{E}_{k,s}=\int d\mathbf{x} \int d\mathbf{v} f_s(\mathbf{R},\mu,v_\parallel,t)\times\left[\mu B_0(\mathbf{R}')+\frac{1}{2}m_sv_\parallel'^2 - \mu B_0(\mathbf{R})+\frac{1}{2}m_sv_\parallel^2\right].
\end{equation}
The time derivation of $\mathcal{E}_k$ is
\begin{equation}\label{eq:Ekchange}
\begin{aligned}
    \frac{d\mathcal{E}_{k,s}}{dt}=&\int d\mathbf{x} \int d\mathbf{v} f_s\times\left[\mu\nabla B_0\cdot\frac{d\mathbf{R}}{dt} + m_sv_\parallel\frac{dv_\parallel}{dt}\right]\\
    =&\int d\mathbf{x} \int d\mathbf{v} f_s\times\left[\left.\frac{d\mathbf{R}}{dt}\right|_{B\perp}\cdot Z_s\left(-\nabla\left\langle\mathbf{\phi}\right\rangle+\nabla\left\langle\delta\mathbf{A}_\perp\cdot\mathbf{v}_\perp\right\rangle\right)-v_\parallel Z_s\frac{\partial\left\langle\delta A_{\parallel}\right\rangle}{\partial t}\right],
\end{aligned}
\end{equation}
We have used the Hamiltonian to the first order in Eq \eqref{eq:Ekchange}. And $\left.\frac{d\mathbf{R}}{dt}\right|_{B\perp}$ is defined as the summation of the linear gyrocenter velocity and the magnetic fluttering velocity,
$$\left.\frac{d\mathbf{R}}{dt}\right|_{B\perp}=v_{\parallel}\frac{\mathbf{B}^{*}}{B_{\parallel}^{*}}+\frac{\mu}{Z_{s}}\frac{\mathbf{b}_{0}\times\nabla B_{0}}{B_{\parallel}^{*}}.
$$
This time derivative of kinetic energy describes the energy transferring
rate from wave to particles, and we can verify the energy conservation by checking
\begin{equation}
    \sum_s \frac{1}{\mathcal{E}_f}\frac{d\mathcal{E}_{k,s}}{dt} = - \frac{1}{\mathcal{E}_f}\frac{d\mathcal{E}_{f}}{dt}.
\end{equation}

\section{Terms in electron momentum equation}\label{sec:EMomt}
\begin{equation}
    \delta\mathbb{P} = \delta P_{\parallel e}\left(\mathbf{b}_{0}+\frac{\delta\mathbf{B}_{\perp}}{B_{0}}\right)+P_{\parallel0e}\frac{\delta\mathbf{B}_{\perp}}{B_{0}}+m_en_{0}\delta u_{\parallel e}\left(\mathbf{v}_{E}+\mathbf{v}_{b\parallel}\right)+ m_en_{0e}u_{e\parallel0}\left(\delta\mathbf{v}_{E}+\mathbf{v}_{b\parallel}\right)+\delta\boldsymbol{\Pi}_{c}+\delta\boldsymbol{\Pi}_{g}.
\end{equation}
\begin{equation}
\begin{split}
\delta\boldsymbol{\Pi}_{c}/m_e \equiv& -\int d\mathbf{v} v_{\parallel}\frac{m_{e}v_{\parallel}^{2}}{eB_{\parallel}^{*}}\bunit\times\left(\nabla\times\bunit\right)\times\hat{b}_{0} \delta f_{e}\\
= -& \frac{m_{e}n_{0e}u_{\parallel0e}}{eB_{0}}\bunit\times\left(\nabla\times\bunit\right)\times\bunit\left\{ \left(\frac{e\phi_{eff}}{T_{e}}-\frac{\delta B_\parallel}{B_0}-\frac{e}{T_{e}}\frac{\partial\phi_{eq}}{\partial\psi_{0}}\delta\psi\right)\left(u_{\parallel0e}^{2}+3v_{th,e}^{2}\right)\right.\\
& + \left(u_{\parallel0e}^{2}+3v_{th,e}^{2}\right)\left(\frac{\partial\ln n_{0e}}{\partial\psi_{0}}\delta\psi+\frac{\partial\ln n_{0e}}{\partial\alpha_{0}}\delta\alpha\right)+3v_{th,e}^{2}\left(\frac{\partial\ln T_{e}}{\partial\psi_{0}}\delta\psi+\frac{\partial\ln T_{e}}{\partial\alpha_{0}}\delta\alpha\right)\\
& +\left.3\left(v_{th,e}^{2}+u_{\parallel0e}^{2}\right)\left(\frac{\partial\ln u_{\parallel0e}}{\partial\psi_{0}}\delta\psi+\frac{\partial\ln u_{\parallel0e}}{\partial\alpha_{0}}\delta\alpha\right)\right\}\\
& - \frac{m_{e}}{eB_{0}}\bunit\times\left(\nabla\times\bunit\right)\times\bunit\int d^{3}vv_{\parallel}^{3}\delta h_{e},
\end{split}
\end{equation}
where $v_{th}\equiv\sqrt{T_e/m_e}$.
\begin{equation}
\begin{split}
    \delta\boldsymbol{\Pi}_{g}/m_e \equiv
    &-\int d\mathbf{v} \frac{\mu v_{\parallel}}{eB_{\parallel}^{*}}\bunit\times\nabla B_{0} \delta f_{e}\\
    =&-\frac{m_{e}}{eB_{0}^{2}}n_{0e}u_{\parallel0e}v_{th,e}^{2}\mathbf{b}_{0}\times\nabla B_{0}\left(\frac{e\phi_{eff}}{T_{e}}-2\frac{\delta B_\parallel}{B_0}-\frac{e}{T_{e}}\frac{\partial\phi_{eq}}{\partial\psi_{0}}\delta\psi\right)\\
    &-\frac{m_{e}n_{0e}u_{\parallel0e}v_{th,e}^{2}}{eB_{0}^{2}}\mathbf{b}_{0}\times\nabla B_{0}\left[\left(\frac{\partial\ln n_{0e}}{\partial\psi_{0}}\delta\psi+\frac{\partial\ln n_{0e}}{\partial\alpha_{0}}\delta\alpha\right)\right.\\
    &\left.+\left(\frac{\partial\ln T_{e}}{\partial\psi_{0}}\delta\psi+\frac{\partial\ln T_{e}}{\partial\alpha_{0}}\delta\alpha\right)+\left(\frac{\partial\ln u_{\parallel0e}}{\partial\psi_{0}}\delta\psi+\frac{\partial\ln u_{\parallel0e}}{\partial\alpha_{0}}\delta\alpha\right)\right]\\
    &-\frac{1}{eB_{0}}\mathbf{b}_{0}\times\nabla B_{0}\int d^{3}v\mu v_{\parallel}\delta h_{e}
\end{split}
\end{equation}
\begin{equation}
\begin{split}
    \delta\Xi =& \left[P_{\perp0e}\frac{\delta\mathbf{B}_{\perp}}{B_{0}}+\delta P_{\perp e}\left(\mathbf{b}_{0}+\frac{\delta\mathbf{B}_{\perp}}{B_{0}}\right)\right]\cdot\frac{\nabla B_{0}}{B_{0}} -e\left[n_{0e}\frac{\delta\mathbf{B}_{\perp}}{B_0}+\delta n_{e}\left(\mathbf{b}_{0}+\frac{\delta\mathbf{B}_{\perp}}{B_{0}}\right)\right]\cdot\nabla\phi_{eq}\\
    &-\left(n_{0e}+\delta n_{e}\right)\left(\mathbf{b}_{0}+\frac{\delta\mathbf{B}_{\perp}}{B_0}\right)\cdot e\nabla\delta\phi +\left(\mathbf{b}_{0} +\frac{\delta\mathbf{B}_{\perp}}{B_0}\right)\cdot\frac{\nabla\delta B_{\parallel}}{B_{0}}\left(P_{\perp0e}+\delta P_{\perp e}\right)\\
    &+e\nabla\times\mathbf{b}_{0}\cdot\nabla\phi\frac{m_{e}}{B_{0}}n_{0}\delta u_{\parallel e}+e\frac{\delta\boldsymbol{\Pi}_{g}/m_e}{\mathbf{b}_{0}\times\nabla B_{0}}\left[\nabla\times\mathbf{b}_{0}\cdot\nabla\left(B_{0}+\delta B_{\parallel}\right)\right]
\end{split}
\end{equation}
The expression of $\delta \mathbb{P}^{na}$  and $\delta\Xi^{na}$ in the equation of $\delta A_\parallel^{na}$ are
\begin{equation}
\begin{split}
    \delta\mathbb{P}^{na}=&\delta P_{\parallel e}^{na}\left(\mathbf{b}_{0}+\frac{\delta\mathbf{B}_{\perp}}{B_{0}}\right)+P_{\parallel0e}\frac{\delta\mathbf{B}_{\perp}}{B_{0}}+m_{e}n_{0}\delta u_{\parallel e}\left(\mathbf{v}_{E}+\mathbf{v}_{b\parallel}\right)+m_{e}n_{0e}u_{e\parallel0}\left(\delta\mathbf{v}_{E}+\mathbf{v}_{b\parallel}\right)\\
    & + \delta\boldsymbol{\Pi}_{c} +\delta\boldsymbol{\Pi}_{g} +\left(\mathbf{b}_{0}+\frac{\delta\mathbf{B}_{\perp}}{B_{0}}\right)\left(\delta P_{\parallel}^{ad}-\frac{e\phi_{eff}}{T_{e}}P_{\parallel0e}+\frac{e\phi_{eff}}{T_{e}}n_{0e}T_{e}\frac{u_{\parallel0e}^{2}}{v_{the}^{2}}\right)
\end{split}
\end{equation}
\begin{equation}
\begin{split}
    \delta\Xi^{na}=&\left[P_{\perp0e}\frac{\delta\mathbf{B}_{\perp}}{B_{0}}+\delta P_{\perp e}\left(\mathbf{b}_{0}+\frac{\delta\mathbf{B}_{\perp}}{B_{0}}\right)\right]\cdot\frac{\nabla B_{0}}{B_{0}}-e\left[n_{0e}\frac{\delta\mathbf{B}_{\perp}}{B_0}+\delta n_{e}\left(\mathbf{b}_{0}+\frac{\delta\mathbf{B}_{\perp}}{B_{0}}\right)\right]\cdot\nabla\phi_{eq}\\
    &+\left(\mathbf{b}_{0}+\frac{\delta\mathbf{B}_{\perp}}{B_{\parallel}^{*}}\right)\cdot\frac{\nabla\delta B_{\parallel}}{B_{0}}\left(P_{\perp0e}+\delta P_{\perp e}\right)\\
	&+en_{0e}\frac{\delta\mathbf{B}_{\perp}}{B_{0}}\cdot\nabla\delta\phi_{ind}+en_{0e}\phi_{eff}\nabla\cdot\left(\mathbf{b}_{0}+\frac{\delta\mathbf{B}_{\perp}}{B_{0}}\right)+e\phi_{eff}\frac{\delta\mathbf{B}_{\perp}}{B_{0}}\cdot\nabla n_{0e}\\
    &- e\delta n_{e}\mathbf{b}_{0}\cdot\nabla\delta\phi_{ff}- e\delta n_{e}\frac{\delta\mathbf{B}_{\perp}}{B_{0}}\cdot\nabla\delta\phi\\
    &+e\nabla\times\mathbf{b}_{0}\cdot\nabla\phi\frac{m_{e}}{B_{0}}n_{0}\delta u_{\parallel e}+e\frac{\delta\boldsymbol{\Pi}_{g}}{\mathbf{b}_{0}\times\nabla B_{0}}\left[\nabla\times\mathbf{b}_{0}\cdot\nabla\left(B_{0}+\delta B_{\parallel}\right)\right]\\
    &-\left(n_{0e}+\delta n_{e}\right)\left(\mathbf{b}_{0}+\frac{\delta\mathbf{B}_{\perp}}{B_{0}}\right)\cdot e\nabla\delta\phi_{00}.\label{eq:NA_stress}
\end{split}
\end{equation}
Note that the last term of $\delta\Xi^{na}$ is non-zero if we define the `zonal' potential as the flux-averaged value along the unperturbed flux surface. In the massless electron limit, we still need to calculate $\delta A_\parallel^{na}$, and this term persists. However, it is impossible to have a parallel acceleration from the zonal potential in the massless limit. The tricky part is that, if we strictly follow the definition of `zonal' part following the non-perturbative flux surface, then even in the massless limit, $\delta h_e$ is not zero. An exact opposite term in $\delta h_e$ will cancel this term in $\delta A_\parallel^{na}$. This mismatch between the massless electron limit and the non-zero $\delta h_e$ is inconvenient and sometimes causes confusion. In practice, we will drop the last term of Eq \eqref{eq:NA_stress} and directly set $\delta h_e$ to 0 for the adiabatic electron response. And the exact form of $\delta\Xi^{na}$ and $\delta h_e$ will be retained for the kinetic electron response.

The equilibrium pressure terms in the equation are defined as
\begin{equation}
\begin{split}
    P_{\perp0e} &\equiv\int d\mathbf{v}\mu B_0 f_{0e}= n_{0e}T_e,\\
P_{\parallel0e} &\equiv \int d\mathbf{v} mv_\parallel^2 f_{0e}= n_{0e}T_e \left(1+\frac{u_{\parallel0e}^2}{v_{th,e}^2}\right),\\
\end{split}
\end{equation}
and the perturbed pressure terms are defined as
\begin{subequations}
\begin{align}
    \delta P_{\perp e} &\equiv \int d\mathbf{v}\mu B_0 \delta f_{e}^{ad} + \int d\mathbf{v}\mu B_0 \delta h_e ,\label{eq:delpperp_orig}\\
    \delta P_{\parallel e} &\equiv \int d\mathbf{v} m_ev_\parallel^2 \delta f_{e}^{ad} + \int d\mathbf{v}m_ev_\parallel^2 \delta h_e.\label{eq:delppara_orig}
\end{align}
\end{subequations}

Where the adiabatic/analytic part of the pressure terms are 
\begin{equation}
\begin{split}
    \delta P_{\perp e}^{ad} \equiv&\int d\mathbf{v}\mu B_0\delta f_e^{ad}\\
    =&en_{0e}\phi_{eff} - 2P_{\perp0e}\frac{\delta B_\parallel}{B_0} + \frac{\partial P_{\perp0e}}{\partial\psi_0}\delta\psi^{ad} + \frac{\partial P_{\perp0e}}{\partial\alpha_0}\delta\alpha^{ad} - en_{0e}\frac{\partial\phi_{eq}}{\partial\psi_0}\delta\psi^{ad},\\
    \delta P_{\parallel e}^{ad}\equiv&\int d\mathbf{v}m_e v_\parallel^2\delta f_e^{ad}\\
    =&\frac{e\phi_{eff}}{T_e}P_{\parallel0e} - P_{\parallel0e}\frac{\delta B_\parallel}{B_0} -\frac{e}{T_e}P_{\parallel0e}\frac{\partial\phi_{eq}}{\partial\psi_0}\delta\psi^{ad} \\
    & + \left[\frac{\partial\left(n_{0e}T_{e}\right)}{\partial\psi_{0}}\delta\psi^{ad}+\frac{\partial\left(n_{0e}T_{e}\right)}{\partial\alpha_{0}}\delta\alpha^{ad}\right]\left(1+\frac{u_{\parallel0e}^{2}}{v_{th}^{2}}\right)\\
    & -n_{0e}\left(\frac{\partial T_{e}}{\partial\psi_{0}}\delta\psi^{ad}+\frac{\partial T_{e}}{\partial\alpha_{0}}\delta\alpha^{ad}\right)\frac{u_{\parallel0e}^{2}}{v_{th}^{2}}\\
    & +2m_{e}n_{0e}u_{\parallel0e}\left(\frac{\partial u_{\parallel0e}}{\partial\psi_{0}}\delta\psi^{ad}+\frac{\partial u_{\parallel0e}}{\partial\alpha_{0}}\delta\alpha^{ad}\right),\\
\end{split}
\end{equation}

\section{Neoclassical tearing mode simulations}\label{sec:NTMmodel}

For simulations of resistive MHD modes like neoclassical tearing mode (NTM), we can add the drag force term in $\dot{v}_{\parallel e}$ to model the resistivity from ion-electron collisions. $\dot{v}_{\parallel e, res} = \nu_{ei} (\mathbf{u}_{\parallel e} - \mathbf{u}_{\parallel i})$, where $\nu_{ei}$ is the collision frequency. This will induce an additional resistivity term in $\phi_{eff}$, and the momentum equation becomes
\begin{equation}
\begin{split}
en_{0e}\frac{\partial}{\partial t}\left[-\frac{c^{2}}{\omega_{pe}^{2}}\nabla_{\perp}^{2}+\frac{n_{0e}+\delta n_{e}}{n_{0e}}\right]\delta A_{\parallel} =& \nabla\cdot\delta\mathbb{P} + \delta\Xi +\frac{m_{e}}{e}\sum_{s\neq e}Z_{s}n_{0s}\frac{\partial\delta u_{\parallel s}}{\partial t} \\
&- \frac{\nu_{ei}m_e}{e} \left(\sum_{s\neq e}Z_s n_{0s}\delta u_{\parallel,s} - e n_{0e} \delta u_{\parallel,e}\right).\label{eq:resist_Apara}
\end{split}
\end{equation}
In GTC simulation, the resistivity term in Eq \eqref{eq:resist_Apara} is directly calculated by $\eta_\parallel\left(\frac{1}{\mu_0}\nabla_\perp^2\delta A_\parallel + \delta j_{\parallel,bs}\right)$, where the parallel resistivity $\eta_\parallel = \nu_{ei}{m_e}/\left(n_{0e}e^2\right)$ can be evaluated by Spitzer formulation. In the code, this term appears in the $\delta\phi_{ind}$, $\phi_{eff}$ calculations, and is considered as part of $\delta A_{\parallel}^{ad}$.

For the same reason, Ampere's law can include the resistivity-induced current, e.g., the bootstrap current, and Eq \eqref{eq:uepara} becomes
\begin{equation}
    e n_e\delta u_{\parallel e} = \frac{1}{\mu_0}\nabla_\perp^2\delta A_{\parallel} + \sum_{s\neq e} Z_s n_s \delta u_{\parallel s} + \delta j_{\parallel,bs}
\end{equation}
In addition, in the simulations of NTM, we use a diffusive model to calculate the perturbed pressure to replace the perturbed pressure in the continuity equation and the momentum equation \cite{WangKJ_NTM_2023},
\begin{equation}
\begin{split}
    \frac{\partial \delta p}{\partial t} =& \chi_\parallel\nabla_\parallel^2\delta p + \chi_\parallel\nabla_\parallel\left(\frac{\delta \mathbf{B}_\perp}{B_0}\cdot\nabla p_0\right) + \chi_\perp\nabla_\perp^2\delta p \\
    &+ \chi_\parallel\left[\frac{\delta \mathbf{B}_\perp}{B_0}\cdot\nabla\left(\nabla_\parallel\delta p + \frac{\delta \mathbf{B}_\perp}{B_0} \cdot \nabla\delta p + \frac{\delta \mathbf{B}_\perp}{B_0}\cdot \nabla p_0\right)\right. \\
    &\left.+ \nabla_\parallel\left(\frac{\delta \mathbf{B}_\perp}{B_0}\cdot\nabla\delta p\right)\right]_{NL},
\end{split}
\end{equation}
where $\chi_\parallel$ and $\chi_\perp$ is the parallel and perpendicular diffusion coefficients for $\delta p$, $\delta p=\sum_{s\neq e}\delta p_s + \delta p_e$ is the total pressure. In \cite{WangKJ_NTM_2023}, $\chi_\parallel$ and $\chi_\perp$ are given as known parameters.

\section{Calculation of equilibrium parallel current}\label{sec:Jpara}

In the GTC ideal MHD simulations in tokamaks, the kink mode can be driven by parallel
current and pressure gradient, the parallel current can be calculated
numerically in several ways.

\subsection{Calculating current directly in Boozer coordinate system}

We directly calculate $J_{\parallel0}$ from Eq \eqref{eq:J_from_delta}. The $\hat{\delta}$-current is given by
\begin{equation}
    \hat{\delta}=-\frac{I\mathsf{g}^{\psi\theta}+g\mathsf{g}^{\psi\zeta}}{\mathsf{g}^{\psi\psi}},\label{eq:delta_calculation}
\end{equation}
where $\mathsf{g}^{\alpha\beta}=\nabla\alpha\cdot\nabla\beta$
is the element of the contra-variant metric tensor. One can note that $\mathsf{g}^{\psi\zeta}/\mathsf{g}^{\psi\theta}\sim O\left(\epsilon^{2}\right)$.
But in the calculation of $\hat{\delta}$, $\mathsf{g}^{\psi\zeta}$ cannot be simply neglected
because $I/g\sim O\left(\epsilon^{2}\right)$, and the two terms are
comparable in Eq (\ref{eq:delta_calculation}). Axisymmetry exists for
2D equilibria, and the toroidal direction $\nabla\phi$ is perpendicular
to $\nabla\psi$ and $\nabla\theta$. However, a transformation has
been applied to $\phi$ to construct the Boozer toroidal angle $\zeta$.
For 2d equilibrium, the difference between $\zeta$ and $\phi$ is
given by \cite{White_Book}
\begin{align}
\begin{split}
    \nu & =\phi-\zeta\\
 & =\int\left(\frac{g\mathcal{J}}{R^{2}}-q\right)\text{d}\theta.
\end{split}
\end{align}
There is one degree of freedom to fully determine $\nu$ function,
and we can simply choose $\nu\left(\theta=0\right)=0$. Now from $\phi\left(\psi,\theta\right)=\zeta+\nu$
we can obtain $\partial_{\psi}\phi=\partial_{\psi}\nu$, $\partial_{\theta}\phi=\partial_{\theta}\nu$,
and $\partial_{\zeta}\phi=1$. Together with $R\left(\psi,\theta\right)$
and $Z\left(\psi,\theta\right)$, the covariant metric tensor is constructed,
\begin{equation}
\mathsf{g}_{\alpha\beta}=\frac{\partial R}{\partial\alpha}\frac{\partial R}{\partial\beta}+R^{2}\frac{\partial\phi}{\partial\alpha}\frac{\partial\phi}{\partial\beta}+\frac{\partial Z}{\partial\alpha}\frac{\partial Z}{\partial\beta},\label{eq:cometric}
\end{equation}
and accordingly, the contra-variant metric tensor is constructed from the covariant one,
\begin{equation}
\left(\begin{array}{ccc}
\mathsf{g}^{\psi\psi} & \mathsf{g}^{\psi\theta} & \mathsf{g}^{\psi\zeta}\\
\mathsf{g}^{\theta\psi} & \mathsf{g}^{\theta\theta} & \mathsf{g}^{\theta\zeta}\\
\mathsf{g}^{\zeta\psi} & \mathsf{g}^{\zeta\theta} & \mathsf{g}^{\zeta\zeta}
\end{array}\right)=\frac{1}{\Delta}\left(\begin{array}{ccc}
\mathsf{g}_{\theta\theta}\mathsf{g}_{\zeta\zeta}-\mathsf{g}_{\theta\zeta}\mathsf{g}_{\theta\zeta} & \mathsf{g}_{\psi\zeta}\mathsf{g}_{\theta\zeta}-\mathsf{g}_{\psi\theta}\mathsf{g}_{\zeta\zeta} & \mathsf{g}_{\psi\theta}\mathsf{g}_{\theta\zeta}-\mathsf{g}_{\theta\theta}\mathsf{g}_{\psi\zeta}\\
\mathsf{g}_{\psi\zeta}\mathsf{g}_{\theta\zeta}-\mathsf{g}_{\psi\theta}\mathsf{g}_{\zeta\zeta} & \mathsf{g}_{\psi\psi}\mathsf{g}_{\zeta\zeta}-\mathsf{g}_{\psi\zeta}\mathsf{g}_{\psi\zeta} & \mathsf{g}_{\psi\theta}\mathsf{g}_{\psi\zeta}-\mathsf{g}_{\psi\psi}\mathsf{g}_{\zeta\zeta}\\
\mathsf{g}_{\psi\theta}\mathsf{g}_{\theta\zeta}-\mathsf{g}_{\theta\theta}\mathsf{g}_{\psi\zeta} & \mathsf{g}_{\psi\theta}\mathsf{g}_{\psi\zeta}-\mathsf{g}_{\psi\psi}\mathsf{g}_{\zeta\zeta} & \mathsf{g}_{\psi\psi}\mathsf{g}_{\theta\theta}-\mathsf{g}_{\psi\theta}\mathsf{g}_{\psi\theta}
\end{array}\right),\label{eq:cometric_contrametric}
\end{equation}
where $\Delta$ is the determinant of the covariant metric tensor,
$\Delta=\mathsf{g}_{\psi\psi}\mathsf{g}_{\theta\theta}\mathsf{g}_{\zeta\zeta}-\mathsf{g}_{\psi\theta}\mathsf{g}_{\psi\theta}\mathsf{g}_{\zeta\zeta}-\mathsf{g}_{\theta\zeta}\mathsf{g}_{\theta\zeta}-\mathsf{g}_{\psi\zeta}\mathsf{g}_{\psi\zeta}+2\mathsf{g}_{\psi\theta}\mathsf{g}_{\theta\zeta}\mathsf{g}_{\psi\zeta}$.
Note that both the covariant and the contravariant metric tensor are symmetric,
$\mathsf{g}_{\alpha\beta}=\mathsf{g}_{\beta\alpha}$, and $\mathsf{g}^{\alpha\beta}=\mathsf{g}^{\beta\alpha}$.
The disadvantage of this method is that $\nu$ itself is a higher-order term compared to $\phi$ or $\zeta$, and often has large numerical errors near the separatrix. The numerical errors in $\hat{\delta}$ term and therefore the error in the parallel current are amplified through Eqs \eqref{eq:cometric} \eqref{eq:cometric_contrametric}, and \eqref{eq:delta_calculation}.

\subsection{Calculating current using force balance}

The total equilibrium current can be expressed as
\begin{align}
\mu_{0}\mathbf{J}_0 & =\mu_{0}j^{\zeta}\hat{\mathbf{e}}_{\zeta}+\mu_{0}j^{\theta}\hat{\mathbf{e}}_{\theta},
\end{align}
where $\hat{\mathbf{e}}_{\alpha}=\partial\mathbf{r}/\partial\alpha$
is the covariant basis vector. Comparing it with the expression of $\nabla\times\mathbf{B}_0$ in Boozer coordinate, we have $\mu_{0}j^{\theta}=-g'(\psi)/\mathcal{J}$, then along with the equilibrium force balance equation $\mathbf{J}_{0}\times\mathbf{B}_0=\nabla p$
we have
\begin{equation}
j^{\theta}q-j^{\zeta}=\frac{d p}{d \psi},
\end{equation}
that is,
\begin{equation}
j^{\zeta} = -\frac{d p}{d \psi} - \frac{q}{\mu_0\mathcal{J}}\frac{d g}{d \psi},
\end{equation}
and the parallel current can be written as
\begin{align}
\begin{split}
    \frac{\mu_{0}J_{\parallel0}}{B_0}=\frac{\mu_0}{B_0^2}\left(gj^\zeta+Ij^\theta\right)=-\frac{dg}{d\psi}-\frac{\mu_{0}g}{B_0^{2}}\frac{dp}{d\psi}.\label{eq:method2}
\end{split}
\end{align}

This method is much more straightforward and numerically friendlier than
the first one. It only requires the two 1-dimensional derivatives
and does not need the calculation of high-order terms like $\hat{\delta}$ or
$\nu$. But it requires a fully consistent pressure profile with other
field quantities like $g$ and $B$ fields.

\section{Derivation of MHD dispersion relation}\label{sec:MHD_DR}

Combining Eqs \eqref{eq:poisson_MHD} and \eqref{eq:Ampere_MHD}, and using the condition $\phi_{ind}=-\phi$ taking $\delta E_\parallel = 0$, we can get the following equation \cite{Deng_2012}
\begin{equation*}
    \frac{\omega^{2}}{v_{A}^{2}}\nabla_{\perp}^{2}\delta\phi+i\mathbf{B}_{0}\cdot\mathbf{\nabla}\left(\frac{\nabla_{\perp}^{2}\left(k_{\parallel}\phi\right)}{B_{0}}\right)+i\omega\mu_{0}\left(i\omega e\delta n+\nabla\cdot\delta\mathbf{J}_{\parallel}\right)=0.
\end{equation*}
Substituting the linear terms of Eq \eqref{eq:MHD_continuity} in the above equation, we obtain the dispersion relation for the system
\begin{equation*}
\begin{aligned}
    0=&\frac{\omega^{2}}{v_{A}^{2}}\nabla_{\perp}^{2}\delta\phi+i\mathbf{B}_{0}\cdot\mathbf{\nabla}\left(\frac{\nabla_{\perp}^{2}\left(k_{\parallel}\phi\right)}{B_{0}}\right)+i\omega\mu_{0}\left[\frac{\mathbf{b}_{0}\times\nabla\left(\delta P_{\parallel}+\delta P_{\perp}\right)}{B_{0}}\cdot\frac{\nabla B_{0}}{B_{0}}-\delta\mathbf{B}_{\perp}\cdot\nabla\left(\frac{J_{\parallel0}}{B_{0}}\right)\right.\\
	&-\frac{\nabla\times\mathbf{B_{0}}}{B_{0}^{2}}\cdot\left(\nabla\delta P_{\parallel}+\frac{\left(\delta P_{\perp}-\delta P_{\parallel}\right)\nabla B_{0}}{B_{0}}\right)+\nabla\cdot\left(\frac{\delta P_{\parallel}\mathbf{b}_{0}\nabla\times\mathbf{b}_{0}\cdot\mathbf{b}_{0}}{B_{0}}\right)-\mathbf{b}_{0}\times\nabla\delta B_{\parallel}\cdot\nabla\left(\frac{P_{\perp0}}{B_{0}^{2}}\right)\\
	&\left.-\frac{\nabla\times\mathbf{b}_{0}\cdot\nabla\delta B_{\parallel}}{B_{0}^{2}}P_{\perp0}+\nabla\cdot\left(\frac{\delta P_{\parallel}\mathbf{b}_{0}\nabla\times\mathbf{b}_{0}\cdot\mathbf{b}_{0}}{B_{0}}\right)\right].
\end{aligned}
\end{equation*}
In the long wavelength limit, we can keep the terms up to the order of $O(\epsilon^2)$, and get the simplified dispersion relation,
\begin{equation*}
\begin{aligned}
    0=&\frac{\omega^{2}}{v_{A}^{2}}\nabla_{\perp}^{2}\delta\phi+i\mathbf{B}_{0}\cdot\mathbf{\nabla}\left(\frac{\nabla_{\perp}^{2}\left(k_{\parallel}\phi\right)}{B_{0}}\right)+i\mathbf{b}_{0}\times\nabla\left(k_{\parallel}\phi\right)\cdot\nabla\left(\frac{\mu_0J_{\parallel0}}{B_{0}}\right)\\
    &-i\omega\mu_{0}\left[\frac{\nabla\times\mathbf{b}_{0}}{B_{0}}\cdot\nabla\delta P_{\parallel}+\frac{\mathbf{b}_{0}\times\nabla B_{0}}{B_{0}^{2}}\cdot\nabla\delta P_{\perp}-\frac{\mathbf{b}_{0}\times\nabla P_{\perp0}}{B_{0}^{2}}\cdot\nabla\delta B_{\parallel} -\frac{(\nabla\times\mathbf{b}_0)_\parallel}{B_0}\mathbf{b}_0\cdot\nabla\delta P_\parallel\right].
\end{aligned}
\end{equation*}

Eq \eqref{eq:bpara_MHD} shows that the compressional magnetic perturbation $\delta B_\parallel/B_0$ is much smaller than the pressure perturbation $\delta P/P_0$ because $\beta_e\ll 1$. However, the $\delta B_\parallel$ drive can correct the $\delta P_\perp$ drive by canceling out the ``drift-reversal" effects from the grad-B drift associated with the perpendicular diamagnetic current\cite{Berk_1977,Tang_1980}. We can use the perpendicular force balance relation given in Eq \eqref{eq:bpara_MHD} to rewrite the $\delta B_\parallel$ drive in the above equation,
$$
\frac{\mathbf{b}_0\times\nabla P_{\perp 0}}{eB_0^2}\cdot\nabla\delta B_\parallel = -\frac{\mu_0\mathbf{J}_{\perp 0}}{eB_0^2}\cdot\nabla\delta P_\perp,
$$
where $\mathbf{J}_{\perp0}=\frac{\mathbf{b}_0\times\nabla P_{\perp0}}{B_0}$ is the perpendicular diamagnetic current. 

And the dispersion relation becomes
\begin{equation}
    \begin{aligned}
        0=&\frac{\omega^{2}}{v_{A}^{2}}\nabla_{\perp}^{2}\delta\phi+i\mathbf{B}_{0}\cdot\mathbf{\nabla}\left(\frac{\nabla_{\perp}^{2}\left(k_{\parallel}\phi\right)}{B_{0}}\right)+i\mathbf{b}_{0}\times\nabla\left(k_{\parallel}\phi\right)\cdot\nabla\left(\frac{\mu_0 J_{\parallel0}}{B_{0}}\right)\\
        &-i\omega\mu_{0}\left[2\frac{\mu_{0}J_{\parallel0}}{B_{0}^{2}}\nabla_{\parallel}\delta P_{\parallel}+\frac{\mathbf{b}_{0}\times\boldsymbol{\kappa}}{B_{0}}\cdot\nabla\left(\delta P_{\perp}+\delta P_{\parallel}\right)\right].
    \end{aligned}
\end{equation}
In the low-beta limit, the perturbed pressure is isotropic $\delta P_\perp=\delta P_\parallel$. And by further assuming $k_\parallel\ll k_\perp$, the equation reduces to the commonly used ideal MHD dispersion relation\cite{Fu_2006, Bao_2021},
\begin{equation*}
    \begin{aligned}
        0=&\frac{\omega^{2}}{v_{A}^{2}}\nabla_{\perp}^{2}\delta\phi+i\mathbf{B}_{0}\cdot\mathbf{\nabla}\left(\frac{\nabla_{\perp}^{2}\left(k_{\parallel}\phi\right)}{B_{0}}\right)+i\mathbf{b}_{0}\times\nabla\left(k_{\parallel}\phi\right)\cdot\nabla\left(\frac{\mu_0J_{\parallel0}}{B_{0}}\right)\\
        &-i\omega\mu_{0}\frac{2\mathbf{b}_{0}\times\boldsymbol{\kappa}}{B_{0}}\cdot\nabla\delta P.
    \end{aligned}
\end{equation*}

\end{appendices}

\bibliography{ref.bib}

@article{Helander2014,
doi = {10.1088/0034-4885/77/8/087001},
url = {https://dx.doi.org/10.1088/0034-4885/77/8/087001},
year = {2014},
month = {jul},
publisher = {IOP Publishing},
volume = {77},
number = {8},
pages = {087001},
author = {Helander, Per},
title = {Theory of plasma confinement in non-axisymmetric magnetic fields},
journal = {Reports on Progress in Physics}
}

@article{Liu2022PRL,
  title = {Regulation of Alfv\'en Eigenmodes by Microturbulence in Fusion Plasmas},
  author = {Liu, P. and Wei, X. and Lin, Z. and Brochard, G. and Choi, G. J. and Heidbrink, W. W. and Nicolau, J. H. and McKee, G. R.},
  journal = {Phys. Rev. Lett.},
  volume = {128},
  issue = {18},
  pages = {185001},
  numpages = {5},
  year = {2022},
  month = {May},
  publisher = {American Physical Society},
  doi = {10.1103/PhysRevLett.128.185001},
  url = {https://link.aps.org/doi/10.1103/PhysRevLett.128.185001}
}

@article{Brochard2024PRL,
  title = {Saturation of Fishbone Instability by Self-Generated Zonal Flows in Tokamak Plasmas},
  author = {Brochard, G. and Liu, C. and Wei, X. and Heidbrink, W. and Lin, Z. and Gorelenkov, N. and Chrystal, C. and Du, X. and Bao, J. and Polevoi, A. R. and Schneider, M. and Kim, S. H. and Pinches, S. D. and Liu, P. and Nicolau, J. H. and L\"utjens, H.},
  journal = {Phys. Rev. Lett.},
  volume = {132},
  issue = {7},
  pages = {075101},
  numpages = {6},
  year = {2024},
  month = {Feb},
  publisher = {American Physical Society},
  doi = {10.1103/PhysRevLett.132.075101},
  url = {https://link.aps.org/doi/10.1103/PhysRevLett.132.075101}
}

@article{Xiao2015,
    author = {Xiao, Yong and Holod, Ihor and Wang, Zhixuan and Lin, Zhihong and Zhang, Taige},
    title = {Gyrokinetic particle simulation of microturbulence for general magnetic geometry and experimental profiles},
    journal = {Physics of Plasmas},
    volume = {22},
    number = {2},
    pages = {022516},
    year = {2015},
    month = {02},
    issn = {1070-664X},
    doi = {10.1063/1.4908275},
    url = {https://doi.org/10.1063/1.4908275},
    eprint = {https://pubs.aip.org/aip/pop/article-pdf/doi/10.1063/1.4908275/16143208/022516\_1\_online.pdf},
}

@article{JARDIN2004133,
title = {A triangular finite element with first-derivative continuity applied to fusion MHD applications},
journal = {Journal of Computational Physics},
volume = {200},
number = {1},
pages = {133-152},
year = {2004},
issn = {0021-9991},
doi = {https://doi.org/10.1016/j.jcp.2004.04.004},
url = {https://www.sciencedirect.com/science/article/pii/S0021999104001366},
author = {S.C. Jardin}
}

@article{GorelenkovNOVA,
    author = {Gorelenkov, N. N and Cheng, C. Z. and Fu, G. Y.},
    title = {Fast particle finite orbit width and Larmor radius effects on low-n toroidicity induced Alfvén eigenmode excitation},
    journal = {Physics of Plasmas},
    volume = {6},
    number = {7},
    pages = {2802-2807},
    year = {1999},
    month = {07},
    issn = {1070-664X},
    doi = {10.1063/1.873545},
    url = {https://doi.org/10.1063/1.873545},
    eprint = {https://pubs.aip.org/aip/pop/article-pdf/6/7/2802/19055052/2802\_1\_online.pdf},
}

@article{Lin1995_NC,
    author = {Lin, Z. and Tang, W. M. and Lee, W. W.},
    title = {Gyrokinetic particle simulation of neoclassical transport},
    journal = {Physics of Plasmas},
    volume = {2},
    number = {8},
    pages = {2975-2988},
    year = {1995},
    month = {08},
    issn = {1070-664X},
    doi = {10.1063/1.871196},
    url = {https://doi.org/10.1063/1.871196},
    eprint = {https://pubs.aip.org/aip/pop/article-pdf/2/8/2975/19189831/2975\_1\_online.pdf},
}

@article{Garbet_2010,
doi = {10.1088/0029-5515/50/4/043002},
url = {https://doi.org/10.1088/0029-5515/50/4/043002},
year = {2010},
month = {mar},
publisher = {},
volume = {50},
number = {4},
pages = {043002},
author = {Garbet, X. and Idomura, Y. and Villard, L. and Watanabe, T.H.},
title = {Gyrokinetic simulations of turbulent transport},
journal = {Nuclear Fusion}
}

@article{Diamond_2005,
doi = {10.1088/0741-3335/47/5/R01},
url = {https://doi.org/10.1088/0741-3335/47/5/R01},
year = {2005},
month = {apr},
publisher = {},
volume = {47},
number = {5},
pages = {R35},
author = {Diamond, P H and Itoh, S-I and Itoh, K and Hahm, T S},
title = {Zonal flows in plasma—a review},
journal = {Plasma Physics and Controlled Fusion}
}

@article{Horton1999,
  title = {Drift waves and transport},
  author = {Horton, W.},
  journal = {Rev. Mod. Phys.},
  volume = {71},
  issue = {3},
  pages = {735--778},
  numpages = {0},
  year = {1999},
  month = {Apr},
  publisher = {American Physical Society},
  doi = {10.1103/RevModPhys.71.735},
  url = {https://link.aps.org/doi/10.1103/RevModPhys.71.735}
}

@article{Tang_1978,
doi = {10.1088/0029-5515/18/8/006},
url = {https://doi.org/10.1088/0029-5515/18/8/006},
year = {1978},
month = {aug},
publisher = {},
volume = {18},
number = {8},
pages = {1089},
author = {Tang, W.M.},
title = {Microinstability theory in tokamaks},
journal = {Nuclear Fusion}
}

@article{Helander2014_3Dgeometry,
doi = {10.1088/0034-4885/77/8/087001},
url = {https://dx.doi.org/10.1088/0034-4885/77/8/087001},
year = {2014},
month = {jul},
publisher = {IOP Publishing},
volume = {77},
number = {8},
pages = {087001},
author = {Helander, Per},
title = {Theory of plasma confinement in non-axisymmetric magnetic fields},
journal = {Reports on Progress in Physics}
}

@article{Brochard2025_NF,
doi = {10.1088/1741-4326/ad8013},
url = {https://dx.doi.org/10.1088/1741-4326/ad8013},
year = {2024},
month = {dec},
publisher = {IOP Publishing},
volume = {65},
number = {1},
pages = {016052},
author = {Brochard, G. and Liu, C. and Wei, X. and Heidbrink, W. and Lin, Z. and Falessi, M.V. and Zonca, F. and Qiu, Z. and Gorelenkov, N. and Chrystal, C. and Du, X. and Bao, J. and Polevoi, A.R. and Schneider, M. and Kim, S.H. and Pinches, S.D. and Liu, P. and Nicolau, J.H. and Lütjens, H. and the ISEP group},
title = {Saturation of fishbone instability through zonal flows driven by energetic particle transport in tokamak plasmas},
journal = {Nuclear Fusion}
}

@article{Wei2025_NF,
doi = {10.1088/1741-4326/ada049},
url = {https://dx.doi.org/10.1088/1741-4326/ada049},
year = {2025},
month = {jan},
publisher = {IOP Publishing},
volume = {65},
number = {2},
pages = {026026},
author = {Wei, Xishuo and Nicolau, Javier H and Choi, Gyungjin and Lin, Zhihong and Yang, Seong-Moo and Kim, SangKyeun and Lee, WooChang and Zhao, Chen and Cote, Tyler and Park, JongKyu and Orlov, Dmitri},
title = {Gyrokinetic simulations of the effects of magnetic islands on microturbulence in KSTAR},
journal = {Nuclear Fusion}
}

@article{LUTJENSXTOR,
title = {The XTOR code for nonlinear 3D simulations of MHD instabilities in tokamak plasmas},
journal = {Journal of Computational Physics},
volume = {227},
number = {14},
pages = {6944-6966},
year = {2008},
issn = {0021-9991},
doi = {https://doi.org/10.1016/j.jcp.2008.04.003},
url = {https://www.sciencedirect.com/science/article/pii/S0021999108002064},
author = {Hinrich Lütjens and Jean-François Luciani}
}

@article{BaoGAMSolver,
    author = {J. Bao and W. L. Zhang and D. Li and Z. Lin},
    title = {Effects of Plasma Diamagnetic Drift on Alfvén Continua and Discrete Eigenmodes in Tokamaks},
    journal = {Journal of Fusion Energy},
    year = {2020},
    doi = {https://doi.org/10.1007/s10894-020-00275-0},
    volume = {39},
    pages = {382-389},
}

@article{WangKJ_NTM_2023,
	author = {Kaijie Wang and Shuying Sun and Wenlu Zhang and Zhihong Lin and Xishuo Wei and Pengfei Liu and Hongying Feng and Xiaogang Wang and Ding Li},
	doi = {10.1088/1361-6587/aceb88},
	journal = {Plasma Physics and Controlled Fusion},
	month = {aug},
	number = {10},
	pages = {105005},
	publisher = {IOP Publishing},
	title = {Verification of gyrokinetic particle simulations of neoclassical tearing modes in fusion plasmas},
	url = {https://dx.doi.org/10.1088/1361-6587/aceb88},
	volume = {65},
	year = {2023}}

@article{Lee1983,
	author = {Lee, W. W.},
	doi = {10.1063/1.864140},
	eprint = {https://pubs.aip.org/aip/pfl/article-pdf/26/2/556/12758385/556\_1\_online.pdf},
	issn = {0031-9171},
	journal = {The Physics of Fluids},
	month = {02},
	number = {2},
	pages = {556-562},
	title = {{Gyrokinetic approach in particle simulation}},
	url = {https://doi.org/10.1063/1.864140},
	volume = {26},
	year = {1983}}

@article{Lin_1998,
	author = {Z. Lin and T. S. Hahm and W. W. Lee and W. M. Tang and R. B. White},
	doi = {10.1126/science.281.5384.1835},
	eprint = {https://www.science.org/doi/pdf/10.1126/science.281.5384.1835},
	journal = {Science},
	number = {5384},
	pages = {1835-1837},
	title = {Turbulent Transport Reduction by Zonal Flows: Massively Parallel Simulations},
	url = {https://www.science.org/doi/abs/10.1126/science.281.5384.1835},
	volume = {281},
	year = {1998}}

@misc{Greader,
	author = {Ben Dudson},
	title = {PyTokamak. https://github.com/bendudson/pyTokamak/blob/master/tokamak/formats/geqdsk.py}}

@article{Scipy,
	adsurl = {https://rdcu.be/b08Wh},
	author = {Virtanen, Pauli and Gommers, Ralf and Oliphant, Travis E. and Haberland, Matt and Reddy, Tyler and Cournapeau, David and Burovski, Evgeni and Peterson, Pearu and Weckesser, Warren and Bright, Jonathan and {van der Walt}, St{\'e}fan J. and Brett, Matthew and Wilson, Joshua and Millman, K. Jarrod and Mayorov, Nikolay and Nelson, Andrew R. J. and Jones, Eric and Kern, Robert and Larson, Eric and Carey, C J and Polat, {\.I}lhan and Feng, Yu and Moore, Eric W. and {VanderPlas}, Jake and Laxalde, Denis and Perktold, Josef and Cimrman, Robert and Henriksen, Ian and Quintero, E. A. and Harris, Charles R. and Archibald, Anne M. and Ribeiro, Ant{\^o}nio H. and Pedregosa, Fabian and {van Mulbregt}, Paul and {SciPy 1.0 Contributors}},
	doi = {10.1038/s41592-019-0686-2},
	journal = {Nature Methods},
	pages = {261--272},
	title = {{{SciPy} 1.0: Fundamental Algorithms for Scientific Computing in Python}},
	volume = {17},
	year = {2020}}

@article{Brizard_2017,
	author = {Alain J. Brizard},
	journal = {Physics of Plasmas},
	pages = {081201},
	title = {Variational principle for the parallel-symplectic representation of electromagnetic gyrokinetic theory},
	volume = {24},
	year = {2017}}

@article{Dong_2019,
	author = {G. Dong and J. Bao and A. Bhattacharjee and Z. Lin},
	journal = {Physics of Plasmas},
	pages = {010701},
	title = {Nonlinear saturation of kinetic ballooning modes by zonal fields in toroidal plasmas},
	volume = {26},
	year = {2019}}

@article{Chen_2001,
	author = {L. Chen and Z. Lin and R.B. White and F. Zonca},
	journal = {Nuclear Fusion},
	number = {6},
	title = {Non-linear zonal dynamics of drift and drift-Alfven turbulence in tokamak plasmas},
	volume = {41},
	year = {2001}}

@article{Deng_2012,
	author = {W. Deng and Z. Lin and I. Holod},
	doi = {10.1088/0029-5515/52/2/023005},
	journal = {Nuclear Fusion},
	month = {jan},
	number = {2},
	pages = {023005},
	publisher = {{IOP} Publishing},
	title = {Gyrokinetic simulation model for kinetic magnetohydrodynamic processes in magnetized plasmas},
	url = {https://doi.org/10.1088/0029-5515/52/2/023005},
	volume = {52},
	year = 2012}

@article{Holod_2009,
	author = {Holod,I. and Zhang,W. L. and Xiao,Y. and Lin,Z.},
	doi = {10.1063/1.3273070},
	eprint = {https://doi.org/10.1063/1.3273070},
	journal = {Physics of Plasmas},
	number = {12},
	pages = {122307},
	title = {Electromagnetic formulation of global gyrokinetic particle simulation in toroidal geometry},
	url = {https://doi.org/10.1063/1.3273070},
	volume = {16},
	year = {2009}}

@article{Dong_2017,
	author = {Dong,Ge and Bao,Jian and Bhattacharjee,Amitava and Brizard,Alain and Lin,Zhihong and Porazik,Peter},
	doi = {10.1063/1.4997788},
	eprint = {https://doi.org/10.1063/1.4997788},
	journal = {Physics of Plasmas},
	number = {8},
	pages = {081205},
	title = {Gyrokinetic particle simulations of the effects of compressional magnetic perturbations on drift-Alfvenic instabilities in tokamaks},
	url = {https://doi.org/10.1063/1.4997788},
	volume = {24},
	year = {2017}}

@article{Brochard_2022,
doi = {10.1088/1741-4326/ac48a6},
url = {https://dx.doi.org/10.1088/1741-4326/ac48a6},
year = {2022},
month = {jan},
publisher = {IOP Publishing},
volume = {62},
number = {3},
pages = {036021},
author = {Brochard, G. and Bao, J. and Liu, C. and Gorelenkov, N. and Choi, G. and Dong, G. and Liu, P. and Mc.Clenaghan, J. and Nicolau, J.H. and Wang, F. and Wang, W.H. and Wei, X. and Zhang, W.L. and Heidbrink, W. and Graves, J.P. and Lin, Z. and Lütjens, H.},
title = {Verification and validation of linear gyrokinetic and kinetic-MHD simulations for internal kink instability in DIII-D tokamak},
journal = {Nuclear Fusion}}

@misc{Dong_2021,
	archiveprefix = {arXiv},
	author = {Ge Dong and Xishuo Wei and Jian Bao and Guillaume Brochard and Zhihong Lin and William Tang},
	eprint = {2106.10849},
	primaryclass = {physics.plasm-ph},
	title = {Deep learning based surrogate model for first-principles global simulations of fusion plasmas},
	year = {2021}}

@article{Brizard_1992,
	author = {Brizard,Alain},
	doi = {10.1063/1.860129},
	eprint = {https://doi.org/10.1063/1.860129},
	journal = {Physics of Fluids B: Plasma Physics},
	number = {5},
	pages = {1213-1228},
	title = {Nonlinear gyrofluid description of turbulent magnetized plasmas},
	url = {https://doi.org/10.1063/1.860129},
	volume = {4},
	year = {1992}}

@article{BrizardHahm_2007,
	author = {Brizard, A. J. and Hahm, T. S.},
	doi = {10.1103/RevModPhys.79.421},
	issue = {2},
	journal = {Rev. Mod. Phys.},
	month = {Apr},
	numpages = {0},
	pages = {421--468},
	publisher = {American Physical Society},
	title = {Foundations of nonlinear gyrokinetic theory},
	url = {https://link.aps.org/doi/10.1103/RevModPhys.79.421},
	volume = {79},
	year = {2007}}

@article{White_1984,
	author = {White,R. B. and Chance,M. S.},
	doi = {10.1063/1.864527},
	eprint = {https://aip.scitation.org/doi/pdf/10.1063/1.864527},
	journal = {The Physics of Fluids},
	number = {10},
	pages = {2455-2467},
	title = {Hamiltonian guiding center drift orbit calculation for plasmas of arbitrary cross section},
	url = {https://aip.scitation.org/doi/abs/10.1063/1.864527},
	volume = {27},
	year = {1984}}

@article{Porazik_2011,
	author = {Porazik, Peter and Lin, Zhihong},
	doi = {10.4208/cicp.241110.280111a},
	journal = {Communications in Computational Physics},
	number = {4},
	pages = {899--911},
	publisher = {Cambridge University Press},
	title = {Gyrokinetic Simulation of Magnetic Compressional Modes in General Geometry},
	volume = {10},
	year = {2011}}

@article{SJWang2024NLT,
  title = {Self-Organized Evolution of the Internal Transport Barrier in Ion-Temperature-Gradient Driven Gyrokinetic Turbulence},
  author = {Wang, Shaojie and Wang, Zihao and Wu, Tiannan},
  journal = {Phys. Rev. Lett.},
  volume = {132},
  issue = {6},
  pages = {065106},
  numpages = {6},
  year = {2024},
  month = {Feb},
  publisher = {American Physical Society},
  doi = {10.1103/PhysRevLett.132.065106},
  url = {https://link.aps.org/doi/10.1103/PhysRevLett.132.065106}
}

@article{GRANDGIRARD2006GYSELA,
title = {A drift-kinetic Semi-Lagrangian 4D code for ion turbulence simulation},
journal = {Journal of Computational Physics},
volume = {217},
number = {2},
pages = {395-423},
year = {2006},
issn = {0021-9991},
doi = {https://doi.org/10.1016/j.jcp.2006.01.023},
url = {https://www.sciencedirect.com/science/article/pii/S0021999106000155},
author = {V. Grandgirard and M. Brunetti and P. Bertrand and N. Besse and X. Garbet and P. Ghendrih and G. Manfredi and Y. Sarazin and O. Sauter and E. Sonnendrücker and J. Vaclavik and L. Villard}
}

@article{ChenZona2016RMP_AE,
  title = {Physics of Alfv\'en waves and energetic particles in burning plasmas},
  author = {Chen, Liu and Zonca, Fulvio},
  journal = {Rev. Mod. Phys.},
  volume = {88},
  issue = {1},
  pages = {015008},
  numpages = {72},
  year = {2016},
  month = {Mar},
  publisher = {American Physical Society},
  doi = {10.1103/RevModPhys.88.015008},
  url = {https://link.aps.org/doi/10.1103/RevModPhys.88.015008}
}

@article{Li2024,
    author = {Li, Jingchun and Lin, Z. and Cheng, J. and Wu, Z. X. and Xu, Jianqiang and He, Y. and Huang, Z. H. and Liang, A. S. and Sun, T. F. and Dong, J. Q. and Shi, Z. B. and Zhong, Wulyv and Xu, M. and HL-2A Team},
    title = {Effects of resonant magnetic perturbations on turbulence and flows in the edge of HL-2A plasmas},
    journal = {Physics of Plasmas},
    volume = {31},
    number = {4},
    pages = {042502},
    year = {2024},
    month = {04},
    issn = {1070-664X},
    doi = {10.1063/5.0191468},
    url = {https://doi.org/10.1063/5.0191468},
    eprint = {https://pubs.aip.org/aip/pop/article-pdf/doi/10.1063/5.0191468/19861797/042502_1_5.0191468.pdf},
}

@article{Ma_2023,
doi = {10.1088/1741-4326/acc116},
url = {https://doi.org/10.1088/1741-4326/acc116},
year = {2023},
month = {mar},
publisher = {IOP Publishing},
volume = {63},
number = {5},
pages = {056014},
author = {Ma, Yuehao and Zhang, Bin and Bao, Jian and Lin, Z. and Zhang, Wenlu and Cai, Huishan and Li, Ding},
title = {Electrostatic turbulence in EAST plasmas with internal transport barrier},
journal = {Nuclear Fusion}
}

@article{Fil2021,
    author = {Fil, N. and Sharapov, S. E. and Fitzgerald, M. and Choi, G. J. and Lin, Z. and Tinguely, R. A. and Oliver, H. J. C. and McClements, K. G. and Puglia, P. G. and Dumont, R. J. and Porkolab, M. and Mailloux, J. and Joffrin, E. and JET Contributors},
    title = {Interpretation of electromagnetic modes in the sub-TAE frequency range in JET plasmas with elevated monotonic q-profiles},
    journal = {Physics of Plasmas},
    volume = {28},
    number = {10},
    pages = {102511},
    year = {2021},
    month = {10},
    issn = {1070-664X},
    doi = {10.1063/5.0057844},
    url = {https://doi.org/10.1063/5.0057844},
    eprint = {https://pubs.aip.org/aip/pop/article-pdf/doi/10.1063/5.0057844/15881149/102511_1_online.pdf},
}

@article{Singh_2025,
doi = {10.1088/1741-4326/ae0801},
url = {https://doi.org/10.1088/1741-4326/ae0801},
year = {2025},
month = {sep},
publisher = {IOP Publishing},
volume = {65},
number = {10},
pages = {106039},
author = {Singh, Tajinder and Rafiq, Tariq and Schuster, Eugenio and Lin, Zhihong and Kuley, Animesh},
title = {Global gyrokinetic simulations of kinetic ballooning mode in NSTX-U plasmas},
journal = {Nuclear Fusion}
}

@article{WangWH2024,
    author = {Wang, W. H. and Wei, X. S. and Lin, Z. and Lau, C. and Dettrick, S. and Tajima, T.},
    title = {A gyrokinetic simulation model for 2D equilibrium potential in the scrape-off layer of a field-reversed configuration},
    journal = {Physics of Plasmas},
    volume = {31},
    number = {7},
    pages = {072507},
    year = {2024},
    month = {07},
    issn = {1070-664X},
    doi = {10.1063/5.0189761},
    url = {https://doi.org/10.1063/5.0189761},
    eprint = {https://pubs.aip.org/aip/pop/article-pdf/doi/10.1063/5.0189761/20076424/072507_1_5.0189761.pdf},
}

@article{WangHY2020,
    author = {Wang, H. Y. and Holod, I. and Lin, Z. and Bao, J. and Fu, J. Y. and Liu, P. F. and Nicolau, J. H. and Spong, D. and Xiao, Y.},
    title = {Global gyrokinetic particle simulations of microturbulence in W7-X and LHD stellarators},
    journal = {Physics of Plasmas},
    volume = {27},
    number = {8},
    pages = {082305},
    year = {2020},
    month = {08},
    issn = {1070-664X},
    doi = {10.1063/5.0014198},
    url = {https://doi.org/10.1063/5.0014198},
    eprint = {https://pubs.aip.org/aip/pop/article-pdf/doi/10.1063/5.0014198/16040963/082305_1_online.pdf},
}

@article{Nicolau_2025,
doi = {10.1088/1741-4326/adf1b7},
url = {https://doi.org/10.1088/1741-4326/adf1b7},
year = {2025},
month = {jul},
publisher = {IOP Publishing},
volume = {65},
number = {8},
pages = {086049},
author = {Nicolau, Javier H. and Wei, Xishuo and Liu, Pengfei and Choi, Gyungjin and Lin, Zhihong},
title = {Trapped electron mode in a quasi-isodynamic stellarator},
journal = {Nuclear Fusion}
}

@article{Huang2025,
    author = {Huang, Handi and Gorelenkov, Nikolai and Wei, Xishuo and Lin, Zhihong and Duarte, Vinicius and Kaye, Stanley and Romanelli, Michele},
    title = {Verification of global gyrokinetic simulation of low frequency mode excited by thermal plasma in spherical tokamak},
    journal = {submitted to Nuclear Fusion} ,
    year = {2015}
}

@article{Wong2024,
    author = {Wong, H. H. and Huang, H. and Liu, P. and Yu, Y. and Wei, X. and Brochard, G. and Fil, N. and Lin, Z. and Podesta, M. and Bonofiglo, P. J. and McClements, K. G. and Michael, C. A. and Crocker, N. A. and Garzotti, L. and Carter, T. A.},
    title = {Linear gyrokinetic simulations of toroidal Alfvén eigenmodes in the Mega-Amp Spherical Tokamak},
    journal = {Physics of Plasmas},
    volume = {31},
    number = {11},
    pages = {112508},
    year = {2024},
    month = {11},
    issn = {1070-664X},
    doi = {10.1063/5.0238302},
    url = {https://doi.org/10.1063/5.0238302},
    eprint = {https://pubs.aip.org/aip/pop/article-pdf/doi/10.1063/5.0238302/20267123/112508_1_5.0238302.pdf},
}

@article{Singh_2024_ADITYA,
doi = {10.1088/1741-4326/ad5a20},
url = {https://doi.org/10.1088/1741-4326/ad5a20},
year = {2024},
month = {jun},
publisher = {IOP Publishing},
volume = {64},
number = {8},
pages = {086038},
author = {Singh, Tajinder and Shah, Kajal and Sharma, Deepti and Ghosh, Joydeep and Jadeja, Kumarpalsinh A. and Tanna, Rakesh L. and Chowdhuri, M.B. and Lin, Zhihong and Sen, Abhijit and Sharma, Sarveshwar and Kuley, Animesh},
title = {Gyrokinetic simulations of electrostatic microturbulence in ADITYA-U tokamak with argon impurity},
journal = {Nuclear Fusion}
}

@article{Liu_2024,
doi = {10.1088/1741-4326/ad4809},
url = {https://doi.org/10.1088/1741-4326/ad4809},
year = {2024},
month = {may},
publisher = {IOP Publishing},
volume = {64},
number = {7},
pages = {076007},
author = {Liu, P. and Wei, X. and Lin, Z. and Heidbrink, W.W and Brochard, G. and Choi, G.J. and Nicolau, J.H. and Zhang, W.},
title = {Cross-scale interaction between microturbulence and meso-scale reversed shear Alfvén eigenmodes in DIII-D plasmas},
journal = {Nuclear Fusion}
}

@article{Taimourzadeh_2019,
doi = {10.1088/1741-4326/ab0c38},
url = {https://doi.org/10.1088/1741-4326/ab0c38},
year = {2019},
month = {apr},
publisher = {IOP Publishing},
volume = {59},
number = {6},
pages = {066006},
author = {Taimourzadeh, S. and Bass, E.M. and Chen, Y. and Collins, C. and Gorelenkov, N.N. and Könies, A. and Lu, Z.X. and Spong, D.A. and Todo, Y. and Austin, M.E. and Bao, J. and Biancalani, A. and Borchardt, M. and Bottino, A. and Heidbrink, W.W. and Kleiber, R. and Lin, Z. and Mishchenko, A. and Shi, L. and Varela, J. and Waltz, R.E. and Yu, G. and Zhang, W.L. and Zhu, Y.},
title = {Verification and validation of integrated simulation of energetic particles in fusion plasmas},
journal = {Nuclear Fusion}
}

@article{WangZX2013PRL,
  title = {Radial Localization of Toroidicity-Induced Alfv\'en Eigenmodes},
  author = {Wang, Zhixuan and Lin, Zhihong and Holod, Ihor and Heidbrink, W. W. and Tobias, Benjamin and Van Zeeland, Michael and Austin, M. E.},
  journal = {Phys. Rev. Lett.},
  volume = {111},
  issue = {14},
  pages = {145003},
  numpages = {5},
  year = {2013},
  month = {Oct},
  publisher = {American Physical Society},
  doi = {10.1103/PhysRevLett.111.145003},
  url = {https://link.aps.org/doi/10.1103/PhysRevLett.111.145003}
}

@InProceedings{ZhangWL2019,
author="Zhang, Wenlu
and Joubert, Wayne
and Wang, Peng
and Wang, Bei
and Tang, William
and Niemerg, Matthew
and Shi, Lei
and Taimourzadeh, Sam
and Bao, Jian
and Lin, Zhihong",
editor="Chandrasekaran, Sunita
and Juckeland, Guido
and Wienke, Sandra",
title="Heterogeneous Programming and Optimization of Gyrokinetic Toroidal Code Using Directives",
booktitle="Accelerator Programming Using Directives",
year="2019",
publisher="Springer International Publishing",
address="Cham",
pages="3--21",
isbn="978-3-030-12274-4"
}

@article{Ethier_2005,
doi = {10.1088/1742-6596/16/1/001},
url = {https://doi.org/10.1088/1742-6596/16/1/001},
year = {2005},
month = {jan},
publisher = {},
volume = {16},
number = {1},
pages = {1},
author = {S Ethier and W M Tang and Z Lin},
title = {Gyrokinetic particle-in-cell simulations of plasma microturbulence on advanced computing platforms},
journal = {Journal of Physics: Conference Series}
}

@article{Lin2002PRL,
  title = {Size Scaling of Turbulent Transport in Magnetically Confined Plasmas},
  author = {Lin, Z. and Ethier, S. and Hahm, T. S. and Tang, W. M.},
  journal = {Phys. Rev. Lett.},
  volume = {88},
  issue = {19},
  pages = {195004},
  numpages = {4},
  year = {2002},
  month = {Apr},
  publisher = {American Physical Society},
  doi = {10.1103/PhysRevLett.88.195004},
  url = {https://link.aps.org/doi/10.1103/PhysRevLett.88.195004}
}

@article{Na2025Nature,
	author = {Na, Yong-Su and Hahm, T. S. and Diamond, P. H. and Di Siena, A. and Garcia, J. and Lin, Z.},
	date = {2025/04/01},
	doi = {10.1038/s42254-025-00814-8},
	id = {Na2025},
	isbn = {2522-5820},
	journal = {Nature Reviews Physics},
	number = {4},
	pages = {190--202},
	title = {How fast ions mitigate turbulence and enhance confinement in tokamak fusion plasmas},
	url = {https://doi.org/10.1038/s42254-025-00814-8},
	volume = {7},
	year = {2025}}

@article{Chen_2023,
doi = {10.1088/1741-4326/acf230},
url = {https://doi.org/10.1088/1741-4326/acf230},
year = {2023},
month = {sep},
publisher = {IOP Publishing},
volume = {63},
number = {10},
pages = {106016},
author = {Chen, Liu and Qiu, Zhiyong and Zonca, Fulvio},
title = {On nonlinear scattering of drift wave by toroidal Alfvén eigenmode in tokamak plasmas},
journal = {Nuclear Fusion}
}

@article{ZhangWL2008PRL,
  title = {Transport of Energetic Particles by Microturbulence in Magnetized Plasmas},
  author = {Zhang, Wenlu and Lin, Zhihong and Chen, Liu},
  journal = {Phys. Rev. Lett.},
  volume = {101},
  issue = {9},
  pages = {095001},
  numpages = {4},
  year = {2008},
  month = {Aug},
  publisher = {American Physical Society},
  doi = {10.1103/PhysRevLett.101.095001},
  url = {https://link.aps.org/doi/10.1103/PhysRevLett.101.095001}
}

@article{Zonca_2015,
doi = {10.1088/1367-2630/17/1/013052},
url = {https://doi.org/10.1088/1367-2630/17/1/013052},
year = {2015},
month = {jan},
publisher = {IOP Publishing},
volume = {17},
number = {1},
pages = {013052},
author = {Zonca, F and Chen, L and Briguglio, S and Fogaccia, G and Vlad, G and Wang, X},
title = {Nonlinear dynamics of phase space zonal structures and energetic particle physics in fusion plasmas},
journal = {New Journal of Physics}
}

@article{Salewski_2025,
doi = {10.1088/1741-4326/adb763},
url = {https://doi.org/10.1088/1741-4326/adb763},
year = {2025},
month = {mar},
publisher = {IOP Publishing},
volume = {65},
number = {4},
pages = {043002},
author = {Salewski, M. and Spong, D.A. and Aleynikov, P. and Bilato, R. and Breizman, B.N. and Briguglio, S. and Cai, H. and Chen, L. and Chen, W. and Duarte, V.N. and Dumont, R.J. and Falessi, M.V. and Fitzgerald, M. and Fredrickson, E.D. and García-Muñoz, M. and Gorelenkov, N.N. and Hayward-Schneider, T. and Heidbrink, W.W. and Hole, M.J. and Kazakov, Ye.O. and Kiptily, V.G. and Könies, A. and Kurki-Suonio, T. and Lauber, Ph. and Lazerson, S.A. and Lin, Z. and Mishchenko, A. and Moseev, D. and Muscatello, C.M. and Nocente, M. and Podestà, M. and Polevoi, A. and Schneider, M. and Sharapov, S.E. and Snicker, A. and Todo, Y. and Qiu, Z. and Vlad, G. and Wang, X. and Zarzoso, D. and Van Zeeland, M.A. and Zonca, F. and Pinches, S.D.},
title = {Energetic particle physics: Chapter 7 of the special issue: on the path to tokamak burning plasma operation},
journal = {Nuclear Fusion}
}

@article{OBREJAN2017GKNET,
title = {Development of a new zonal flow equation solver by diagonalisation and its application in non-circular cross-section tokamak plasmas},
journal = {Computer Physics Communications},
volume = {216},
pages = {8-17},
year = {2017},
issn = {0010-4655},
doi = {https://doi.org/10.1016/j.cpc.2017.02.010},
url = {https://www.sciencedirect.com/science/article/pii/S0010465517300620},
author = {Kevin Obrejan and Kenji Imadera and Jiquan Li and Yasuaki Kishimoto}
}

@article{Watanabe_2006,
doi = {10.1088/0029-5515/46/1/003},
url = {https://doi.org/10.1088/0029-5515/46/1/003},
year = {2005},
month = {dec},
publisher = {},
volume = {46},
number = {1},
pages = {24},
author = {Watanabe, T.-H. and Sugama, H.},
title = {Velocity–space structures of distribution function in toroidal ion temperature gradient turbulence},
journal = {Nuclear Fusion}
}

@article{IDOMURA2008GT5D,
title = {Conservative global gyrokinetic toroidal full-f five-dimensional Vlasov simulation},
journal = {Computer Physics Communications},
volume = {179},
number = {6},
pages = {391-403},
year = {2008},
issn = {0010-4655},
doi = {https://doi.org/10.1016/j.cpc.2008.04.005},
url = {https://www.sciencedirect.com/science/article/pii/S0010465508001409},
author = {Yasuhiro Idomura and Masato Ida and Takuma Kano and Nobuyuki Aiba and Shinji Tokuda}
}

@article{Jenko2000GENE,
    author = {Jenko, F. and Dorland, W. and Kotschenreuther, M. and Rogers, B. N.},
    title = {Electron temperature gradient driven turbulence},
    journal = {Physics of Plasmas},
    volume = {7},
    number = {5},
    pages = {1904-1910},
    year = {2000},
    month = {05},
    issn = {1070-664X},
    doi = {10.1063/1.874014},
    url = {https://doi.org/10.1063/1.874014},
    eprint = {https://pubs.aip.org/aip/pop/article-pdf/7/5/1904/19273118/1904_1_online.pdf},
}

@article{CANDY2003GYRO,
title = {An Eulerian gyrokinetic-Maxwell solver},
journal = {Journal of Computational Physics},
volume = {186},
number = {2},
pages = {545-581},
year = {2003},
issn = {0021-9991},
doi = {https://doi.org/10.1016/S0021-9991(03)00079-2},
url = {https://www.sciencedirect.com/science/article/pii/S0021999103000792},
author = {J. Candy and R.E. Waltz}
}

@article{KLEIBER2024EUTERPE,
title = {EUTERPE: A global gyrokinetic code for stellarator geometry},
journal = {Computer Physics Communications},
volume = {295},
pages = {109013},
year = {2024},
issn = {0010-4655},
doi = {https://doi.org/10.1016/j.cpc.2023.109013},
url = {https://www.sciencedirect.com/science/article/pii/S0010465523003582},
author = {R. Kleiber and M. Borchardt and R. Hatzky and A. Könies and H. Leyh and A. Mishchenko and J. Riemann and C. Slaby and J.M. García-Regaña and E. Sánchez and M. Cole}
}

@article{JOLLIET2007ORB5,
title = {A global collisionless PIC code in magnetic coordinates},
journal = {Computer Physics Communications},
volume = {177},
number = {5},
pages = {409-425},
year = {2007},
issn = {0010-4655},
doi = {https://doi.org/10.1016/j.cpc.2007.04.006},
url = {https://www.sciencedirect.com/science/article/pii/S0010465507002251},
author = {S. Jolliet and A. Bottino and P. Angelino and R. Hatzky and T.M. Tran and B.F. Mcmillan and O. Sauter and K. Appert and Y. Idomura and L. Villard}
}

@article{Choi_2021,
doi = {10.1088/1741-4326/abf0dd},
url = {https://doi.org/10.1088/1741-4326/abf0dd},
year = {2021},
month = {apr},
publisher = {IOP Publishing},
volume = {61},
number = {6},
pages = {066007},
author = {Choi, G.J. and Liu, P. and Wei, X.S. and Nicolau, J.H. and Dong, G. and Zhang, W.L. and Lin, Z. and Heidbrink, W.W. and Hahm, T.S.},
title = {Gyrokinetic simulation of low-frequency Alfvénic modes in DIII-D tokamak},
journal = {Nuclear Fusion}
}

@inproceedings{LinIAEA2023,
  title = {Prediction of Energetic Particle Confinement in {{ITER}} Operation Scenarios},
  booktitle = {Proceedings of the 29th {{International Conference}} on {{Plasma Physics}} and {{Controlled Nuclear Fusion Research}}},
  author = {Lin, Z. and Bass, E. and Brochard, G. and Ghai, Y. and Gorelenkov, N. and Idouakass, M. and Liu, C. and Liu, P. and Podesta, M. and Spong, D. and Wei, X. and Heidbrink, W. and McKee, G. and Waltz, R. E. and Bao, J. and Cornille, B. and Duarte, V. N. and Falgout, R. and Gorelenkova, M. and {Hayward-Schneider} and Kim, S.H. and Joubert, W. and Klasky, S. and Lyngaas, I. and Mehta, K. and Nicolau, J. H. and Pinches, S. D. and Polevoi, A.R. and Schneider, M. and Tang, W. and Wang, P. and Williams, S.},
  year = {2023},
  publisher = {International Atomic Energy Agency},
  address = {London}
}

@article{Bass2010PoP,
    author = {Bass, E. M. and Waltz, R. E.},
    title = {Gyrokinetic simulations of mesoscale energetic particle-driven Alfvénic turbulent transport embedded in microturbulence},
    journal = {Physics of Plasmas},
    volume = {17},
    number = {11},
    pages = {112319},
    year = {2010},
    month = {11},
    issn = {1070-664X},
    doi = {10.1063/1.3509106},
    url = {https://doi.org/10.1063/1.3509106},
    eprint = {https://pubs.aip.org/aip/pop/article-pdf/doi/10.1063/1.3509106/16066011/112319_1_online.pdf},
}

@misc{Ku2018XGC,
title = {XGC},
author = {Ku, Seung-Hoe and Hager, Robert and Scheinberg, Aaron and Dominski, Julien and Sharma, Amil and Churchill, Michael and Choi, Jong and Sturdevant, Ben and Mollén, Albert and Wilkie, George and Chang, Choong-Seock and Yoon, Eisung and Adams, Mark and Seo, Janghoon and Koh, Sehoon and D'Azevedo, Eduardo and Abbott, Steve and Worley, Patrick H. and Ethier, Stephane and Park, Gunyoung and Lang, Jianying and MacKie-Mason, Brian and Germaschewski, Kai and Suchyta, Eric and Carey, Varis and Cole, Michael and Trivedi, Pallavi and Chowdhury, Jugal},
abstractNote = {XGC (X-point included gyrokinetic code), Version 3 Primary Author: Seunghoe Ku (PPPL, sku@pppl.gov), Robert Hager (PPPL, rhager@pppl.gov) 31 July 2017 XGC is a gyrokinetic particle-in-cell code, which is sepeciallized in tokamak edge simulatin. The simulation domain can include the magnetic separatrix, magnetic axis and the biased material wall. XGC can run in full-f, total-delta-f, and conventional delta-f mode. The ion is kinetic always except ETG simulation. The electron can be adiabatic, fluid, drift kinetic, or gyrokinetic (for ETG). XGC is written in Fortran 90 and is designed for HPCs utilizing MPI, OpenMP, CUDA (GPU), OpenACC (GPU), and vectorization (Intel MIC-KNL). The weak scaling is roughly linear to the maximal nodes of leading HPCs in US. There several versions of XGCs for different purpose: XGC0 is earlist version of XGC0 which is designed for neoclassical transport in the tokamak edge. It uses full-f method, and 00-mode electrostatic field is solved only. RMP and current response can be calculated with coupling with M3D. XGC1 is for turbulence simulation with low parallel wavenumber. Piecewise field following coordinates are used to handle low k-parallel perturbation with small (~64) number of toroidal resolution. In total-delta-f method, it utilizes phase space grid in addition to particles, to handel non-maxwellian distribution in the tokamak edge. XGCa is axisymmetric version of XGC1 to simulate neoclassical transport. References: [1] S. Ku et al., Nuclear Fusion 49, 115021 (2009) [2] S. Ku, R. Hager, C.S. Chang et al., J. Comp. Physics, 315, 467 (2016) https://doi.org/10.1016/j.jcp.2016.03.062 [2] R. Hager. E.S. Yoon. S.Ku et al., J. Comp. Physics, 315, 644 (2016) https://doi.org/10.1016/j.jcp.2016.03.064 [3] R. Hager, J. Lang et al., Phys. Plasmas 24,054508 (2017) http://dx.doi.org/10.1063/1.4983320},
doi = {10.11578/dc.20180627.11},
url = {https://doi.org/10.11578/dc.20180627.11},
howpublished = {[Computer Software] \url{https://doi.org/10.11578/dc.20180627.11}},
year = {2018},
month = {jun}
}

@article{PEETERS2009GKW,
title = {The nonlinear gyro-kinetic flux tube code GKW},
journal = {Computer Physics Communications},
volume = {180},
number = {12},
pages = {2650-2672},
year = {2009},
note = {40 YEARS OF CPC: A celebratory issue focused on quality software for high performance, grid and novel computing architectures},
issn = {0010-4655},
doi = {https://doi.org/10.1016/j.cpc.2009.07.001},
url = {https://www.sciencedirect.com/science/article/pii/S0010465509002112},
author = {A.G. Peeters and Y. Camenen and F.J. Casson and W.A. Hornsby and A.P. Snodin and D. Strintzi and G. Szepesi}
}

@article{CHEN2007GEM,
title = {Electromagnetic gyrokinetic $\delta f$ particle-in-cell turbulence simulation with realistic equilibrium profiles and geometry},
journal = {Journal of Computational Physics},
volume = {220},
number = {2},
pages = {839-855},
year = {2007},
issn = {0021-9991},
doi = {https://doi.org/10.1016/j.jcp.2006.05.028},
url = {https://www.sciencedirect.com/science/article/pii/S0021999106002634},
author = {Yang Chen and Scott E. Parker}
}

@article{MaYuehao2025,
    author = {Ma, Yuehao and Liu, Pengfei and Bao, Jian and Lin, Zhihong and Cai, Huishan},
    title = {Electromagnetic turbulence in EAST plasmas with internal transport barrier}, 
    journal = {Submitted to Nulcear Fusion},
    year = 2026
}

@article{REWOLDT2007,
title = {Linear comparison of gyrokinetic codes with trapped electrons},
journal = {Computer Physics Communications},
volume = {177},
number = {10},
pages = {775-780},
year = {2007},
issn = {0010-4655},
doi = {https://doi.org/10.1016/j.cpc.2007.06.017},
url = {https://www.sciencedirect.com/science/article/pii/S0010465507003268},
author = {G. Rewoldt and Z. Lin and Y. Idomura}
}

@article{Singh_2022,
doi = {10.1088/1741-4326/ac906d},
url = {https://doi.org/10.1088/1741-4326/ac906d},
year = {2022},
month = {oct},
publisher = {IOP Publishing},
volume = {62},
number = {12},
pages = {126006},
author = {Singh, Tajinder and Nicolau, Javier H. and Lin, Zhihong and Sharma, Sarveshwar and Sen, Abhijit and Kuley, Animesh},
title = {Global gyrokinetic simulations of electrostatic microturbulent transport using kinetic electrons in LHD stellarator},
journal = {Nuclear Fusion}
}

@article{Lee_2016,
    author = {Lee, W. W.},
    title = {Magnetohydrodynamics for collisionless plasmas from the gyrokinetic perspective},
    journal = {Physics of Plasmas},
    volume = {23},
    number = {7},
    pages = {070705},
    year = {2016},
    month = {07},
    issn = {1070-664X},
    doi = {10.1063/1.4960029},
    url = {https://doi.org/10.1063/1.4960029},
    eprint = {https://pubs.aip.org/aip/pop/article-pdf/doi/10.1063/1.4960029/14126779/070705_1_online.pdf},
}

@article{Frieman_1966,
    author = {Frieman, Edward and Davidson, Ronald and Langdon, Bruce},
    title = {Higher‐Order Corrections to the Chew‐Goldberger‐Low Theory},
    journal = {The Physics of Fluids},
    volume = {9},
    number = {8},
    pages = {1475-1482},
    year = {1966},
    month = {08},
    issn = {0031-9171},
    doi = {10.1063/1.1761881},
    url = {https://doi.org/10.1063/1.1761881},
    eprint = {https://pubs.aip.org/aip/pfl/article-pdf/9/8/1475/12549736/1475_1_online.pdf},
}

@article{CGL_1956,
    author = {Chew, G. F. and Goldberger, M. L. and Low, F. E.},
    title = {The Boltzmann equation an d the one-fluid hydromagnetic equations in the absence of particle collisions},
    journal = {Proceedings of the Royal Society of London. A. Mathematical and Physical Sciences},
    volume = {236},
    number = {1204},
    pages = {112-118},
    year = {1956},
    month = {07},
    issn = {0080-4630},
    doi = {10.1098/rspa.1956.0116},
    url = {https://doi.org/10.1098/rspa.1956.0116},
    eprint = {https://royalsocietypublishing.org/rspa/article-pdf/236/1204/112/49767/rspa.1956.0116.pdf},
}

@techreport{Kulsrud_1980,
  author       = {Kulsrud, R M},
  title        = {MHD description of plasma:  handbook of plasma physics},
  institution  = {Princeton Univ., NJ (USA). Plasma Physics Lab.},
  doi          = {10.2172/7072860},
  url          = {https://www.osti.gov/biblio/7072860},
  place        = {United States},
  year         = {1980},
  month        = {10}}

@article{Lee_Qin_2003,
    author = {Lee, W. W. and Qin, H.},
    title = {Alfvén waves in gyrokinetic plasmas},
    journal = {Physics of Plasmas},
    volume = {10},
    number = {8},
    pages = {3196-3203},
    year = {2003},
    month = {08},
    issn = {1070-664X},
    doi = {10.1063/1.1590666},
    url = {https://doi.org/10.1063/1.1590666},
    eprint = {https://pubs.aip.org/aip/pop/article-pdf/10/8/3196/19271127/3196_1_online.pdf},
}

@article{CANDY2016CGYRO,
title = {A high-accuracy Eulerian gyrokinetic solver for collisional plasmas},
journal = {Journal of Computational Physics},
volume = {324},
pages = {73-93},
year = {2016},
issn = {0021-9991},
doi = {https://doi.org/10.1016/j.jcp.2016.07.039},
url = {https://www.sciencedirect.com/science/article/pii/S0021999116303400},
author = {J. Candy and E.A. Belli and R.V. Bravenec}
}

@article{KOTSCHENREUTHER1995GS2,
title = {Comparison of initial value and eigenvalue codes for kinetic toroidal plasma instabilities},
journal = {Computer Physics Communications},
volume = {88},
number = {2},
pages = {128-140},
year = {1995},
issn = {0010-4655},
doi = {https://doi.org/10.1016/0010-4655(95)00035-E},
url = {https://www.sciencedirect.com/science/article/pii/001046559500035E},
author = {Mike Kotschenreuther and G. Rewoldt and W.M. Tang}
}

@article{Startsev_GTS,
    author = {Startsev, Edward A. and Wang, Weixing and Yoo, Min-Gu and Chen, Jin and Ethier, Stephane},
    title = {Verification of electromaENEnetic simulation capabilities in global gyrokinetic particle-in-cell code GTS},
    journal = {Physics of Plasmas},
    volume = {31},
    number = {11},
    pages = {113902},
    year = {2024},
    month = {11},
    issn = {1070-664X},
    doi = {10.1063/5.0217324},
    url = {https://doi.org/10.1063/5.0217324},
    eprint = {https://pubs.aip.org/aip/pop/article-pdf/doi/10.1063/5.0217324/20247047/113902_1_5.0217324.pdf},
}

@article{BARNES2019STELLA,
title = {stella: An operator-split, implicit–explicit $\delta$f-gyrokinetic code for general magnetic field configurations},
journal = {Journal of Computational Physics},
volume = {391},
pages = {365-380},
year = {2019},
issn = {0021-9991},
doi = {https://doi.org/10.1016/j.jcp.2019.01.025},
url = {https://www.sciencedirect.com/science/article/pii/S002199911930066X},
author = {M. Barnes and F.I. Parra and M. Landreman}
}

@article{EUTERPE_2025_GK_IN_STELLARATOR, title={Gyrokinetic simulations of magnetohydrodynamic modes in stellarator plasmas}, volume={91}, DOI={10.1017/S0022377825000492}, number={4}, journal={Journal of Plasma Physics}, author={Nührenberg, Carolin and Kleiber, R. and Mishchenko, A. and Könies, A. and Borchardt, M. and Hatzky, R.}, year={2025}, pages={E93}}

@article{Li_2025_dbpara,
doi = {10.1088/1741-4326/ae06b3},
url = {https://doi.org/10.1088/1741-4326/ae06b3},
year = {2025},
month = {sep},
publisher = {IOP Publishing},
volume = {65},
number = {10},
pages = {106040},
author = {Li, P.-Y. and Hatch, D.R. and Parisi, J.F. and Lampert, M. and Belli, E.A. and Kotschenreuther, M. and Mahajan, S.M.},
title = {Importance of $\delta B_{||}$ on ETG stability, turbulence, and transport in NSTX},
journal = {Nuclear Fusion}
}

@article{MISHCHENKO2019,
title = {Pullback scheme implementation in ORB5},
journal = {Computer Physics Communications},
volume = {238},
pages = {194-202},
year = {2019},
issn = {0010-4655},
doi = {https://doi.org/10.1016/j.cpc.2018.12.002},
url = {https://www.sciencedirect.com/science/article/pii/S0010465518304181},
author = {A. Mishchenko and A. Bottino and A. Biancalani and R. Hatzky and T. Hayward-Schneider and N. Ohana and E. Lanti and S. Brunner and L. Villard and M. Borchardt and R. Kleiber and A. Könies}
}

@article{GENE_CANCELLATION,
title = {The global version of the gyrokinetic turbulence code GENE},
journal = {Journal of Computational Physics},
volume = {230},
number = {18},
pages = {7053-7071},
year = {2011},
issn = {0021-9991},
doi = {https://doi.org/10.1016/j.jcp.2011.05.034},
url = {https://www.sciencedirect.com/science/article/pii/S0021999111003457},
author = {T. Görler and X. Lapillonne and S. Brunner and T. Dannert and F. Jenko and F. Merz and D. Told}
}

@article{GENE_TEM_NTM,
  title = {Turbulent Multiscale Interactions between Tearing Modes, Trapped-Electron Modes, and Zonal Flows},
  author = {Jitsuk, T. and Pueschel, M. J. and Terry, P. W. and Di Siena, A.},
  journal = {Phys. Rev. Lett.},
  volume = {136},
  issue = {1},
  pages = {015101},
  numpages = {7},
  year = {2026},
  month = {Jan},
  publisher = {American Physical Society},
  doi = {10.1103/dlct-5vwy},
  url = {https://link.aps.org/doi/10.1103/dlct-5vwy}
}

@article{GEM_resplitting,
title = {Re-splitting $\delta$f method for electro-magnetic gyrokinetic particle-in-cell (PIC) simulation of tokamak plasmas},
journal = {Computer Physics Communications},
volume = {250},
pages = {107050},
year = {2020},
issn = {0010-4655},
doi = {https://doi.org/10.1016/j.cpc.2019.107050},
url = {https://www.sciencedirect.com/science/article/pii/S0010465519303790},
author = {Lei Ye and Yang Chen}
}

@article{Kennedy_2024,
doi = {10.1088/1741-4326/ad58f3},
url = {https://doi.org/10.1088/1741-4326/ad58f3},
year = {2024},
month = {jul},
publisher = {IOP Publishing},
volume = {64},
number = {8},
pages = {086049},
author = {Kennedy, D. and Roach, C.M. and Giacomin, M. and Ivanov, P.G. and Adkins, T. and Sheffield, F. and Görler, T. and Bokshi, A. and Dickinson, D. and Dudding, H.G. and Patel, B.S.},
title = {On the importance of parallel magnetic-field fluctuations for electromagnetic instabilities in STEP},
journal = {Nuclear Fusion}
}

@article{Sheffield_2025,
doi = {10.1088/1361-6587/ad9fde},
url = {https://doi.org/10.1088/1361-6587/ad9fde},
year = {2024},
month = {dec},
publisher = {IOP Publishing},
volume = {67},
number = {1},
pages = {015028},
author = {Sheffield, F and Görler, T and Wilms, F and Merlo, G and Jenko, F},
title = {Implementation of a long-wavelength model for parallel magnetic fluctuations in the global GENE code},
journal = {Plasma Physics and Controlled Fusion}
}

@article{Evans_2005_RMP,
doi = {10.1088/0029-5515/45/7/007},
url = {https://doi.org/10.1088/0029-5515/45/7/007},
year = {2005},
month = {jun},
publisher = {},
volume = {45},
number = {7},
pages = {595},
author = {Evans, T.E. and Moyer, R.A. and Watkins, J.G. and Osborne, T.H. and Thomas, P.R. and Becoulet, M. and Boedo, J.A. and Doyle, E.J. and Fenstermacher, M.E. and Finken, K.H. and Groebner, R.J. and Groth, M. and Harris, J.H. and Jackson, G.L. and La Haye, R.J. and Lasnier, C.J. and Masuzaki, S. and Ohyabu, N. and Pretty, D.G. and Reimerdes, H. and Rhodes, T.L. and Rudakov, D.L. and Schaffer, M.J. and Wade, M.R. and Wang, G. and West, W.P. and Zeng, L.},
title = {Suppression of large edge localized modes with edge resonant magnetic fields in high confinement DIII-D plasmas},
journal = {Nuclear Fusion}
}

@article{Mishchenko2004MixtureVariable,
    author = {Mishchenko, Alexey and Könies, Axel and Kleiber, Ralf and Cole, Michael},
    title = {Pullback transformation in gyrokinetic electromagnetic simulations},
    journal = {Physics of Plasmas},
    volume = {21},
    number = {9},
    pages = {092110},
    year = {2014},
    month = {09},
    issn = {1070-664X},
    doi = {10.1063/1.4895501},
    url = {https://doi.org/10.1063/1.4895501},
    eprint = {https://pubs.aip.org/aip/pop/article-pdf/doi/10.1063/1.4895501/13778409/092110_1_online.pdf},
}

@article{Manuilskiy2000PoPSplitWeight,
    author = {Manuilskiy, Igor and Lee, W. W.},
    title = {The split-weight particle simulation scheme for plasmas},
    journal = {Physics of Plasmas},
    volume = {7},
    number = {5},
    pages = {1381-1385},
    year = {2000},
    month = {05},
    issn = {1070-664X},
    doi = {10.1063/1.873955},
    url = {https://doi.org/10.1063/1.873955},
    eprint = {https://pubs.aip.org/aip/pop/article-pdf/7/5/1381/19273221/1381_1_online.pdf},
}

@article{YChen_2020,
	author = {Yang Chen and Wenlu Zhang and Jian Bao and Zhihong Lin and Chao Dong and Jintao Cao and Ding Li},
	doi = {10.1088/0256-307X/37/9/095201},
	eid = {095201},
	journal = {Chinese Physics Letters},
	number = {9},
	pages = {095201},
	publisher = {Chin. Phys. Lett.},
	title = {Verification of Energetic-Particle-Induced Geodesic Acoustic Mode in Gyrokinetic Particle Simulations},
	url = {http://cpl.iphy.ac.cn/EN/abstract/article_105730.shtml},
	volume = {37},
	year = {2020}}

@article{Fang2019a,
    author = {Kaisheng Fang and Zhihong Lin},
    title = {Global gyrokinetic simulation of microturbulence with kinetic electrons in the presence of magnetic island in tokamak},
    journal = {Physics of Plasmas},
    year = {2019}
}

@article{Dimits1993,
title = {Partially Linearized Algorithms in Gyrokinetic Particle Simulation},
journal = {Journal of Computational Physics},
volume = {107},
number = {2},
pages = {309-323},
year = {1993},
issn = {0021-9991},
doi = {https://doi.org/10.1006/jcph.1993.1146},
url = {https://www.sciencedirect.com/science/article/pii/S0021999183711460},
author = {A.M. Dimits and W.W. Lee}
}

@article{WWLee1987,
title = {Gyrokinetic particle simulation model},
journal = {Journal of Computational Physics},
volume = {72},
number = {1},
pages = {243-269},
year = {1987},
issn = {0021-9991},
doi = {https://doi.org/10.1016/0021-9991(87)90080-5},
url = {https://www.sciencedirect.com/science/article/pii/0021999187900805},
author = {W.W Lee}
}

@article{Parker1993,
    author = {Parker, S. E. and Lee, W. W.},
    title = {A fully nonlinear characteristic method for gyrokinetic simulation},
    journal = {Physics of Fluids B: Plasma Physics},
    volume = {5},
    number = {1},
    pages = {77-86},
    year = {1993},
    month = {01},
    issn = {0899-8221},
    doi = {10.1063/1.860870},
    url = {https://doi.org/10.1063/1.860870},
    eprint = {https://pubs.aip.org/aip/pfb/article-pdf/5/1/77/12343795/77\_1\_online.pdf},
}

@article{Gaffey_1976,
	author = {Gaffey, John D.},
	doi = {10.1017/S0022377800020134},
	journal = {Journal of Plasma Physics},
	number = {2},
	pages = {149--169},
	publisher = {Cambridge University Press},
	title = {Energetic ion distribution resulting from neutral beam injection in tokamaks},
	volume = {16},
	year = {1976}}

@article{Lin_1995,
	author = {Lin, Z. and Lee, W. W.},
	doi = {10.1103/PhysRevE.52.5646},
	issue = {5},
	journal = {Phys. Rev. E},
	month = {Nov},
	numpages = {0},
	pages = {5646--5652},
	publisher = {American Physical Society},
	title = {Method for solving the gyrokinetic Poisson equation in general geometry},
	url = {https://link.aps.org/doi/10.1103/PhysRevE.52.5646},
	volume = {52},
	year = {1995}}

@article{Lin_Chen_2001,
	author = {Lin,Zhihong and Chen,Liu},
	doi = {10.1063/1.1356438},
	eprint = {https://doi.org/10.1063/1.1356438},
	journal = {Physics of Plasmas},
	number = {5},
	pages = {1447-1450},
	title = {A fluid--kinetic hybrid electron model for electromagnetic simulations},
	url = {https://doi.org/10.1063/1.1356438},
	volume = {8},
	year = {2001}}

@article{Lin_2007,
	author = {Z Lin and Y Nishimura and Y Xiao and I Holod and W L Zhang and L Chen},
	doi = {10.1088/0741-3335/49/12b/s15},
	journal = {Plasma Physics and Controlled Fusion},
	month = {nov},
	number = {12B},
	pages = {B163--B172},
	publisher = {{IOP} Publishing},
	title = {Global gyrokinetic particle simulations with kinetic electrons},
	url = {https://doi.org/10.1088/0741-3335/49/12b/s15},
	volume = {49},
	year = 2007}

@article{Bao_2017,
	author = {Bao,J. and Liu,D. and Lin,Z.},
	doi = {10.1063/1.4995455},
	eprint = {https://doi.org/10.1063/1.4995455},
	journal = {Physics of Plasmas},
	number = {10},
	pages = {102516},
	title = {A conservative scheme of drift kinetic electrons for gyrokinetic simulation of kinetic-MHD processes in toroidal plasmas},
	url = {https://doi.org/10.1063/1.4995455},
	volume = {24},
	year = {2017}}

@article{Fu_2006,
	author = {Fu,G. Y. and Berk,H. L.},
	doi = {10.1063/1.2196246},
	eprint = {https://doi.org/10.1063/1.2196246},
	journal = {Physics of Plasmas},
	number = {5},
	pages = {052502},
	title = {Effects of pressure gradient on existence of Alfv{\'e}n cascade modes in reversed shear tokamak plasmas},
	url = {https://doi.org/10.1063/1.2196246},
	volume = {13},
	year = {2006}}

@article{Bao_2021,
	author = {Bao,J. and Zhang,W. L. and Li,D. and Lin,Z.},
	doi = {10.1007/s10894-020-00275-0},
	journal = {Journal of Fusion Energy},
	pages = {382-389},
	title = {Effects of plasma diamagnetic drift on Alfv{\'e}n continua and discrete eigenmodes in tokamaks},
	url = {https://doi.org/10.1007/s10894-020-00275-0},
	volume = {39},
	year = {2020}}

@article{Berk_1977,
	author = {Berk, H. L. and Dominguez, R. R.},
	doi = {10.1017/S0022377800020869},
	journal = {Journal of Plasma Physics},
	number = {1},
	pages = {31--48},
	publisher = {Cambridge University Press},
	title = {Variational method for electromagnetic waves in a magneto-plasma},
	volume = {18},
	year = {1977}}

@article{Tang_1980,
	author = {W.M. Tang and J.W. Connor and R.J. Hastie},
	doi = {10.1088/0029-5515/20/11/011},
	journal = {Nuclear Fusion},
	month = {nov},
	number = {11},
	pages = {1439--1453},
	publisher = {{IOP} Publishing},
	title = {Kinetic-ballooning-mode theory in general geometry},
	url = {https://doi.org/10.1088/0029-5515/20/11/011},
	volume = {20},
	year = 1980}

@book{White_Book,
	author = {Roscoe B White},
	edition = 3,
	isbn = {1783263636},
	publisher = {IMPERIAL COLLEGE PRESS},
	title = {Theory of toroidally confined plasmas},
	year = 2013}
\end{document}